\documentclass[a4paper, 12pt]{article}

\usepackage{tikz}
\usepackage{latexsym,amsmath,amsfonts,amssymb}
\usepackage{amsthm}
\usepackage{mathrsfs}
\usepackage{multirow}
\usepackage[latin1]{inputenc}
\usepackage[american]{babel}
\usepackage{bbm}
\usepackage[nosort]{cite}
\usepackage{hyperref}
\hypersetup{colorlinks, citecolor=[rgb]{.7,0,0}, linkcolor=[rgb]{0,0,0.7}, urlcolor=[rgb]{0,0,0.5}}
\renewcommand{\baselinestretch}{1.2}
\newcommand{\Vol}{\mathrm{Vol}}
\newcommand{\ib}{{\bar \imath}}
\newcommand{\jb}{{\bar \jmath}}
\newcommand{\parfrac}[2]{\frac{\partial #1}{\partial #2}}

\newcommand{\ud}[2]{^{#1}_{\phantom{#1}#2}}
\newcommand{\du}[2]{_{#1}^{\phantom{#1}#2}}

\newcommand{\wt}{\widetilde}
\newcommand{\wh}{\widehat}
\newcommand{\wb}{\overline}
\newcommand{\matht}[1]{\texorpdfstring{\ensuremath{\boldsymbol{#1}}}{#1}}

\newcommand{\ds}{\displaystyle}

\newcommand{\ie}{\textit{i.e.}}

\numberwithin{equation}{section}

\newcommand{\nn}{\nonumber}
\newcommand{\mat}[1]{\begin{pmatrix} #1 \end{pmatrix}}

\newcommand{\be}{\begin{equation}} \newcommand{\ee}{\end{equation}}
\newcommand{\bea}{\begin{equation} \begin{aligned}} \newcommand{\eea}{\end{aligned} \end{equation}}
\newcommand{\ba}{\begin{array}} \newcommand{\ea}{\end{array}}

\newcommand{\cA}{\mathcal{A}}

\newcommand{\cC}{\mathcal{C}}
\newcommand{\cD}{\mathcal{D}}

\newcommand{\cF}{\mathcal{F}}
\newcommand{\cG}{\mathcal{G}}

\newcommand{\cI}{\mathcal{I}}

\newcommand{\cK}{\mathcal{K}}
\newcommand{\cL}{\mathcal{L}}
\newcommand{\ccL}{\mathscr{L}}
\newcommand{\cM}{\mathcal{M}}
\newcommand{\cN}{\mathcal{N}}
\newcommand{\cO}{\mathcal{O}}
\newcommand{\cP}{\mathcal{P}}
\newcommand{\cQ}{\mathcal{Q}}
\newcommand{\cR}{\mathcal{R}}
\newcommand{\cS}{\mathcal{S}}

\newcommand{\cU}{\mathcal{U}}
\newcommand{\cV}{\mathcal{V}}
\newcommand{\cW}{\mathcal{W}}

\newcommand{\cZ}{\mathcal{Z}}

\newcommand{\bB}{\mathbb{B}}
\newcommand{\bC}{\mathbb{C}}

\newcommand{\bH}{\mathbb{H}}

\newcommand{\bN}{\mathbb{N}}

\newcommand{\bP}{\mathbb{P}}
\newcommand{\bQ}{\mathbb{Q}}
\newcommand{\bR}{\mathbb{R}}

\newcommand{\bZ}{\mathbb{Z}}
\newcommand{\fJ}{\mathfrak{J}}
\newcommand{\fK}{\mathfrak{K}}

\newcommand{\fT}{\mathfrak{T}}

\newcommand{\sT}{{\sf{T}}}
\newcommand{\unit}{\mathbbm{1}}

\newcommand{\rSO}{\mathrm{SO}}

\newcommand{\rSU}{\mathrm{SU}}
\newcommand{\rU}{\mathrm{U}}

\newcommand{\fsu}{\mathfrak{su}}

\newcommand{\fsp}{\mathfrak{sp}}

\DeclareMathOperator{\Tr}{Tr}

\DeclareMathOperator{\re}{\mathbb{R}e}
\DeclareMathOperator{\im}{\mathbb{I}m}

\makeatletter
\def\blfootnote{\gdef\@thefnmark{}\@footnotetext}
\makeatother

\begin{document}

\thispagestyle{empty}
\begin{flushright}
SISSA  11/2020/FISI
\end{flushright}
\vspace{13mm}  %\vspace{10mm}
\begin{center}
{\huge  Superconformal indices at large $N$ \\[.5em] and the entropy of AdS$_5 \times \mathrm{SE}_5$ black holes
}
\\[13mm]
{\large Francesco Benini$^{1,2,3}$, Edoardo Colombo$^4$, Saman Soltani$^{1,2}$, \\[.5em]
Alberto Zaffaroni$^{4,5}$, and Ziruo Zhang$^{1,2}$}
 
\bigskip
{\it
$^1$ SISSA, Via Bonomea 265, 34136 Trieste, Italy \\[.2em]
$^2$ INFN, Sezione di Trieste, Via Valerio 2, 34127 Trieste, Italy \\[.2em]
$^3$ ICTP, Strada Costiera 11, 34151 Trieste, Italy \\[.2em]
$^4$ Dipartimento di Fisica, Universit\`a di Milano-Bicocca, I-20126 Milano, Italy \\[.2em]
$^5$ INFN, Sezione di Milano-Bicocca, I-20126 Milano, Italy \\[.2em]
}

\bigskip
\bigskip

{\bf Abstract}\\[8mm]
{\parbox{16cm}{\hspace{5mm}
The large $N$ limit of the four-dimensional superconformal index  was computed and successfully compared to the entropy of a class of AdS$_5$ black holes only  in the particular case of  equal angular momenta.
Using the Bethe Ansatz formulation, we compute the index at large $N$ with arbitrary chemical potentials for all charges and angular momenta, for general $\mathcal{N}=1$ four-dimensional conformal  theories with a holographic dual. We conjecture and bring some evidence that a particular universal contribution to the  sum over Bethe vacua  dominates the index at large $N$. For $\mathcal{N}=4$ SYM,  this contribution correctly leads to the entropy of BPS Kerr-Newman black holes in AdS$_5 \times S^5$ for arbitrary values of the conserved charges, thus completing the microscopic derivation  of their microstates. We  also consider theories dual to AdS$_5 \times \mathrm{SE}_5$, where SE$_5$ is a Sasaki-Einstein  manifold. We first check our results against the so-called universal black hole. We then  explicitly construct the near-horizon geometry of BPS Kerr-Newman black holes in AdS$_5 \times T^{1,1}$,  charged under the baryonic symmetry of the conifold theory and with equal angular momenta. We compute the entropy of these black holes using the attractor mechanism and find complete agreement with the field theory predictions.
 }}
\end{center}

\newpage
\pagenumbering{arabic}
\setcounter{page}{1}
\setcounter{footnote}{0}
\renewcommand{\thefootnote}{\arabic{footnote}}

{\renewcommand{\baselinestretch}{1} \parskip=0pt
\setcounter{tocdepth}{2}
\tableofcontents}

%%%%%%%%%%%%%%%%%%%%%%%%%%%%%%%%%%%%%%%%%%%%%%%%%%
%%%%%%%%%%%%%%%%%%%%%%%%%%%%%%%%%%%%%%%%%%%%%%%%%%

\section{Introduction}
\label{sec: intro}

There has been some progress in the microscopic explanation of the entropy of BPS asymp\-tot\-ical\-ly-AdS black holes, initiated with the counting of microstates of static magnetically-charged black holes in AdS$_4\times S^7$ \cite{Benini:2015eyy, Benini:2016hjo, Benini:2016rke} and continued, more recently, with the counting for Kerr-Newman black holes in AdS$_5\times S^5$ \cite{Cabo-Bizet:2018ehj, Choi:2018hmj, Benini:2018ywd}.%
\footnote{The microstate counting for rotating black holes in AdS$_4$ was performed in \cite{Choi:2019zpz, Bobev:2019zmz, Benini:2019dyp}. Those results have been extended to other compactifications and other dimensions. See \cite{Zaffaroni:2019dhb} for a more complete list of references.}
The latter result, in particular, shed light on a long-standing puzzle. The holographic description of electrically-charged and rotating BPS black holes in AdS$_5\times S^5$ is in terms of $1/16$ BPS states of the dual four-dimensional ${\cal N}=4$ super-Yang-Mills (SYM) boundary theory on $S^3$. These states are counted (with sign) by the superconformal index \cite{Romelsberger:2005eg, Kinney:2005ej, Bhattacharya:2008zy}, and one would expect that the contribution from black holes saturates it at large $N$. However, the large $N$ computation of the superconformal index performed in \cite{Kinney:2005ej} gave a result of order one, while the entropy for the black holes is of order $N^2$, suggesting a large cancellation between bosonic and fermionic BPS states. On the other hand, it has been argued in \cite{Choi:2018hmj, Benini:2018ywd} that non-trivial complex phases of the fugacities for flavor symmetries can obstruct such a cancellation between bosonic and fermionic states --- as already observed in \cite{Benini:2015eyy, Benini:2016rke} --- and that the entropy of Kerr-Newman black holes is indeed correctly captured by the index for complex values of the chemical potentials associated with electric charges and angular momenta.

The family of AdS$_5\times S^5$ supersymmetric black holes found in \cite{Gutowski:2004ez, Gutowski:2004yv, Chong:2005da, Chong:2005hr, Kunduri:2006ek} depends on three charges $Q_a$ associated with the Cartan subgroup of the internal isometry $\rSO(6)$, and two angular momenta $J_i$ in AdS$_5$, subject to a non-linear constraint.%
\footnote{Supersymmetric hairy black holes depending on all charges have been recently found in \cite{Markeviciute:2018yal, Markeviciute:2018cqs}, but their entropy seems to be parametrically smaller in the range of parameters where our considerations apply.}
The entropy can be written as the value at the critical point (\ie, as a Legendre transform) of the function \cite{Hosseini:2017mds}
\be\label{HHZ}
\cS(X_a,\tau,\sigma)= - i \pi N^2 \, \frac{X_1 X_2 X_3}{\tau \, \sigma} - 2 \pi i \left (\sum_{a=1}^3 X_a Q_a +  \tau J_1 + \sigma J_2\right)
\ee
with the constraint $X_1+ X_2+ X_3 -\tau-\sigma = \pm 1$, where $N$ is the number of colors of the dual 4d $\cN=4$ $\rSU(N)$ SYM theory.  The same entropy function can also be obtained by computing the zero-temperature limit of the on-shell action of a class of supersymmetric but non-extremal complexified Euclidean black holes \cite{Cabo-Bizet:2018ehj, Cassani:2019mms}. The two constraints with $\pm$ sign lead to the same value for the entropy, which is real precisely when the non-linear constraint on the black hole charges is imposed. The parameters $X_a$, $\tau$ and $\sigma$ are chemical potentials for the conserved charges $Q_a$ and $J_i$ and can also be identified with the parameters the superconformal index depends on. With this identification, we expect that the entropy $S(Q_a, J_1, J_2)$ is just the constrained Legendre transform of $\log \cI(X_a,\tau,\sigma)$,  where $\cI(X_a,\tau,\sigma)$ is  the superconformal index.

Up to now, the entropy of AdS$_5\times S^5$ Kerr-Newman black holes has been derived from the superconformal index  and shown to be in agreement with \eqref{HHZ} only in particular limits. In \cite{Choi:2018hmj}, the entropy was derived for large black holes (whose size is much larger than the AdS radius) using a Cardy limit of the superconformal index where $\im(X_a), \tau, \sigma \ll 1$. In \cite{Benini:2018ywd}, the entropy was instead derived in the large $N$ limit in the case of black holes with equal angular momenta, $J_1=J_2$.%
\footnote{The same result has been later reproduced with a different approach in \cite{Cabo-Bizet:2019eaf}.}
The large $N$ limit has been evaluated  by writing the index as a sum over Bethe vacua \cite{Benini:2018mlo}, an approach that has been successful for AdS black holes in many other contexts.  

It is one of the purposes of this paper to extend the derivation of \cite{Benini:2018ywd} to the case of unequal angular momenta, thus providing a large $N$ microscopic counting  of the microstates  of BPS Kerr-Newman black holes in AdS$_5\times S^5$ for arbitrary values of the conserved charges. We will make use of the Bethe Ansatz formulation of the superconformal index derived for $\tau=\sigma$ in \cite{Closset:2017bse} and generalized to unequal  angular chemical potentials in  \cite{Benini:2018mlo}. This formulation allows us to write the index as a sum over the solutions to a set of Bethe Ansatz Equations (BAEs) --- whose explicit form and solutions have been studied in  \cite{Benini:2018ywd, Hosseini:2016cyf, Hong:2018viz, Lanir:2019abx, ArabiArdehali:2019orz, Lezcano:2019pae} --- and over some auxiliary integer parameters $m_i$. We expect that, in the large $N$ limit, one particular solution dominates the sum.%
\footnote{It is argued in \cite{ArabiArdehali:2019orz} that there exist families of continuous solutions. This does not affect our argument provided the corresponding contribution to the index is subleading.}
We will show that the ``basic solution" to the BAEs, already used in \cite{Benini:2018ywd}, correctly reproduces the entropy of black holes  in the form \eqref{HHZ} for a choice of integers $m_i$.   We stress that our result comes from a single contribution  to the index, which is an infinite sum. Such a contribution might not be the dominant one --- and so our estimate of the index might be incorrect --- in some regions of the space of chemical potentials.  It is known from the analysis in \cite{Benini:2018ywd} that when the  charges become smaller than  a given threshold, new solutions take over and dominate the asymptotic behavior of the index. This suggests the existence of a  rich structure   where other black holes might also contribute. However, we conjecture and we will bring some evidence that the contribution of the basic solution is the dominant one in the region of the space of chemical potentials corresponding to sufficiently large charges. 

We will also extend the large $N$ computation of the index to a general class of superconformal theories dual to AdS$_5\times {\rm SE}_5$, where ${\rm SE}_5$ is a five-dimensional Sasaki-Einstein manifold. The analysis for $J_1=J_2$ was already performed in \cite{Lanir:2019abx}. For toric holographic quiver gauge theories, we find a prediction for the entropy of black holes in AdS$_5\times {\rm SE}_5$ in the form of the entropy function
\be
\label{HHZ2_intro}
\cS(X_a,\tau,\sigma)= - \frac{i \pi N^2}6 \, \sum_{a,b,c}^D C_{abc} \, \frac{X_a X_b X_c}{\tau \, \sigma} - 2 \pi i \left (\sum_{a=1}^D X_a Q_a +  \tau J_1 + \sigma J_2\right) \;,
\ee
with the constraint $\sum_{a=1}^D X_a -\tau-\sigma = \pm 1$, in terms of chemical potentials $X_a$ for a basis of independent R-symmetries $R_a$. The coefficients $C_{abc} \, N^2 = \frac14 \Tr R_a R_b R_c$ are the 't~Hooft anomaly coefficients for this basis of R-symmetries. The form of the entropy function \eqref{HHZ2_intro} was conjectured in \cite{Hosseini:2018dob} and reproduced for various toric models in the special case  $\tau=\sigma$ in \cite{Lanir:2019abx}.  We will give a general derivation, valid for all toric quivers and even more. We will also show that both constraints in \eqref{HHZ2_intro}, which lead to the same value for the entropy,  naturally arise from the index in different regions of the space of chemical potentials. The function \eqref{HHZ2_intro} was also derived in the Cardy limit in \cite{Amariti:2019mgp}.

In the last part of the paper we will provide some evidence that \eqref{HHZ2_intro} correctly reproduces the entropy of black holes in AdS$_5\times {\rm SE}_5$. We first check that our formula correctly reproduce the entropy of the \emph{universal} black hole that arises as a solution in five-dimensional minimal gauged supergravity, and, as such, can be embedded in any AdS$_5\times {\rm SE}_5$ compactification. It corresponds to a black hole with electric charges aligned with the exact R-symmetry of the dual superconformal field theory and with arbitrary angular momenta $J_1$ and $J_2$. Since the solution is universal, the computation can be reduced to that of $\cN=4$ SYM  and it is almost trivial. More interesting are black holes with general electric charges. Unfortunately, to the best of our knowledge, there are no available such black hole solutions in compactifications based on  Sasaki-Einstein manifolds ${\rm SE}_5$ other than $S^5$.  To overcome this obstacle, we will   explicitly construct the near-horizon geometry of supersymmetric black holes in AdS$_5\times T^{1,1}$  with equal angular momenta and charged under the baryonic symmetry of the dual Klebanov-Witten theory \cite{Klebanov:1998hh}. Luckily, the background AdS$_5\times T^{1,1}$  admits a consistent truncation to a five-dimensional gauged supergravity containing the massless gauge field associated to the baryonic symmetry \cite{Cassani:2010na, Bena:2010pr, Halmagyi:2011yd}.  We then use the strategy suggested in \cite{Hosseini:2017mds}: a rotating black hole in five dimensions with $J_1=J_2$ can be dimensionally reduced along the Hopf fiber of the horizon three-sphere to a static solution of four-dimensional $\cN=2$ gauged supergravity.  We will explicitly solve the BPS equations \cite{DallAgata:2010ejj, Halmagyi:2013sla, Klemm:2016wng} for the horizon of static black holes with the appropriate electric and magnetic charges in $\cN=2$ gauge supergravity in four dimensions. The main complication is the presence of hypermultiplets. By solving the hyperino equations at the horizon, we will be able to recast all other supersymmetric conditions as a set of attractor equations, and we will show that these are equivalent to the extremization of \eqref{HHZ2_intro} for the Klebanov-Witten theory with $\tau=\sigma$. This provides a highly non-trivial check of our result, and the conjecture that the basic solution to the BAEs dominates the index. 

The paper is organized as follows. In Section~\ref{sec: N=4 SYM} we review the setting introduced in \cite{Benini:2018ywd} and we evaluate the large $N$ contribution of the ``basic solution"  to the BAEs to the index for generic angular fugacities.
We show that it correctly captures the semiclassical Bekenstein-Hawking entropy of BPS black holes in AdS$_5\times S^5$.
In Section~\ref{sec: toric} we discuss the generalization of this result to general toric quiver theories and find agreement with the entropy function prediction \eqref{HHZ2_intro} in certain corners of the space of chemical potentials.
In Section~\ref{sec: universal} we discuss the particular case of the universal black hole, which can be embedded in all string and M-theory supersymmetric compactifications with an AdS$_5$ factor.
In Section~\ref{sec: conifold} we match formula \eqref{HHZ2_intro} with the entropy of a supersymmetric black hole in AdS$_5\times T^{1,1}$, whose near-horizon geometry we explicitly construct.
Technical computations as well as some review material can be found in several appendices.

\bigbreak

\noindent
\textit{Note added:} while this work was ready to be posted on the arXiv, the preprint \cite{Cabo-Bizet:2020nkr} appeared, which discusses the index in the particular case $\tau=\sigma$ using a different approach.

%%%%%%%%%%%%%%%%%%%%%%%%%%%%%%%%%%%%%%%%%%%%%%%%%%
%%%%%%%%%%%%%%%%%%%%%%%%%%%%%%%%%%%%%%%%%%%%%%%%%%

\section{The index of \matht{\cN=4} SYM at large \matht{N}}
\label{sec: N=4 SYM}

We are interested in evaluating the large $N$ limit of the superconformal index of 4d $\cN=1$ holographic theories. We will consider in this section the simplest example, namely $\cN=4$ $\rSU(N)$ SYM.  The superconformal index counts (with sign) the 1/16 BPS states of the theory on $\bR\times S^3$ that preserve one complex supercharge $\cQ$. These states  are characterized by two angular momenta $J_{1,2}$ on $S^3$ and three R-charges for $\rU(1)^3 \subset \rSO(6)_R$. We write $\cN=4$ SYM in $\cN=1$ notation in terms of a vector multiplet and three chiral multiplets $\Phi_I$ and introduce a symmetric basis of R-symmetry generators $R_{1,2,3}$ such that $R_I$ assigns R-charge $2$ to $\Phi_I$ and zero to $\Phi_J$ with $J\ne I$.  The index is  defined by the trace \cite{Romelsberger:2005eg, Kinney:2005ej}
\be
\label{eq:traceindex}
\cI(p,q,v_1, v_2) = \Tr\,(-1)^F e^{-\beta\{\cQ,\cQ^\dagger\}} \, p^{J_1+\frac{r}{2}} \, q^{J_2+\frac{r}{2}} \, v_1^{q_1} \, v_2^{q_2} \;,
\ee
in terms of two flavor generators \mbox{$q_{1,2} = \frac{1}{2}(R_{1,2}-R_3)$} commuting with $\cQ$, and the R-charge $r=\frac{1}{3}(R_1+R_2+R_3)$. Notice that $(-1)^F = e^{2\pi i J_{1,2}} = e^{i\pi R_{1,2,3}}$.
Here $p,q,v_I$ with $I=1,2$ are complex fugacities associated to the various quantum numbers, while the corresponding chemical potentials $\tau, \sigma, \xi_I$ are defined by
\be
p = e^{2\pi i\tau} \;,\qquad\qquad q = e^{2\pi i\sigma} \;,\qquad\qquad v_I = e^{2\pi i \xi_I} \;.
\ee
The index is well-defined for $|p|, |q|<1$.

It is convenient to redefine the flavor chemical potentials in terms of
\be
\Delta_I = \xi_I + \frac{\tau+\sigma}3 \qquad\qquad\text{for}\qquad I=1,2 \;.
\ee
It is also convenient to introduce an auxiliary chemical potential $\Delta_3$ such that
\be
\label{delta3}
\tau + \sigma - \Delta_1 - \Delta_2 - \Delta_3 \in 2\bZ+1 \;,
\ee
and use the corresponding fugacities
\be
y_I = e^{2\pi i \Delta_I} \;.
\ee
The index then takes the more transparent form
\be
\label{identification charges}
\cI = \Tr_\text{BPS}  \, p^{J_1} \, q^{J_2} \, y_1^{R_1/2} \, y_2^{R_2/2} \, y_3^{R_3/2} \;.
\ee
It shows that the constrained fugacities $p,q, y_I$ with $I=1,2,3$ are associated to the angular momenta $J_{1,2}$ and the charges $Q_I \equiv \frac12 R_I$.

Our starting point is the so-called Bethe Ansatz formulation of the superconformal index \cite{Closset:2017bse, Benini:2018mlo}. The special case that the two angular chemical potentials are equal, $\tau=\sigma$, was already studied in \cite{Benini:2018ywd} (see also \cite{ArabiArdehali:2019orz}).  Here we take them unequal. The formula of \cite{Benini:2018mlo} can be applied when the ratio between the two angular chemical potentials is a rational number.%
\footnote{This might sound like a strong limitation. However, the index (\ref{identification charges}) is invariant under integer shifts of $\tau$ and $\sigma$ compatible with (\ref{delta3}). As proven in \cite{Benini:2018mlo}, the set of complex number pairs $\{\tau,\sigma\}\in\bH^2$ (two copies of the upper half-plane) whose ratio becomes a (real) rational number after some integer shifts of $\tau$ and $\sigma$, is dense in $\bH^2$. Thus, by continuity, the formula of \cite{Benini:2018mlo} fixes the large $N$ limit of the superconformal index for generic complex chemical potentials.}
We thus set
\be
\tau = a \omega \;,\qquad \sigma = b\omega \qquad\text{with}\qquad \im\omega > 0
\ee
and with $a,b \in \bN$ coprime positive integers. We call $\bH = \{\omega \,|\, \im\omega>0\}$  the upper half-plane. We then have the fugacities
\be
h = e^{2\pi i\omega} \;,\qquad p = h^a = e^{2\pi i \tau} \;,\qquad q = h^b = e^{2\pi i \sigma} \qquad\text{with}\qquad |h|, |p|, |q|<1 \;.
\ee

The formula in \cite{Benini:2018mlo} allows us to write the superconformal index as a sum over the solutions to a set of Bethe Ansatz Equations (BAEs). Explicitly, the index reads 
\be
\label{BA formula}
\cI = \kappa_N \sum_{\hat u \,\in\, \text{BAE}} \cZ_\text{tot} \, H^{-1} \Big|_{\hat u} \;.
\ee
The expressions of $\kappa_N$, $H$ and $\cZ_\text{tot}$ for a generic $\cN=1$ theory are given in \cite{Benini:2018mlo}.
Here, we specialize them to $\cN=4$ $\rSU(N)$ SYM. The quantity
\be
\kappa_N = \frac1{N!} \biggl( \frac{ (p;p)_\infty \, (q;q)_\infty \, \wt\Gamma(\Delta_1; \tau,\sigma) \, \wt\Gamma(\Delta_2; \tau, \sigma) }{ \wt\Gamma(\Delta_1 + \Delta_2; \tau, \sigma) } \biggr)^{N-1}
\ee
is a prefactor written in terms of the elliptic gamma function $\wt\Gamma$ and the Pochhammer symbol:
\be
\wt\Gamma(u;\tau,\sigma) \equiv \Gamma(z; p,q) = \prod_{m,n=0}^\infty \frac{1-p^{m+1} q^{n+1}/z}{1-p^m q^nz} \;,\qquad (z;q)_\infty = \prod_{n=0}^\infty (1-zq^n) \;,
\ee
where $z = e^{2\pi i u}$.
The sum in (\ref{BA formula}) is over the solution set to the following BAEs:%
\footnote{The Bethe operators $Q_i$ should not be confused with the charges $Q_I$ introduced before.}
\be\label{BA_op_SYM}
1 = Q_i(u; \Delta, \omega) \equiv e^{2\pi i \left( \lambda + 3\sum_j u_{ij} \right)} \prod_{j=1}^N \frac{ \theta_0 \bigl( u_{ji} + \Delta_1; \omega\bigr) \, \theta_0 \bigl( u_{ji} + \Delta_2; \omega\bigr) \, \theta_0 \bigl( u_{ji} - \Delta_1 - \Delta_2; \omega\bigr) }{ \theta_0 \bigl( u_{ij} + \Delta_1; \omega\bigr) \, \theta_0 \bigl( u_{ij} + \Delta_2; \omega\bigr) \, \theta_0 \bigl( u_{ij} - \Delta_1 - \Delta_2; \omega\bigr) }
\ee
written in terms of $u_{ij} = u_i - u_j$ with $i,j=1, \ldots, N$ and the theta function
\be
\theta_0(u;\omega) = (z;h)_\infty (h/z;h)_\infty \;.
\ee
The unknowns are the ``complexified $\rSU(N)$ holonomies'', which are expressed here in terms of $\rU(N)$ holonomies $u_i$ further constrained by
\be
\label{SU constraint}
\sum_{i=1}^N u_i = 0 \pmod{\bZ} \;,
\ee
as well as a ``Lagrange multiplier'' $\lambda$. The $\rSU(N)$ holonomies are to be identified with the first $N-1$ variables $u_{i=1,\dots,N-1}$. As unknowns in the BAEs, they are subject to the identifications
\be
u_i \,\sim\, u_i+1 \,\sim\, u_i + \omega \;,
\ee
meaning that each one of them naturally lives on a torus of modular parameter $\omega$. Instead, the last holonomy $u_N$ is determined by the constraint \eqref{SU constraint}. The relation between $\rSU(N)$ and $\rU(N)$ holonomies will be further clarified in Appendix~\ref{subapp: m sum rule}. The prescription in \eqref{BA formula} is to sum over all the inequivalent solutions on the torus \cite{Benini:2018mlo}. The function $H$ is the Jacobian
\be
H = \det \left[ \frac1{2\pi i} \, \parfrac{(Q_1,\dots,Q_N)}{(u_1,\dots,u_{N-1},\lambda)} \right] \;.
\ee
Finally, the function $\cZ_\text{tot}$ is the following sum over a set of integers $m_i=1,\ldots, ab$:
\be
\label{Z_tot_def}
\cZ_\text{tot} = \sum_{\{m_i\}=1}^{ab} \cZ\bigl(u - m \omega; \tau, \sigma \bigr) \;,\qquad
\ee
where $\cZ$, for $\cN=4$ $\rSU(N)$ SYM, reads
\be
\label{z_def}
\cZ = \prod_{\substack{i,j=1 \\ i\neq j }}^N \frac{ \wt\Gamma(u_{ij} + \Delta_1; \tau,\sigma) \, \wt\Gamma(u_{ij} + \Delta_2; \tau,\sigma) }{ \wt\Gamma(u_{ij} + \Delta_1 + \Delta_2; \tau,\sigma) \, \wt\Gamma(u_{ij}; \tau,\sigma) } \;.
\ee
The sum in \eqref{Z_tot_def} freely varies over the first $N-1$ integers $m_{i=1, \dots, N-1}$ as indicated, while $m_N$ is determined by the constraint
\be
\label{m_N_constraint}
\sum_{i=1}^N m_i=0\;.
\ee
More details can be found in  \cite{Benini:2018mlo, Benini:2018ywd}. In the following, when a double sum starts from 1 we will leave it implicit.

\subsection{The building block}
\label{subsec: building block}

We will show that one particular contribution to the sums in (\ref{BA formula}) and (\ref{Z_tot_def}) alone reproduces the entropy function of \cite{Hosseini:2017mds} and therefore it captures the Bekenstein-Hawking entropy of BPS black holes in AdS$_5 \times S^5$. To that aim, we are interested in the contribution from the so-called ``basic solution'' to the BAEs \cite{Hosseini:2016cyf, Hong:2018viz, Benini:2018ywd}, namely
\be
\label{basic_solution}
u_{i} = \frac{N-i}N \;\omega + \wb u\;, \qquad\qquad u_{ij} \equiv u_i - u_j = \frac{j-i}N \;\omega \;, \qquad\qquad \lambda = \frac{N-1}{2} \;.
\ee
Here $\wb u$ is fixed by enforcing the constraint \eqref{SU constraint}. We also consider the contribution from a particular choice for the integers $\{m_j\}$:
\be
\label{integersm} 
m_j \in \{1, \dots, ab\} \qquad \text{such that} \qquad m_j = j \mod ab \;.
\ee
Note that this choice for $\{m_j\}$ does not satisfy the constraint \eqref{m_N_constraint}. Nevertheless, we show in Appendix~\ref{subapp: m sum rule} that this does not affect the contribution to leading order in $N$, in the sense that changing the single entry $m_N$ has a subleading effect.

Now, the crucial technical point is to evaluate the following basic building block:
\be
\label{building block def}
\Psi = \sum_{ i \neq j }^N \log \wt\Gamma \left( \Delta + \omega \, \frac{j-i}N + \omega \bigl( m_j - m_i \bigr); a\omega, b\omega \right)
\ee
for $N \to \infty$. Here $\Delta$ plays the role of an electric chemical potential.
In order to simplify the discussion, we assume that $N$ is a multiple of $ab$, \ie, we take $N = ab\wt N$. As we show in Appendix~\ref{subapp: generic N} this assumption can be removed without affecting the leading behavior at large $N$.

We make use of the following identity \cite{Felder:2002}:
\be
\label{product of Gamma's fund}
\wt\Gamma(u; \tau, \sigma) = \prod_{r=0}^{a-1} \prod_{s=0}^{b-1} \, \wt\Gamma\Bigl( u + \bigl( r\tau + s\sigma \bigr); a\tau, b\sigma \Bigr)
\ee
for any $\tau, \sigma \in \bH$ and any $a,b \in \bN$. This is immediate to prove exploiting the infinite product expression of $\wt\Gamma$. Now, exchanging $a\leftrightarrow b$ and $r\leftrightarrow s$ in the formula, and then setting $\tau \to a\omega$, $\sigma \to b\omega$, we obtain the formula of \cite{Felder:1999}:
\be
\wt\Gamma(u; a\omega, b\omega) = \prod_{r=0}^{a-1} \prod_{s=0}^{b-1} \, \wt\Gamma\Bigl( u + \bigl( as + br \bigr)\omega; ab\omega, ab\omega \Bigr) \;.
\ee

Going back to $\Psi$, we can thus write
\be
\Psi = \sum_{r=0}^{a-1} \sum_{s=0}^{b-1} \sum_{i\neq j}^N \log \wt\Gamma \left( \Delta + \omega\, \frac{j-i}N + \omega\bigl( m_j - m_i + as + br \bigr); ab\omega, ab\omega \right) \;.
\ee
Let us set $i = \gamma ab + c$, $j = \delta ab + d$ with $\gamma, \delta = 0, \dots, \wt N-1$ and $c,d = 1, \dots, ab$. Then
\be
\label{before simplification}
\raisebox{0pt}[\height][2.2em]{$\ds
\Psi = \sum_{r=0}^{a-1} \sum_{s=0}^{b-1} \underbrace{\sum_{\gamma,\delta=0}^{\wt N-1} \sum_{c,d=1}^{ab} }_{\text{s.t. } i\neq j} \log\wt\Gamma \left( \Delta + \omega \frac{\delta-\gamma}{\wt N} + \omega \frac{d-c}N + \omega\bigl( d - c + as + br \bigr) ; ab\omega, ab\omega \right) \;.
$}
\ee
We will now perform two simplifications, and prove in Appendix \ref{subapp: simplifications building block} that their effect is of subleading order at large $N$. More precisely, $\Psi$ is of order $N^2$ while the two simplifications modify it at most at order $N$ if $\im\bigl( \Delta/\omega \bigr) \not\in \bZ \times \im\bigl( 1/\omega \bigr)$, or at most at order $N \log N$ if $\Delta=0$. First, we substitute the condition $i\neq j$ with the condition $\gamma \neq \delta$ in the summation. Second and more importantly, we drop the term $\omega(d-c)/N$ in the argument. We then redefine $c \to ab-c$, $d \to d+1$, $\gamma \to \gamma-1$, $\delta \to \delta -1$ and obtain
\be
\label{after simplification}
\Psi \simeq \sum_{r=0}^{a-1} \sum_{s=0}^{b-1} \sum_{\gamma \neq \delta}^{\wt N} \sum_{c,d=0}^{ab-1} \log\wt\Gamma \left( \Delta + \omega \frac{\delta - \gamma}{\wt N} + \omega\bigl( d + c + 1 - ab + as + br \bigr); ab\omega, ab\omega \right)
\ee
where $\simeq$ means equality at leading order in $N$. At this point we can resum over $c,d$ using (\ref{product of Gamma's fund}) (with $\tau,\sigma \to \omega$ and $a,b \to ab$):
\be
\Psi \simeq \sum_{r=0}^{a-1} \sum_{s=0}^{b-1} \sum_{\gamma \neq \delta}^{\wt N} \log\wt\Gamma \left( \Delta + \omega \frac{\delta - \gamma}{\wt N} + \omega\bigl( 1-ab + as + br \bigr); \omega, \omega \right) \;.
\ee

\begin{figure}[t]
\centering
\begin{tikzpicture}
\filldraw [yellow!15!white] (-2.16,-1) to (-.66,-1) to (1.33,2) to (-.16,2) to cycle;
\filldraw [blue!10!white] (-.66,-1) to (1.33,2) to (2.82,2) to (.83, -1) to cycle;
\draw [->] (-3,0) to (3,0);
\draw [->] (0,-1) to (0,2);
\draw [very thick, blue!80!black] (-.66,-1) to (1.33,2);
\draw [very thick, blue!80!black] (-2.16,-1) to (-.16,2);
\draw [very thick, blue!80!black] (.83, -1) to (2.82,2);
\filldraw [red!80!black] (0, 0) circle [radius=.07] node [below right, black] {\small $0$};
\filldraw [red!80!black] (1, 1.5) circle [radius=.07] node [below right, black] {\small $\omega$};
\filldraw [red!80!black] (-1.5, 0) circle [radius=.07] node [above left, black] {\small $-1$};
\filldraw [red!80!black] (1.5, 0) circle [radius=.07] node [below right, black] {\small $1$};
\filldraw [red!80!black] (-.5, 1.5) circle [radius=.07] node [above left, black] {\small $\omega-1$};
\filldraw [red!80!black] (2.5, 1.5) circle [radius=.07] node [below right, black] {\small $\omega + 1$};
\end{tikzpicture}
\caption{Fundamental strips for $[\Delta]_\omega$ and $[\Delta]'_\omega$. The function $[\Delta]_\omega$ is the restriction of $\Delta$ mod $1$ to the region $\im(-1/\omega) > \im(\Delta/\omega)>0$ (in yellow, on the left), while $[\Delta]'_\omega$ is the restriction of $\Delta$ mod $1$ to the region $0 > \im(\Delta/\omega) > \im(1/\omega)$ (in blue, on the right).
\label{fig: strip}}
\end{figure}
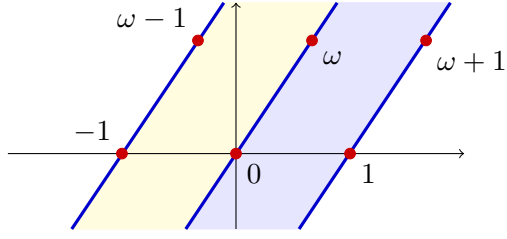

We recall the large $N$ limit computed in \cite{Benini:2018ywd}:
\be
\label{single_fugacity_index}
\sum_{i\neq j}^N \log\wt\Gamma\left( \Delta + \omega \, \frac{j-i}N; \omega, \omega \right) = -\pi i N^2 \, \frac{B_3\bigl( [\Delta]_\omega' - \omega \bigr)}{3\omega^2} + \cO(N)
\ee
valid for $\im\bigl( \Delta/\omega \bigr) \not\in \bZ \times \im\bigl( 1/\omega \bigr)$. Here $B_3(x)$ is a Bernoulli polynomial:
\be
B_3(x) = x \, \bigl( x - \tfrac12\bigr) \bigl(x-1 \bigr) \;.
\ee
It has the property that $B_3(1-x) = -B_3(x)$. The function $[\Delta]'_\omega$ was defined in \cite{Benini:2018ywd} in the following way:
\be{}
[\Delta]'_\omega = \biggl\{ z \,\biggm|\, z = \Delta \text{ mod } 1 \,,\;\; 0 > \im \left( \frac z\omega \right) > \im \Bigl( \frac1\omega \Bigr) \biggr\} \;.
\ee
This function is only defined for $\im\bigl( \Delta/\omega \bigr) \not\in \bZ \times \im\bigl( 1/\omega \bigr)$, it is continuous in each open connected domain, and it is periodic by construction under $\Delta \to \Delta+1$. In the following we will  also use the function $[\Delta]_\omega = [\Delta]'_\omega -1$,
\be{}
[\Delta]_\omega = \biggl\{ z \,\biggm|\, z = \Delta \text{ mod } 1 \,,\;\; \im \Bigl( - \frac1\omega \Bigr) > \im \left( \frac z\omega \right) > 0 \biggr\} \;.
\ee
The functions $[\Delta]_\omega$ and $[\Delta]'_\omega$ are the mod $1$ reductions of $\Delta$ to the fundamental strips shown in Figure~\ref{fig: strip}.
Then we use the following formula:
\begin{multline}
\label{bernoulli_property}
\frac1{ab} \sum_{r=0}^{a-1} \sum_{s=0}^{b-1} B_3\bigl( x + \omega(as+br-ab) \bigr) = {} \\
{} = B_3\left( x - \frac{a+b}2\omega \right) + \frac{2a^2 b^2 - a^2 - b^2}4 \, \omega^2 \, B_1\left( x - \frac{a+b}2 \omega \right) \;,
\end{multline}
where $B_1(x) = x - \frac12$ is another Bernoulli polynomial --- and $B_1(1-x) = -B_1(x)$.
Thus
\be
\label{Psi large N final}
\Psi = - \pi i N^2 \, \frac{ B_3\Bigl( [\Delta]'_\omega - \tfrac{\tau+\sigma}2 \Bigr) }{ 3\tau \sigma} - \frac{\pi i N^2}{12} \Bigl( 2ab - \frac ab - \frac ba \Bigr) \, B_1\Bigl( [\Delta]'_\omega - \tfrac{\tau+\sigma}2 \Bigr) + \cO(N)
\ee
for $\im\bigl( \Delta/\omega \bigr) \not\in \bZ \times \im\bigl( 1/\omega \bigr)$. As a check, notice that $\bigl[ \tau + \sigma - \Delta \bigr]'_\omega = \tau + \sigma + 1 - [\Delta]'_\omega$. From the properties of $B_{1,3}(x)$ noticed above, it follows
\be
\Psi(\tau + \sigma - \Delta) = - \Psi(\Delta)
\ee
at leading order in $N$. This is in accordance with the inversion formula of the elliptic gamma function:
\be
\label{inversion formula}
\wt \Gamma(u;\tau,\sigma) = 1/\wt \Gamma(\tau+\sigma -u; \tau,\sigma) \;.
\ee

The case $\Delta=0$ requires some care, because $[0]_\omega$ is undefined. Taking the limit of $\Psi$ as $\Delta \to 0$ from the left or the right, one obtains two values that differ by an imaginary quantity. The limit from the right corresponds to taking $[\Delta]'_\omega \to 0$ in (\ref{Psi large N final}), while the limit from the left corresponds to $[\Delta]_\omega \to 0$ (\ie, $[\Delta]'_\omega \to 1$). The difference is
\be
\label{v1}
\Psi \Big|_{[\Delta]'_\omega \to 0} - \Psi\Big|_{[\Delta]_\omega \to 0} = \frac{i \pi N^2}6 \left( 3 + ab + \frac ab + \frac ba \right) \;.
\ee
Since $\Psi$ is in any case ambiguous by shifts of $2\pi i$ because it is a logarithm, only the remainder modulo $2\pi i$ is meaningful but this is an order 1 quantity which can be neglected. In fact it turns out that, with $N = ab\wt N$, the quantity on the right-hand-side of (\ref{v1}) is always an integer multiple of $i\pi \wt N$, and so its exponential is a sign. We should also notice that, for $\Delta=0$, our approximation gets corrections at order $N\log N$.

\subsection{The index and the entropy function}
\label{sec:N4entropy}

We are now ready to put all the ingredients together. Our working assumption is that, in the large $N$ limit, the index \eqref{BA formula} is dominated by the basic solution \eqref{basic_solution} and the choice of integers \eqref{integersm}. Some evidence that the basic solution dominates the index  for $\tau=\sigma$ has been given in \cite{Benini:2018ywd} (see also \cite{ArabiArdehali:2019orz}).

The leading contribution to \eqref{BA formula} originates from $\cZ_\text{tot}$ that can be evaluated using \eqref{Psi large N final}. Indeed, the term $\kappa_N$ is manifestly sub-leading. That the contribution of  $H$ is also subleading follows from the analysis in \cite{Benini:2018ywd} for $\tau=\sigma$, since $H$ only depends on the solutions to the BAEs  and not explicitly  on $\tau$ and $\sigma$.  The large $N$ limit of the index at leading order is then
\be
\label{III}
\log\cI = \Psi(\Delta_1) + \Psi(\Delta_2) - \Psi(\Delta_1 + \Delta_2) - \Psi(0) \;,
\ee
where the definition of the last term has an ambiguity of order 1.  

Recall that in (\ref{delta3}) we introduced the auxiliary chemical potential $\Delta_3$. 
Notice in particular that the chemical potentials are defined modulo 1. Using the basic properties
\be
\label{properties of [ ]}
[\Delta + 1]_\omega =[\Delta]_\omega \;,\qquad [\Delta + \omega]_\omega = [\Delta]_\omega+ \omega \;,\qquad [-\Delta]_\omega= - [\Delta]_\omega -1 \;,
\ee
we find
\be
[\Delta_3]_\omega = \tau + \sigma -1 - [\Delta_1 + \Delta_2]_\omega \;.
\ee
It follows from the definition of the function $[\Delta]_\omega$ that $[\Delta_1 + \Delta_2]_\omega = [\Delta_1]_\omega + [\Delta_2]_\omega+n$ where $n=0$ or $n=1$. 
The result then breaks into two cases. 

If $[\Delta_1 + \Delta_2]_\omega = [\Delta_1]_\omega + [\Delta_2]_\omega$ then
\be
[\Delta_1]_\omega + [\Delta_2]_\omega + [\Delta_3]_\omega - \tau - \sigma  = - 1 \;,
\ee
and, using \eqref{III} and \eqref{Psi large N final},
\bea
\log\cI &= - \pi i N^2 \, \frac{[\Delta_1]_\omega \, [\Delta_2]_\omega \, \bigl( \tau + \sigma - 1 - [\Delta_1]_\omega - [\Delta_2]_\omega \bigr) }{\tau \, \sigma} \\
&= -i \pi N^2 \, \frac{[\Delta_1]_\omega \, [\Delta_2]_\omega \, [\Delta_3]_\omega }{\tau\, \sigma} \;.
\eea
To obtain this formula we used $\Psi(0) = \Psi\big|_{[\Delta]_\omega \to 0}$.
Notice that the contributions from $B_1$ cancel out. As we will see in Section~\ref{sec: toric}, this is a consequence of the relation $a=c$ among the two four-dimensional central charges in the large $N$ limit.

If $[\Delta_1 + \Delta_2]_\omega = [\Delta_1]_\omega + [\Delta_2]_\omega + 1$, namely $[\Delta_1 + \Delta_2]'_\omega = [\Delta_1]'_\omega + [\Delta_2]'_\omega$, then
\be
 [\Delta_1]'_\omega + [\Delta_2]'_\omega + [\Delta_3]'_\omega - \tau - \sigma= 1 \;,
\ee
and
\bea
\log\cI &= - \pi i N^2 \, \frac{[\Delta_1]'_\omega \, [\Delta_2]'_\omega \, \bigl( \tau + \sigma + 1 - [\Delta_1]'_\omega - [\Delta_2]'_\omega \bigr) }{\tau\,\sigma} \\
&= -i \pi N^2 \, \frac{[\Delta_1]'_\omega \, [\Delta_2]'_\omega \, [\Delta_3]'_\omega }{\tau\,\sigma} \;.
\eea
This time we used $\Psi(0) = \Psi\big|_{[\Delta]'_\omega \to 0}$. 

As in \cite{Benini:2018ywd}, we can extract the entropy of the dual black holes by taking the Legendre transform of the logarithm of the index. The precise identification of the charges associated with the chemical potentials follows from (\ref{identification charges}). The  prediction for the entropy can then be combined into two constrained entropy functions
\begin{multline}
\label{HHZ2}
\cS_\pm(X_I,\tau,\sigma, \Lambda)= - i \pi N^2 \, \frac{X_1 X_2 X_3}{\tau \, \sigma} - 2 \pi i \left (\sum_{I=1}^3 X_I Q_I +  \tau J_1 + \sigma J_2\right)\\
-2\pi i \Lambda\bigg(X_1+ X_2+ X_3 -\tau-\sigma \pm 1\bigg)\;,
\end{multline}
where  we used a neutral variable $X_I$ to denote either $[\Delta_I]_\omega$ or $[\Delta_I]^\prime_\omega$, we introduced a Lagrange multiplier $\Lambda$ to enforce the constraint, and we recall that $Q_I = \frac12 R_I$. This completes our derivation of the entropy of supersymmetric black holes in AdS$_5\times S^5$ for general angular momenta and electric charges. The expression \eqref{HHZ2} represents indeed the two entropy functions derived in \cite{Hosseini:2017mds}, where it was shown  that the (constrained) extremization of \eqref{HHZ2} reproduces the entropy of a black hole of angular momenta $J_1$ and $J_2$ and charges $Q_I$. The two results correspond to the two entropy functions that reproduce the same black hole entropy, and are associated to two Euclidean complex solutions that regularize the black hole horizon \cite{Cabo-Bizet:2018ehj}.

%%%%%%%%%%%%%%%%%%%%%%%%%%%%%%%%%%%%%%%%%%%%%%%%%%
%%%%%%%%%%%%%%%%%%%%%%%%%%%%%%%%%%%%%%%%%%%%%%%%%%

\section{The index of quiver theories with a holographic dual}
\label{sec: toric}

We want to generalize the large $N$ computation of the superconformal index to theories dual to AdS$_5\times {\rm SE}_5$ compactifications, where ${\rm SE}_5$ is a five-dimensional Sasaki-Einstein manifold. We can write general formul\ae{} with very few assumptions. We consider 4d $\cN=1$ theories with $\rSU(N)$ gauge groups as well as adjoint and bi-fundamental chiral multiplet fields. To cancel gauge anomalies, the total number of fields transforming in the fundamental representation of a group must be the same as the number of anti-fundamentals. We also require equality of the conformal central charges  $c=a$  in the large $N$ limit, as dictated by holography. Our analysis extends the results found in \cite{Lanir:2019abx} for equal angular momenta.

We then assume  that in the large $N$ limit, as for ${\cal N}=4$ SYM, the leading contribution to the superconformal index comes from the basic solution and the choice of integers $\{m_i\}$ discussed in (\ref{integersm}). As already shown in \cite{Lanir:2019abx, Lezcano:2019pae}, the basic solution to the BAEs for $\cN=4$ SYM \cite{Hosseini:2016cyf, Hong:2018viz, Benini:2018ywd} can easily be extended to quiver gauge theories by setting
\be
\label{basic_solution2}
u^{\alpha\beta}_{ij} \equiv u_i^\alpha - u_j^\beta = \frac{j-i}N \, \omega  \qquad\qquad \alpha, \beta = 1,\ldots, G \;,
\ee
where $\alpha,\beta$ run over the various gauge groups in the theory and $G$ is the number of gauge groups. Similarly, we choose the integers
\be
\label{m in basic solution 2}
m_j^\alpha \in \{1, \dots, ab\} \qquad\text{such that}\qquad m_j^\alpha = j \mod ab \;.
\ee
Notice in particular that neither $u^{\alpha\beta}_{ij}$ nor $m_j^\alpha$ depend on $\alpha,\beta$.
As for $\cN=4$ SYM, the contribution of the  determinant $H$ to the Bethe Ansatz expansion \eqref{BA formula}  is subleading \cite{Lanir:2019abx}.

Using the general expressions given in \cite{Benini:2018mlo} and following the logic of Section~\ref{sec: N=4 SYM}, it is easy to write the large $N$ limit of the leading contribution to the superconformal index of a holographic theory, with adjoint and bi-fun\-da\-men\-tal chiral fields.  We find
\be
\label{index}
\log \cI = \sum_{i\ne j}^N \Biggl[ \sum_{I_{\alpha\beta}} \log \wt\Gamma \Bigl( u_{ij}^{\alpha\beta} - \omega \bigl( m_i^\alpha - m_j^\beta \bigr) + \Delta_{I_{\alpha\beta}}; \tau,\sigma \Bigr) - \sum_{\alpha=1}^{G} \log \wt\Gamma \Bigl( u^{\alpha\alpha}_{ij} - \omega \bigl( m_i^\alpha - m_j^\beta \bigr) ; \tau,\sigma \Bigr) \Biggr]
\ee
where $z_i^\alpha = e^{2\pi i u_i^\alpha}$ are the gauge fugacities, $u_i^\alpha$ represent the basic solution \eqref{basic_solution2} and $m_i^\alpha$ are given in \eqref{m in basic solution 2}. The sum over $I_{\alpha\beta}$ is over all adjoint (if $\alpha=\beta$) and bi-fundamental (if $\alpha\neq\beta$) chiral multiplets in the theory. The second sum is the contribution of vector multiplets. When no confusion is possible, we will keep the gauge group indices implicit and just write $\Delta_{I_{\alpha\beta}} \equiv \Delta_I$. In the previous formula,
\be
\label{def variables Delta_I}
\Delta_I =\xi_I +  r_I \, \frac{\tau+\sigma}{2} \;,
\ee
where $r_I$ is the exact R-charge of the field and $\xi_I$ are the flavor chemical potentials. The R-charges satisfy
\be
\sum_{I\in W} r_I = 2
\ee
for each superpotential term $W$ in the Lagrangian. In this notation, the index $W$ runs over the monomials in the superpotential, while $I \in W$ indicates all chiral fields appearing in a given monomial. Using that each superpotential term must be invariant under the flavor symmetries, but chemical potentials are only defined up to integers, we also require
\be
\sum_{I\in W} \xi_I = n_W \qquad\text{for some}\qquad n_W \in \mathbb{Z} \;.
\ee 
The values $n_W \equiv n_0 =\pm 1$ have been used in \cite{Kim:2019yrz, Cabo-Bizet:2019osg} to study the Cardy limit.
As a consequence of the previous formul\ae, for each superpotential term we have
\bea
\label{constr} 
\sum_{I\in W} \Delta_I = \tau+\sigma +n_W \;.
\eea
Hence, we stress that the chemical potentials $\Delta_I$ are \emph{not} independent.
Notice that the expression  \eqref{index}  correctly reduces to the one for $\cN=4$ SYM, Eqn.~\eqref{z_def}, once we use the definition \eqref{delta3} as well as the inversion formula for the elliptic gamma function \eqref{inversion formula}. We also need to use the exact R-charges $r_I=2/3$ of the chiral fields $\Phi_I$.
 
Applying \eqref{Psi large N final}, we can evaluate the large $N$ limit of \eqref{index}  and obtain 
\begin{align}
\label{index2}
\log \cI &\;\simeq\;  - \frac{\pi i N^2}{ 3\tau \sigma}  \,   \sum_{I} \left[     B_3\Bigl( [\Delta_{I} ]_\omega +1 - \tfrac{\tau+\sigma}2 \Bigr) +\frac{\tau\sigma}{4} \Bigl( 2ab - \frac ab - \frac ba \Bigr) \, B_1\Bigl( [\Delta_{I}  ]_\omega +1 - \tfrac{\tau+\sigma}2 \Bigr) \right] \nn\\ 
&\quad\;\;\; + \frac{\pi i G N^2}{ 3\tau \sigma}  \,   \left[     B_3\Bigl( 1 - \tfrac{\tau+\sigma}2 \Bigr) +\frac{\tau\sigma}{4} \Bigl( 2ab - \frac ab - \frac ba \Bigr) \, B_1\Bigl( 1 - \tfrac{\tau+\sigma}2 \Bigr) \right] \;.
\end{align}
The corrections are of order $N\log N$ or smaller. The formula is obtained by summing \eqref{Psi large N final} for each chiral multiplet, as well as \eqref{Psi large N final} with $[\Delta]_\omega \to 0$ (and opposite sign) for each vector multiplet.
We stress that \eqref{index2} comes from a \emph{single} contribution --- in the Bethe Ansatz expansion --- to the index. Such a contribution might not be the dominant one, and so our estimate of the index might be incorrect, in some regions of the space of chemical potentials. However, we conjecture and we will bring some evidence that this contribution always captures the semiclassical Bekenstein-Hawking entropy of BPS black holes.
 
Due to the presence of the brackets $[\Delta_{I} ]_\omega$, the  expression \eqref{index2} assumes different analytic forms in different regions of the space of chemical potentials $\Delta_{I}$. There are two regions where the expression greatly simplifies. They correspond to the natural generalization of the two regions for ${\cal N}=4$ SYM discussed in Section~\ref{sec:N4entropy} and are expected to lead to the correct black hole entropy. In particular, they smoothly reduce to the results obtained in the Cardy limit  \cite{Kim:2019yrz, Cabo-Bizet:2019osg, Amariti:2019mgp} and match the previous analysis done for equal angular momenta \cite{Lanir:2019abx}. The first region corresponds to chemical potentials $\Delta_I$  satisfying 
\bea
\label{constr2}
\sum_{I\in W} [\Delta_I]_\omega = \tau+\sigma -1 \;.
\eea
As we will discuss later, many models --- in particular all toric ones --- exhibit a corner in the space of chemical potentials where this constraint is satisfied.
We can define the rescaled variables
\be
\label{rescaled Delta variables}
\wh \Delta_{I}  =  2 \, \frac{[\Delta_{I} ]_\omega}{\tau+\sigma-1}
\ee
which, under the assumption \eqref{constr2}, satisfy
\be\label{cos} 
\sum_{I\in W} \wh\Delta_I = 2
\ee
and can be interpreted as an assignment of R-charges to the chiral fields in the theory. In terms of $\wh\Delta_I$ the  contributions in \eqref{index2} combine into 
\begin{align}
\log \cI &\;\simeq\; - \frac{ \pi i N^2 }{24} \, \frac{(\tau +\sigma -1)^3}{\tau \sigma} \biggl[ \sum\nolimits_{I}   \bigl( \wh\Delta_{I} - 1 \bigr)^3 + G \biggr] \\ 
&\quad\;\;\; {} + \frac{ \pi i N^2 }{24} \, \frac{(\tau +\sigma  -1)}{\tau \sigma} \left( 1 - \tau\sigma \Bigl( 2 a b  - \frac{a}{b} - \frac{b}{a} \Bigr)\right )  \biggl[ \sum\nolimits_{I}  \bigl( \wh\Delta_{I}  - 1 \bigr) + G \biggr]  \;. \nn
\end{align}
Introducing the charge operator $R(\wh\Delta)$ of R-charges parametrized by $\wh\Delta_I$ and indicating with $\Tr$ the sum over all fermions in the theory, we can also write
\be
\log \cI \;\simeq\;  -\frac{ \pi i  }{24 } \Biggl[  \frac{(\tau +\sigma -1)^3}{\tau \sigma}  \Tr R(\wh\Delta)^3 - \frac{(\tau +\sigma  -1)}{\tau \sigma} \left ( 1 - \tau\sigma \Bigl( 2 a b  - \frac{a}{b}-\frac{b}{a} \Bigr) \right) \Tr R(\wh\Delta) \Biggr] \;,
\ee
valid at leading order in $N$.

In the large $N$ limit, theories with a holographic dual satisfy $c=a$. Using standard formul\ae{} for the central charges $a$ and $c$ in terms of the fermion R-charges \cite{Anselmi:1997am}, one finds $\Tr R = \cO(1)$ and $a= \frac{9}{32} \Tr R^3 + \cO(1)$ from which we obtain the final expression
\be
\label{result}
\log \cI \;\simeq\; -\frac{ 4 \pi i  }{27} \, \frac{(\tau +\sigma -1)^3}{\tau \sigma} \, a(\wh\Delta) \;,
\ee
where
\be
a = \frac9{32} \, N^2 \biggl( \sum\nolimits_I \bigl( \wh\Delta_I -1 \bigr)^3 + G \biggr)
\ee
at leading order in $N$. The result \eqref{result} was conjectured in \cite{Hosseini:2018dob} --- see Eqn.~(A.7). It is also compatible with the Cardy limit performed in \cite{Kim:2019yrz, Cabo-Bizet:2019osg}.

We can find an analogous result in a second region of chemical potentials where
\be
\label{constr22}
\sum_{I\in W} [\Delta_I]^\prime_\omega = \tau+\sigma +1 \;,
\ee
written in terms of the primed bracket $[\Delta]^\prime_\omega = [\Delta]_\omega+1$. As discussed at the end of Section~\ref{sec: N=4 SYM}, the contribution of vector multiplets can be written, up to subleading terms, as minus the contribution of a chiral multiplet with $[\Delta_I]'_\omega \to 0$. After defining another set of normalised R-charges,
\be\label{paramet}
\wh\Delta_I^\prime = 2 \, \frac{ [\Delta_I]^\prime_\omega}{\tau+\sigma+1}
\ee
which satisfy
\be
\sum_{I\in W} \wh\Delta^\prime_I = 2
\ee
under the assumption \eqref{constr22}, we can rewrite the index as
\be
\log\cI \;\simeq\; -\frac{\pi i}{24} \Biggl[  \frac{(\tau +\sigma +1)^3}{\tau \sigma} \Tr R(\wh\Delta^\prime)^3 - \frac{(\tau +\sigma  +1)}{\tau \sigma} \left( 1 - \tau\sigma \Bigl( 2 a b  - \frac{a}{b}-\frac{b}{a} \Bigr) \right) \Tr R(\wh\Delta^\prime) \Biggr]
\ee
at leading order in $N$. This reduces to the simple expression
\be
\label{result2a}
\log \cI \;\simeq\; -\frac{ 4 \pi i  }{27 } \, \frac{(\tau +\sigma +1)^3}{\tau \sigma} \, a(\wh\Delta^\prime)
\ee
for holographic theories.

In the remainder of this section we will interpret the general results \eqref{result}  and \eqref{result2a} and provide examples. In particular, we will show that both regions \eqref{constr2} and \eqref{constr22} in the space of chemical potentials always exist in toric quiver gauge theories. We will also see that the two expressions \eqref{result}  and \eqref{result2a}  lead to the very same result for the semiclassical entropy of dual black holes, generalizing what happens for ${\cal N}=4$ SYM.

\subsection{Example: the conifold}
\label{subsec:conifold}

We start with the example of the Klebanov-Witten theory dual to AdS$_5\times T^{1,1}$, the near-horizon limit of a set of $N$ D3-branes sitting at a conifold singularity \cite{Klebanov:1998hh}.  This example was already studied for equal angular momenta in \cite{Lanir:2019abx} and our results are consistent with those found there when we set $\tau=\sigma$.

The theory has gauge group $\rSU(N)\times \rSU(N)$, bi-fundamental chiral multiplets $A_1,A_2$ transforming in the representation $(N,\wb N)$ and $B_1,B_2$ transforming in the representation $(\wb N , N)$, and a superpotential
\be
W = \Tr \bigl( A_1 B_1 A_2 B_2 - A_1 B_2 A_2 B_1 \bigr) \;.
\ee
The global symmetry of the theory is $U(1)_R \times SU(2)_{F_1} \times SU(2)_{F_2} \times U(1)_B$, where the first factor is the superconformal R-symmetry with charge $r$, while the other three factors are flavor symmetries. The charge assignments of chiral multiplets under the maximal torus are in Table~\ref{tab: charges}.
\begin{table}[t]
\centering
$\displaystyle
\begin{array}{|c||c|c|c|c||c|c|c|c|}
\hline
\text{Field} & r & Q_{F_1} & Q_{F_2} & Q_B & R_1 & R_2 & R_3 & R_4 \\
\hline\hline
A_1 & \frac12 & 1 & 0 & 1 & 2 & 0 & 0 & 0 \\
A_2 & \frac12 & -1 & 0 & 1 & 0 & 2 & 0 & 0 \\
B_1 & \frac12 & 0 & 1 & -1 & 0 & 0 & 2 & 0 \\
B_2 & \frac12 & 0 & -1 & -1 & 0 & 0 & 0 & 2 \\
\hline
\end{array}
$
\caption{Charges of chiral multiplets in the Klebanov-Witten theory, under the maximal torus of the global symmetry $U(1)_R \times SU(2)_{F_1} \times SU(2)_{F_2} \times U(1)_B$. In the table we indicate two useful basis. Notice that $r$ and $R_I$ are R-charges, while $Q_{F_{1,2}}$ and $Q_B$ are flavor charges.
\label{tab: charges}}
\end{table}
The index is defined as
\be
\cI = \Tr\, (-1)^F \, e^{-\beta\{\cQ, \cQ^\dag\}} \, p^{J_1 + r/2} \, q^{J_2 + r/2} \, v_{F_1}^{Q_{F_1}} \, v_{F_2}^{Q_{F_2}} \, v_B^{Q_B} \;.
\ee
It is convenient to introduce an alternative basis of R-charges $R_I$ with $I=1,2,3,4$, such that each of them assigns R-charge 2 to one of the chiral multiplets and zero to the other ones. Correspondingly, we associate a variable $\Delta_I$ to each chiral multiplet. Notice that $(-1)^F = e^{2\pi i J_{1,2}} = e^{\pi i R_{1,2,3,4}}$.
According to (\ref{def variables Delta_I}) and up to integer ambiguities, the variables $\Delta_I$ are related to the chemical potentials for the charges in Table~\ref{tab: charges} by
\bea
\Delta_1 &= \xi_{F_1} + \xi_B + \frac{\tau+\sigma}4 \;,\qquad\qquad& \Delta_3 &= \xi_{F_2} - \xi_B + \frac{\tau+\sigma}4 \;, \\
\Delta_2 &= -\xi_{F_1} + \xi_B + \frac{\tau+\sigma}4 \;,\qquad& \Delta_4 &= -\xi_{F_2} - \xi_B + \frac{\tau+\sigma}4 + (2\bZ+1) \;.
\eea
Then, the constraint \eqref{constr} reads
\be
\Delta_1+\Delta_2+\Delta_3+\Delta_4 = \tau+\sigma + n_W
\ee
and the index takes the more transparent form
\be
\cI = \Tr_\text{BPS} p^{J_1} \, q^{J_2} \, y_1^{R_1/2} \, y_2^{R_2/2} \, y_3^{R_3/2} \, y_4^{R_4/2} \;.
\ee
This shows that $\Delta_I$ are the chemical potentials associated to the charges $Q_I \equiv R_I/2$.

We select three independent variables, say $\Delta_1, \Delta_2$ and  $\Delta_3$. Then, using \eqref{properties of [ ]} we find that
\be{}
[ \Delta_4]_\omega = \tau+\sigma -1 -[\Delta_1+\Delta_2+\Delta_3]_\omega \;.
\ee 
In general there are three possible cases:
\be
\label{cases}
[\Delta_1+\Delta_2+\Delta_3]_\omega = [ \Delta_1]_\omega + [ \Delta_2]_\omega +[ \Delta_3]_\omega  + n \qquad\text{with}\qquad n=0,1,2
\ee
that we call Case I, II and III, respectively.%
\footnote{For the sake of comparison, the  notation is the same as in \cite{Lanir:2019abx}.} 

Case I corresponds to the corner of moduli space \eqref{constr2} where 
\be
\label{constrcon}
[ \Delta_1]_\omega + [ \Delta_2]_\omega +[ \Delta_3]_\omega +[ \Delta_4]_\omega = \tau+ \sigma -1 \;.
\ee
In this corner, we can use \eqref{result}. One can explicitly compute, at leading order in $N$,
\be
\Tr R(\wh\Delta)^3 = N^2 \biggl( 2 + \sum_{I=1}^4 \bigl(\wh\Delta_I -1 \bigr)^3 \biggr) = 3 N^2 \Bigl( \wh\Delta_1 \wh\Delta_2 \wh\Delta_3 +\wh\Delta_1 \wh\Delta_2 \wh\Delta_4 +\wh\Delta_1 \wh\Delta_3 \wh\Delta_4 + \wh\Delta_2 \wh\Delta_3 \wh\Delta_4 \Bigr)
\ee 
imposing $\sum_{I=1}^4 \wh\Delta_I = 2$.  Using \eqref{rescaled Delta variables},  we can write the index \eqref{result} as
\be
\label{conifoldSCI}
\log \cI \;\simeq\; - \frac{\pi i N^2} {\tau \sigma}  \Bigl( [\Delta_1]_\omega  [\Delta_2]_\omega [\Delta_3]_\omega +[\Delta_1]_\omega [\Delta_2]_\omega [\Delta_4]_\omega +[\Delta_1]_\omega [\Delta_3]_\omega [\Delta_4]_\omega + [\Delta_2]_\omega [\Delta_3]_\omega [\Delta_4]_\omega \Bigr)
\ee
with the constraint \eqref{constrcon}.%
\footnote{For toric models, discussed in detail in Section~\ref{subsec:toric}, we can  compute the index using formula \eqref{result200}. The 't~Hooft coefficients  are expressed in terms of toric data as $C_{abc}= \bigl| \det \{ v_a, v_b, v_c \} \bigr|$, where $v_a$ are the integer vectors defining the toric fan \cite{Benvenuti:2006xg}. For the conifold: $v_1=(1,0,0), \, v_2=(1,1,0),\, v_3=(1,1,1)\, , v_4= (1,0,1)$ and thus $C_{123}=C_{124}=C_{134}=C_{234}=1$ (and symmetrizations), recovering the expression above.}

Case III corresponds to the corner of moduli space \eqref{constr22}. Indeed
\be
\label{constrcon2}
[ \Delta_1]^\prime_\omega + [ \Delta_2]^\prime_\omega +[ \Delta_3]^\prime_\omega +[ \Delta_4]^\prime_\omega = \tau+ \sigma +1 \;.
\ee
In this corner, we can use \eqref{result2a} and \eqref{paramet} and find
\be
\label{conifoldSCI2}
\log \cI \;\simeq\;  - \frac{\pi i N^2} {\tau \sigma}  \Bigl( [\Delta_1]^\prime_\omega  [\Delta_2]^\prime_\omega [\Delta_3]^\prime_\omega +[\Delta_1]^\prime_\omega [\Delta_2]^\prime_\omega [\Delta_4]^\prime_\omega +[\Delta_1]^\prime_\omega [\Delta_3]^\prime_\omega [\Delta_4]^\prime_\omega + [\Delta_2]^\prime_\omega [\Delta_3]^\prime_\omega [\Delta_4]^\prime_\omega \Bigr)
\ee
with the constraint \eqref{constrcon2}.

The entropy, which is the logarithm of the number of states, is given by the Legendre transform of the index, \ie, by the critical value of the entropy function
\bea
\label{conifold entropy function general}
\cS &= - \frac{\pi i N^2}{\tau\sigma} \Bigl( X_1X_2X_3 + X_1X_2X_4 + X_1X_3X_4 + X_2X_3X_4 \Bigr) \\
&\quad -2\pi i \biggl( \tau J_1 + \sigma J_2 + \sum_{I=1}^4 X_I Q_I \biggr) - 2\pi i \Lambda \biggl( \sum_{I=1}^4 X_I - \tau - \sigma \pm 1 \biggr) \;.
\eea
Here the variables $X_I$ stand for $[\Delta_I]_\omega$ or $[\Delta_I]'_\omega$ depending on whether we are in case I or III, respectively, and the $\pm$ sign is chosen accordingly. One can check that the two signs lead to the same entropy. We will give a general argument in Section~\ref{subsec:entropyfunctional_generic}.

In Section~\ref{sec: conifold} we will compare the field theory result (\ref{conifold entropy function general}) with the entropy of black holes in AdS$_5 \times T^{1,1}$, in the special case that $J_1 = J_2 \equiv J$ and the $SU(2)_{F_1} \times SU(2)_{F_2}$ symmetry is unbroken. To that purpose, let us specialize the index to the case that $\tau=\sigma$ and $\xi_{F_1} = \xi_{F_2} = 0$, which corresponds to $X_1=X_2$ and $X_3=X_4$. It is then useful to define the new variables
\be
X_R = X_1 + X_3 \;,\qquad\qquad X_B = \frac{X_1 - X_3}2 \;,
\ee
associated to R-symmetry and baryonic symmetry, respectively. The entropy function takes the simplified form
\be
\label{conifold entropy function special}
\cS = - \frac{ \pi i N^2}{2\tau^2} \, X_R \bigl( X_R^2 - 4 X_B^2 \bigr) - 2\pi i \Bigl( 2\tau J + X_R \, r + X_B Q_B \Bigr) - 2\pi i \Lambda \Bigl( 2X_R - 2\tau \pm 1 \Bigr) \;.
\ee

\subsection{Example: toric models}
\label{subsec:toric}

In this section we consider the gauge theory dual to an AdS$_5\times {\rm SE}_5$ geometry, where SE$_5$ is a toric Sasaki-Einstein manifold. The theory lives on a stack of $N$ D3-branes sitting at the toric Calabi-Yau singularity $C({\rm SE}_5)$ obtained by taking the cone over SE$_5$ \cite{Klebanov:1998hh, Morrison:1998cs}. There is a general construction to extract gauge theory data from the geometry of the Calabi-Yau singularity \cite{Hanany:2005ve, Franco:2005rj, Franco:2005sm, Feng:2005gw}. The main complication compared to the $\bC^3$ and the conifold cases is that there is no one-to-one correspondence between bi-fundamental fields $\Phi_I$ (and associated variables $\Delta_I$) and R-symmetries $R_a$. 
However, we will argue in general that there  always exist two corners of the space of chemical potentials where  \eqref{constr2} and \eqref{constr22} are satisfied and the results \eqref{result}  and \eqref{result2a} are valid. There are also other corners that should be analyzed separately for every specific model. Our findings are consistent with the case-by-case analysis performed in  \cite{Lanir:2019abx} for equal angular momenta.

We first need to understand how to write the trial central charges $a(\wh\Delta)$ and $a(\wh\Delta^\prime)$  that enter in the expressions \eqref{result}  and \eqref{result2a}. Since the quantities $\wh\Delta_I$ and $\wh\Delta_I^\prime$ satisfy the constraints \eqref{cos},  they can be interpreted as a set of   trial R-charges for the chiral fields in the quiver.     In the toric case, we can find an efficient parametrization  of the trial R-charges of fields using the data of the toric diagram. Let us review how this is done.

A toric Calabi-Yau threefold singularity can be specified by a fan, \ie, a convex cone in $\mathbb{R}^3$ defined by $D$ integer vectors $v_a=(1,{\vec v}_a)$ lying on a plane. The restrictions ${\vec v}_a$ of those vectors to the plane define a regular convex polygon with integer vertices called the toric diagram. In the list $\{v_a\}$ we should include all integer vectors such that $\vec v_a$ is along the perimeter of the polygon, \ie, we should include all integer points along the edges of the toric diagram. Moreover, we take the points $\vec v_a$ to be ordered in a counterclockwise fashion.
The number of vectors in the fan is associated with the total rank of the global symmetry of the dual field theory \cite{Franco:2005sm}: for a toric model with $D$ vectors in the fan (including integer points along the edges of the toric diagram) there is a flavor symmetry of rank $D-1$, besides the R-symmetry $U(1)_R$.%
\footnote{The distinction between R- and flavor symmetries changes in the case of extended supersymmetry.}
This allows us to parametrize flavor and R-symmetries in terms of variables  associated with the vertices of (and integer points along) the toric diagram.
In particular, the possible R-charges of fields in a toric theory can be parametrized using $D$ variables $\delta_a$ satisfying the constraint
\be
\label{constraint delta_a}
\sum_{a=1}^D \delta_a = 2 \;,
\ee
and the corresponding R-charge can be written as
\be
\label{trial R-charge}
R(\delta) = \sum_{a=1}^D \frac{\delta_a}2 \, R_a
\ee
in terms of a basis $\{R_a\}$.
This is done as follows \cite{Butti:2005vn}. In a minimal toric phase,%
\footnote{There are many different quiver theories that describe the same IR SCFT. They are called ``phases'', and are related by Seiberg dualities. The toric phases are the quiver theories where all gauge groups are $SU(N)$ with the same rank $N$. It turns out that all toric phases have the same number $G$ of gauge groups, but have different matter content. The ``minimal'' phases correspond to the quivers with the smallest number of chiral fields. There could be one or more minimal toric phases, for a given IR SCFT.}
the theory contains a number $G$ of gauge group factors $\rSU(N)$ equal to twice the area of the toric diagram.
Moreover, defining the vectors $\vec w_a = \vec v_{a+1} - \vec v_a$ lying in the plane (we identify indices modulo $D$, so that, for example, $\vec v_{D+1} \equiv \vec v_1$), for each pair $(a,b)$ such that $\vec w_a$ can be rotated counterclockwise into $\vec w_b$ in the plane with an angle smaller than $\pi$, there are precisely%
\footnote{The condition on the angle guarantees that the formula for the number of fields gives a non-negative integer.}
$\det \{ \vec w_a, \vec w_b\}$ bi-fundamental chiral fields $\Phi_{ab}$ with R-charge 
\be
\label{aa}
R[\Phi_{ab}] = \delta_{a+1}+ \delta_{a+2}+\ldots+ \delta_{b} \;.
\ee
Interestingly, for all toric models the trial central charge $a(\delta)$ is a homogeneous function of degree three at large $N$:
\be\label{a-charge}
a(\delta) = \frac9{32} \Tr R(\delta)^3 = \frac{9 N^2}{64} \sum_{a,b,c=1}^{D} C_{abc} \, \delta_a \, \delta_b \, \delta_c \;.
\ee
Here $N^2 C_{abc} = \frac14 \Tr R_a R_b R_c$ are the 't~Hooft anomaly coefficients, which can be read from the toric data through $C_{abc}= \bigl| \det \{ v_a,v_b,v_c\} \bigr|$ \cite{Benvenuti:2006xg}. Another important property of toric models that we will use in the following is that the constraints
\be
\sum_{I\in W} R[\Phi_I] = 2 \;,
\ee
that must be satisfied for each monomial term $W$ in the superpotential, always reduce to (\ref{constraint delta_a}).
Indeed, it follows from tiling techniques  \cite{Hanany:2005ve, Franco:2005rj, Franco:2005sm, Feng:2005gw, Butti:2005vn}
that the R-charges $R[\Phi_I]$, $ I\in W$, of the chiral fields entering in a superpotential monomial $W$  correspond to a partition of the $D$ elementary R-charges $\{\delta_1,\ldots,  \delta_D\}$ into sums of the form \eqref{aa}, with each $\delta_a$ entering in just one $R[\Phi_I]$. 
 
We can similarly parametrize the chemical potentials $\Delta[\Phi]$ entering the superconformal index in terms of $D$ basic quantities $\Delta_a$, $a=1,\ldots, D$. For the chiral fields $\Phi_{ab}$ we have
\be
\label{aaaa}
\Delta[\Phi_{ab}] = \Delta_{a+1}+ \Delta_{a+2}+\ldots+ \Delta_{b} \;.
\ee
The conditions
\be
\sum_{I\in W} \Delta[\Phi_I] = \tau+\sigma +n_W \;,
\ee
to be imposed for each monomial term $W$ in the superpotential (and where $n_W$ is the same for all monomial terms),
are then equivalent to 
\be
\sum_{a=1}^D \Delta_a = \tau+\sigma +n_W \;.
\ee
Independently of the value of $n_W$, we have
\be{}
[\Delta_D]_\omega =  \tau +\sigma  -1 - \biggl[ \sum\nolimits_{a=1}^{D-1} \Delta_a \biggr]_\omega  \;.
\ee
In general 
\be{}
\biggl[ \sum\nolimits_{a=1}^{D-1} \Delta_a \biggr]_\omega = \sum_{a=1}^{D-1} [  \Delta_a]_\omega + n
\ee
where $n=0, \ldots, D-2$, thus dividing the space of parameters into $D-1$ regions.

Two regions are particularly important for our analysis. The region $n =0$ corresponds to 
\be
\label{c1}
\sum_{a=1}^D [\Delta_a]_\omega = \tau+\sigma -1 \;,
\ee
while $n=D-2$ corresponds to
\be
\label{c2}
\sum_{a=1}^D [\Delta_a]^\prime_\omega = \tau+\sigma +1 \;.
\ee
We can argue that the two regions \eqref{c1} and \eqref{c2} are always realized somewhere in the space of parameters.
For example, we can choose one  elementary variable, say $\Delta_1$, to live in the fundamental strip $\im (-1/\omega) > \im \bigl(\Delta_1/\omega \bigr) > 0$ (see Fig.~\ref{fig: strip}) and slightly on the right of the vertical line passing through $\tau+\sigma-1$, while all the other $\Delta_a$ to live in the fundamental strip and slightly on the left of the vertical line passing through zero. One easily verifies that they can be arranged to satisfy \eqref{c1}.
A similar construction gives parameters satisfying \eqref{c2}.
We now argue that \eqref{c1} and \eqref{c2} imply \eqref{constr2} and \eqref{constr22}, respectively. 
We start noticing that
\be
\sum_{a=1}^D [\Delta_a]_\omega = \tau+\sigma -1  \qquad \Rightarrow \qquad \im \left( \frac1\omega \sum\nolimits_{a=1}^D [\Delta_a]_\omega \right) = \im \left ( -\frac{1}{\omega}\right ) \;.
\ee
Since each of the $[\Delta_a]_\omega$ lives in the fundamental strip $\im (-1/\omega) > \im \bigl( [\Delta_a]_\omega/\omega \bigr) > 0 $,  the previous equation implies that $\im (-1/\omega) > \im \bigl( \sum_{a\in S}[\Delta_a]_\omega/\omega \bigr) > 0$ for any proper subset $S$ of the indices $\{1,\dots,D\}$. Thus \eqref{c1} implies that
\be
\label{bb}
\left[ \sum\nolimits_{a\in S} \Delta_a\right ]_\omega = \sum\nolimits_{a\in S} [\Delta_a]_\omega
\ee
for any proper subset $S\subsetneq \{1,\dots,D\}$.
This implies that all charges in \eqref{aaaa} \emph{split}, in the sense that $\bigl[ \Delta_{a+1}+\ldots+ \Delta_{b} \bigr]_\omega = [\Delta_{a+1}]_\omega+\ldots+ [\Delta_{b}]_\omega$.  At this point, since all  $\bigl[ \Delta[\Phi_I] \bigr]_\omega$ split and each $\Delta_a$ enters precisely once in every superpotential constraint,  the  condition \eqref{constr2} is a consequence of \eqref{c1}.%
\footnote{There is an alternative algorithm that produces potentials $\Delta_I$ satisfying (\ref{constr2}). Choose a perfect matching $p_\alpha$ of the dimer model of the theory \cite{Franco:2005sm}. It divides the chiral fields into two groups: those $\Phi_\text{P}$ appearing in the perfect matching, and those $\Phi_\text{NP}$ not doing so. Choose the potentials $\Delta_\text{NP}$ to be in the fundamental strip and slightly on the left of the origin. Each superpotential term $W$ contains one and only one of the fields $\Phi_\text{P}$ (by definition of perfect matching): choose the corresponding $\Delta_\text{P}$ to be in the fundamental strip and slightly on the right of the point $\tau+\sigma-1$, in such a way that (\ref{constr2}) for that particular $W$ is satisfied. The drawback of this construction is that it does not tell us what the independent variables $\Delta_a$ are.}
A similar argument shows  that  \eqref{c2} implies \eqref{constr22}.  
Notice that the region specified by \eqref{constr2} can be larger than \eqref{c1} and, similarly, the region specified by \eqref{constr22} can be larger than \eqref{c2}. This, in particular, happens for Calabi-Yau cones with codimension-one orbifold singularities. This is the case of the models SPP and dP$_4$ discussed in \cite{Lanir:2019abx}.%
\footnote{Models with codimension-one orbifold singularities are characterized by toric diagrams where at least one vector $\vec v_a$ lies in the interior of an edge. The parameters  $\delta_a$  associated with integer points lying in the interior of an edge of the polygon enter in the parametrization \eqref{aa} of the R-charges of chiral fields, but no elementary field carries precisely charge  $\delta_a$.
In order to recover the region \eqref{constr2}, we can require the following. Construct a set $M$ by grouping the points $\{1, \dots, D\}$ along the toric diagram in the following way: Break each edge in two pieces at a non-integer point, and then for each vertex form a group (that will be an element of $M$) that contains the vertex itself and all other integer points (if any) along the two pieces of edges on the two sides. (In the absence of orbifold singularity, $M$ necessarily coincides with $\{1,\dots, D\}$.)
Then require that the sums split over the groups in $M$ for every proper subgroup $S' \subsetneq M$, and for every possible choice of $M$.
% it is enough to require that the sums $\sum_{a\in S} \Delta_a$ split when $S$ a proper subset of the set of vertices of the toric diagram (excluding points lying inside the edges).
This region is typically larger than \eqref{c1}.} 
For all the cones without orbifold singularities that we checked, the two regions \eqref{constr2} and \eqref{c1} coincide. It would be interesting to see if this is a general result.

We are now ready to evaluate the index. Consider region \eqref{constr2} first. Since the chemical potentials $[\Delta_I]_\omega$ split, the rescaled quantities
\be
\label{rescaled Delta variables2}
\wh \Delta_{a}  =  2 \, \frac{[\Delta_{a} ]_\omega}{\tau+\sigma-1} \qquad\text{with}\qquad \sum_{a=1}^D \wh\Delta_a =2
\ee
provide a parametrization of the R-charges of chiral fields in the quiver in the sense discussed above. Using the general formula \eqref{a-charge} we can then write
\be\label{a-charge2}
a(\wh\Delta) = \frac{9 N^2}{64} \sum_{a,b,c=1}^{D} C_{abc} \, \wh\Delta_a \, \wh\Delta_b \, \wh\Delta_c \;.
\ee
Plugging it into \eqref{result} and re-expressing the result in terms of the chemical potentials $[\Delta_{a} ]_\omega$, we find the large $N$ limit of the superconformal index in region \eqref{constr2}:
\be
\label{result200}
\log \cI \;\simeq\;  - \pi i N^2  \sum_{a, b, c=1}^D \frac{C_{abc}}{6} \, \frac{ [\Delta_a]_\omega [\Delta_b]_\omega [\Delta_c]_\omega}{\tau \sigma}  \;, \qquad \qquad \sum_{a=1}^D [\Delta_a]_\omega =  \tau+\sigma -1 \;.
\ee
A similar argument shows that, in region \eqref{constr22}, 
\be
\label{result2200}
\log \cI \;\simeq\;  - \pi i N^2  \sum_{a, b, c=1}^D \frac{C_{abc}}{6} \, \frac{ [\Delta_a]^\prime_\omega [\Delta_b]^\prime_\omega[\Delta_c]^\prime_\omega}{\tau \sigma}  \;, \qquad \qquad \sum_{a=1}^D [\Delta_a]^\prime_\omega =  \tau+\sigma +1 \;.
\ee
We will show in the next section that both \eqref{result200} and \eqref{result2200} lead to the same entropy.

\subsection{The entropy function}
\label{subsec:entropyfunctional_generic} 

For toric holographic quivers, we have found two different expressions, \eqref{result200} and \eqref{result2200}, for the large $N$ limit of the superconformal index that are valid in two different regions in the space of chemical potentials. 
The two expressions  differ only for the constraint and give rise to the very same entropy. This generalizes an observation made in \cite{Cabo-Bizet:2018ehj} for ${\cal N}=4$ SYM and holds for general quivers. 

To show that, we  define two entropy functions 
\begin{multline}
\label{Spm}
S_{\pm} = - \pi i N^2  \sum_{a, b, c=1}^D \frac{C_{abc}}{6} \, \frac{ X_a X_b X_c}{\tau\sigma}  - 2\pi i \biggl( \tau J_1 +\sigma J_2 + \sum_{a=1}^D X_a Q_a \biggr) \\
{} - 2 \pi i \Lambda \biggl( \sum_{a=1}^D X_a -  \tau -\sigma  \pm 1 \biggr) \;,
\end{multline}
where $\Lambda$ is a Lagrange multiplier and we used neutral variables $X_a$ to denote either $[\Delta_a]_\omega$ or $[\Delta_a]^\prime_\omega$. Each of the electric charges $Q_a \equiv R_a/2$ is defined in terms of an R-charge $R_a$ that assigns charge 2 to all chiral multiplets $\Phi_{ab}$ such that $\delta_a$ appears in the decomposition (\ref{aa}), and zero to all the other ones. The 't~Hooft anomaly coefficients are defined by
\be
C_{abc} \, N^2 = \frac14 \Tr R_a R_b R_c \;.
\ee
Above, $S_+$ is the prediction for the entropy of the dual black hole based on  the superconformal index in the region of parameters \eqref{constr2} while $S_-$ in the region \eqref{constr22}. The form of the entropy function \eqref{Spm} was first conjectured in \cite{Hosseini:2018dob}.
 
Observe that, since $S_\pm \pm 2 \pi i \Lambda$ are homogeneous functions of degree one in $(X_a, \tau,\sigma)$, the values of the functions $S_\pm(X_a, \tau,\sigma,\Lambda)$ at the critical point are related to the Lagrange multiplier by
\be
S_\pm \big|_\text{crit} = \mp 2 \pi i \Lambda \;.
\ee
Observe also that, if $Q_a, J_i$ are real (as charges should be), then the two functions are related by $\wb{S_+(X_a, \tau,\sigma, \Lambda)} = S_- \bigl( -\wb X_a, -\wb\tau, -\wb\sigma, \wb\Lambda \bigr)$. Hence, if $(X_a,\tau,\sigma , \Lambda)$ is a critical point of $S_+$, then $\bigl( - \wb X_a, - \wb\tau, - \wb\sigma, \wb\Lambda \bigr)$  is a critical point of $S_-$ with critical value
\be
\label{entropycrit}
S_- \big | _\text{crit} = \wb S_+ \big|_\text{crit} \;.
\ee
For arbitrary and general real charges  $Q_a$ and $J_i$, the critical value of $S_+$ is not real.  For ${\cal N}=4$ SYM, however, it becomes real and equal to the entropy  when imposing the non-linear constraint on conserved charges that characterizes supersymmetric black holes \cite{Hosseini:2017mds, Cabo-Bizet:2018ehj}. The same phenomenon was already observed in AdS$_4$ in \cite{Benini:2016rke}. We expect the same to be true for general  black holes in  Sasaki-Einstein compactifications. Even if this were wrong and $S_+$ were not real, it would still makes sense to identify the entropy with $\re S_+$. In all cases, we see from \eqref{entropycrit} that both constraints in \eqref{Spm} lead to the very same result for the entropy.

The entropy functions \eqref{Spm} give our general result for the entropy of black holes in AdS$_5\times  {\rm SE}_5$. We derived it for toric quiver gauge theories, but the very same argument can be extended to a class of more general non-toric quivers. In particular, the expression \eqref{Spm} only depends on the 't~Hooft anomaly coefficients $C_{abc}$ for a basis of R-symmetries 
and, as such, we expect that it is the correct result for generic holographic quiver theories.

%%%%%%%%%%%%%%%%%%%%%%%%%%%%%%%%%%%%%%%%%%%%%%%%%%
%%%%%%%%%%%%%%%%%%%%%%%%%%%%%%%%%%%%%%%%%%%%%%%%%%

\section{The universal rotating black hole}
\label{sec: universal}

In this section we discuss the case of the {\it universal rotating black hole} which has electric charge aligned with the exact R-symmetry of the theory. The black hole arises as a solution of minimal gauged supergravity in five dimensions and, as such, it can be embedded in any AdS$_5\times {\rm SE}_5$ compactification of type IIB and, more generally, in any  AdS$_5$ solution of type II or M theory.%
\footnote{It is believed and checked in many cases that the effective theory for all such compactifications can be consistently truncated to minimal gauged supergravity.}
Due to its universal character, most of the analysis is identical to the one for AdS$_5\times S^5$. It is however interesting to see how the details work.

The universal black hole in AdS$_5$ was found in \cite{Chong:2005hr} in minimal gauged supergravity in five dimensions. It has charge $Q$ under the graviphoton  and angular momenta $J_1$ and $J_2$ in AdS$_5$.%
\footnote{To compare with the notations of  \cite{Chong:2005hr}: $ Q_{{\rm there}} = -\sqrt{3} g Q_{{\rm here}}$ and $G_{{\text{N}}}^{(5)}=1$, $\ell_5=1/g$.}
The entropy can be compactly written as \cite{Kim:2006he}
\be
\label{KN5}
S(Q,J)= 2 \pi \sqrt{3 Q^2 -2 a (J_1+J_2)}
\ee
where we introduced the quantity
\be
\label{relation a GNewton}
a=\frac{\pi \ell_5^3}{8 G_{{\text{N}}}^{(5)}} \;,
\ee
where $G_{{\text{N}}}^{(5)}$ is the five-dimensional Newton constant and $\ell_5$ is the radius of AdS$_5$. The conserved charges must satisfy the  nonlinear constraint
\be
\label{chargesKN5}
8 Q^3 +6 a Q^2 -6 a (J_1+J_2) Q -2 a J_1 J_2 -4 a^2 (J_1+J_2) =0
\ee
for the BPS black hole to have a smooth horizon.

Consider now the uplift of the universal black hole to  AdS$_5\times {\rm SE}_5$, where ${\rm SE}_5$ is a Sasaki-Einstein manifold. In such an embedding, the standard holographic dictionary identifies $a$ with the central charge of the dual CFT$_4$. The  black hole carries angular momenta $J_1$ and $J_2$ and an electric charge aligned with the exact R-symmetry of the dual CFT$_4$. We need to  check that its entropy is reproduced by our result \eqref{result} (the same result can be similarly obtained using  \eqref{result2a} instead). It is convenient to parametrize the chemical potentials as
\be
\Delta_a = \frac{\tau+ \sigma - 1}{2} \left ( \wh \Delta_a^{(0)} + \wh\delta_a \right ) \;,
\ee
where $\wh\Delta_a^{(0)}$ is the \emph{exact} superconformal R-symmetry of the dual CFT$_4$ while $\wh\delta_a$ parametrize a basis of flavor symmetries. These quantities satisfy   
\be
\label{cc}
\sum_{a=1}^D \wh \Delta_a^{(0)}=2 \;,\qquad\qquad\qquad  \sum_{a=1}^D \wh \delta_a=0 \;.
\ee
The entropy of the universal black hole is given by the Legendre transform of \eqref{result}. 
Using \eqref{c1} we can write the entropy function as
\be
\cS = -\frac{ 4 \pi i }{27}  \, \frac{(\tau +\sigma - 1)^3}{\tau \sigma} \, a\Bigl( \wh \Delta^{(0)} +\wh \delta \, \Bigr)  - 2\pi i \Bigl( (\tau+\sigma - 1) Q + \tau J_1 +\sigma J_2 \Bigr)  \;,
\ee
where  we introduced a charge $Q = \frac 12 \sum_{a=1}^D \wh\Delta^{(0)}_a \, Q_a $ in the direction of the exact R-symmetry, and set all other charges to zero.  We need to extremize the function $\cS$ with respect to $\tau$, $\sigma$ and $\wh\delta_a$ subject to the constraint \eqref{cc}. By $a$-maximization, since $\wh\Delta^{(0)}_a$ is the exact R-symmetry, the function is extremized at $\wh\delta_a=0$.  We can then restrict the  entropy function to
\be
\cS = -\frac{ 4 \pi i  a }{27 }  \, \frac{(\tau +\sigma - 1)^3}{\tau \sigma}  - 2\pi i \Bigl( ( \tau+\sigma - 1) Q +  \tau  J_1 +\sigma J_2 \Bigr) \;,
\ee
where $a \equiv a\bigl(\wh \Delta^{(0)} \bigr)$ is the  central charge of the CFT$_4$, or, introducing a Lagrange multiplier $\Lambda$, 
\be
\label{efs}
\cS = - 4\pi i  a  \, \frac{\Delta^3}{\tau \sigma}  - 2\pi i \Bigl( 3 \Delta Q +  \tau  J_1 +\sigma J_2 \Bigr)- 2 \pi i \Lambda \bigl( 3 \Delta - \tau-\sigma +1 \bigr) \;.
\ee
If we set $a=a_{{\cal N}=4} = \frac14 N^2$, the function \eqref{efs} becomes identical to the entropy function of ${\cal N}=4$ SYM for  equal charges $Q_1=Q_2=Q_3 \equiv Q$, which is known to correctly reproduce \eqref{KN5} \cite{Hosseini:2017mds}. 
An analytic derivation of \eqref{KN5} and \eqref{chargesKN5}  for ${\cal N}=4$ SYM is
explicitly  discussed in \cite{Cabo-Bizet:2018ehj} and for equal angular momenta in \cite{Benini:2018ywd}. 
The charge constraint \eqref{chargesKN5} is obtained as the requirement that the extremum of $\cS$ be real. 

At this point,  the result for the universal black hole simply follows from the homogeneity properties of \eqref{efs}:
\be
S(Q,J_1,J_2)= \frac{a}{a_{{\cal N}=4} } \, S_{{\cal N}=4}\left (  \frac{a_{{\cal N}=4} }{a}  \,Q ,\; \frac{a_{{\cal N}=4} }{a} \, J_1 ,\; \frac{a_{{\cal N}=4} }{a} \, J_2  \right ) \;.
\ee
It is then immediate to derive the relations \eqref{KN5} and \eqref{chargesKN5}, thus completing our derivation.

%%%%%%%%%%%%%%%%%%%%%%%%%%%%%%%%%%%%%%%%%%%%%%%%%%
%%%%%%%%%%%%%%%%%%%%%%%%%%%%%%%%%%%%%%%%%%%%%%%%%%

\section{AdS\matht{_5} Kerr-Newman black holes in \matht{T^{1,1}}}
\label{sec: conifold}

We would like to compare the entropy function we obtained in Section~\ref{sec: toric} from the large $N$ limit of the superconformal index of generic (toric) quiver gauge theories, with the Bekenstein-Hawking entropy of BPS black holes in the corresponding 5d gauged supergravities. In particular, the setup we would like to analyze is that of type IIB supergravity on asymptotically AdS$_5 \times \mathrm{SE}_5$ spacetimes, where $\mathrm{SE}_5$ is a toric Sasaki-Einstein manifold,%
\footnote{More precisely, the cone over $\mathrm{SE}_5$ is a toric Calabi-Yau threefold.}
reduced and truncated to a 5d $\cN=2$ gauged supergravity on AdS$_5$. Unfortunately, with the exception of the case of $S^5$ truncated to the so-called 5d STU model, and the case of any $\mathrm{SE}_5$ truncated to minimal $\cN=2$ gauged supergravity (that we analyzed in Section~\ref{sec: universal}), all other known consistent truncations are to gauged supergravities with hypermultiplets (besides vector multiplets), and no supersymmetric black hole solutions have been constructed in such theories to date.

The strategy we propose to perform a test of our field theory results is as in \cite{Hosseini:2017mds}. We assume that a 5d BPS rotating black hole solution exists. Such a solution has the topology of a fibration of AdS$_2$ over $S^3$ (the three-sphere being the topology of the event horizon), and thus we can reduce it along the Hopf fiber of $S^3$. This gives a (putative) 4d BPS rotating black hole solution, with the same entropy.%
\footnote{The 4d solution has an exotic asymptotic behavior, that follows from the reduction of AdS$_5$ \cite{Hristov:2014eza}. Nonetheless, it has a regular extremal horizon, whose area determines the entropy.}
The reduction generates an extra vector field $A^0$, corresponding to the isometry along the Hopf fiber. The 4d black hole has one unit of magnetic charge under $A^0$, corresponding to the first Chern class of the Hopf fibration. Calling $J_1$ and $J_2$ the 5d angular momenta along two orthogonal planes, the quantity $J_1 + J_2$ appears in 4d as the electric charge under $A^0$, while $J_1 - J_2$ becomes the angular momentum of the 4d black hole. Constructing such a 4d rotating black hole solution is still a difficult task, and an attractor mechanism is not known in general.\footnote{There are however some general results for theories with vector multiplets \cite{Hristov:2018spe,Hristov:2019mqp}.} However, if we restrict to 5d black holes with two equal angular momenta $J_1 = J_2$ (so that the isometry of the squashed $S^3$ is enhanced from $U(1)^2$ to $U(1) \times SU(2)$), then the 4d black hole is static: in this case we can determine its entropy by exploiting the attractor mechanism in the near-horizon geometry \cite{DallAgata:2010ejj, Halmagyi:2013sla, Klemm:2016wng}, without actually constructing the whole solution.

The simplest non-trivial example is when SE$_5$ is $T^{1,1}$, the base of the conifold Calabi-Yau threefold, whose holographic dual is the Klebanov-Witten gauge theory \cite{Klebanov:1998hh}. We already presented the field theory analysis in Section~\ref{subsec:conifold}. On the other hand, starting from 10d type IIB supergravity on $T^{1,1}$, we can exploit a consistent truncation that preserves $SU(2)^2 \times U(1)$ isometry, down to a 5d $\cN=2$ gauged supergravity with the graviton multiplet, two vector multiplets and two hypermultiplets. This is the second truncation presented in Section~7 of \cite{Cassani:2010na} (see also \cite{Bena:2010pr, Halmagyi:2011yd}). On the AdS$_5$ vacuum, one vector multiplet (sometimes called ``Betti multiplet'') is massless and is associated to the baryonic symmetry, while the other vector multiplet is massive.

Hence, with the simplification that $J_1 = J_2$ and only the R-symmetry and baryonic symmetry charges are turned on (while the $SU(2)^2$ isometry of $T^{1,1}$ is unbroken), we will be able to match the Legendre transform of the superconformal index at large $N$ with the extremization problem that comes from the attractor mechanism in supergravity. It follows that the bulk and boundary computations of the entropy exactly match.

\subsection{Reduction from 5d to 4d and the attractor mechanism}
\label{subsec:reduction}

A 5d $\cN=2$ Abelian gauged supergravity with $n_V$ vector multiplets and $n_H$ hypermultiplets --- whose main building blocks we summarize in Appendix~\ref{app:5d_sugra} --- is specified by the following data \cite{Gunaydin:1983bi, Gunaydin:1984ak, Ceresole:2000jd}:
\begin{enumerate}

\item A very special real manifold $\cS\cM$ of real dimension $n_V$, specified by a symmetric tensor of Chern-Simons couplings $C_{IJK}$ with $I, J, K=1,\ldots, n_V+1$. The coordinates are $\Phi^I$ with the cubic constraint
\be
\label{5d constraint}
\cV(\Phi) \,\equiv\, \frac16 C_{IJK} \Phi^I \Phi^J \Phi^K = 1 \;.
\ee

\item A quaternionic-K\"ahler manifold $\cQ\cM$ of real dimension $4n_H$ with coordinates $q^u$.

\item A set of $n_V+1$ Killing vectors $k^u_I$ (that could be linearly dependent, or vanish) on $\cQ\cM$, compatible with the quaternionic-K\"ahler structure, representing the isometries to be gauged by the vector fields $A^I$. Each Killing vector comes equipped with a triplet of moment maps $\vec P_I$.%
\footnote{If $n_H=0$, instead, one has to specify $n_V$ Fayet-Iliopoulos parameters $\zeta^I$, not all vanishing. \label{foo: FI}}
\end{enumerate}
On the other hand, a 4d $\cN=2$ Abelian gauged supergravity with $n_V+1$ vector multiplets and $n_H$ hypermultiplets --- that we summarize in Appendix~\ref{app:4d_sugra} --- is specified by the following data (see for instance \cite{Andrianopoli:1996cm, Craps:1997gp}):
\begin{enumerate}

\item A special K\"ahler manifold $\cK\cM$ of complex dimension $n_V+1$, with coordinates $z^I$ and $I=1, \dots, n_V+1$. We will work  in a duality frame in which the geometry is specified by holomorphic sections $X^\Lambda(z)$, with $\Lambda = 0, \ldots, n_V+1$, and a holomorphic prepotential $F(X)$, homogeneous of degree two.

\item A quaternionic-K\"ahler manifold $\cQ\cM$ of real dimension $4n_H$ with coordinates $q^u$.

\item In duality frames in which all gaugings are purely electric, a set of $n_V+2$ Killing vectors $k^u_\Lambda$ (that could be linearly dependent, or vanish) on $\cQ\cM$, compatible with the quaternionic-K\"ahler structure, representing the isometries to be electrically gauged by the vector fields $A^\Lambda$. Each Killing vector comes equipped with a triplet of moment maps $\vec P_\Lambda$ (see footnote~\ref{foo: FI}).
\end{enumerate}

We reduce the 5d theory on a circle, that will eventually be the Hopf fiber of $S^3$. Following \cite{Andrianopoli:2004im, Behrndt:2005he, Cardoso:2007rg, Looyestijn:2010pb, Klemm:2016kxw, Hosseini:2017mds} we use the ansatz
\bea
\label{ansatz}
ds_{(5)}^2 &= e^{2\wt \phi}ds_{(4)}^2 + e^{-4\wt \phi} \bigl( dy - A^0_{(4)} \bigr)^2 \\
\Phi^I &= - e^{2\wt \phi} \im z^I \;.
\eea
Here $y$ is the direction of the circular fiber, that we take with range $4\pi/g$ in terms of the coupling $g = \ell_5^{-1}$ inversely proportional to the AdS$_5$ radius $\ell_5$, therefore the size of the circle is $e^{-2\wt\phi}\, 4\pi/g$. Because of the constraint $\cV(\Phi)=1$ in (\ref{5d constraint}), the field $\wt\phi$ is redundant and can be eliminated with $e^{-6\wt\phi} = - \cV( \im z^I)$. On the other hand, $A^0_{(4)}$ is the Kaluza-Klein vector. As noted in \cite{Hristov:2014eba,Hosseini:2017mds}, a Scherk-Schwarz twist for the gravitino as in \cite{Looyestijn:2010pb} is necessary to satisfy the BPS conditions in 4d. We prefer to work in a gauge in which all bosonic fields are periodic around the circle, but there are flat gauge connections $\xi^I$ turned on along $y$. This corresponds to the ansatz
\be
\label{modified_ansatz}
A^I_{(5)} = A^I_{(4)} + \re z^I \, \bigl( dy -A^0_{(4)}\bigr) + \xi^I dy \;,
\ee
together with no $y$-dependence for any field. Notice that this ansatz is invariant under the redefinitions
\be
\label{ansatz redundancy}
z^I \to z^I + \delta \xi^I \;,\qquad A_{(4)}^I \to A_{(4)}^I + \delta \xi^I A^0_{(4)} \;,\qquad \xi^I \to \xi^I - \delta\xi^I
\ee
where $\delta\xi^I$ are real parameters. We will fix this redundancy below. The reduction of the 5d theory can be found in Appendix~\ref{app:SS_reduction}. The resulting 4d data in terms of 5d ones are as follows.
\begin{enumerate}

\item The special K\"ahler manifold in 4d is described by the prepotential
\be
\label{prepot}
F(X) = \frac{1}{6} \, C_{IJK} \, \frac{\check X^I \check X^J \check X^K}{X^0} \qquad\text{with}\qquad \check X^I = X^I + \xi^I X^0 \;.
\ee
The holomorphic sections $X^\Lambda$ can be used as homogeneous coordinates, and the physical scalars are identified with the special coordinates $z^I = X^I/X^0$.

\item The quaternionic-K\"ahler manifold in 4d is the same as in 5d.

\item The 4d Killing vectors $k^u_I$ are inherited from 5d, while the additional Killing vector is
\be
k_0^u = \xi^I k_I^u \qquad\Rightarrow\qquad \vec P_0 = \xi^I \vec P_I \;,
\ee
and is gauged by the Kaluza-Klein vector field $A^0_{(4)}$.

\end{enumerate}

Next, we study the attractor equations for the near-horizon limit of 4d BPS static black hole solutions \cite{DallAgata:2010ejj, Halmagyi:2013sla, Klemm:2016wng}. Our goal is to use the BPS equations to fix the VEVs in massive vector multiplets and hypermultiplets, and be left with an extremization principle for the scalars in massless vector multiplets, similarly to \cite{Benini:2017oxt, Hosseini:2017fjo}. We consider the near-horizon geometry AdS$_2 \times S^2$:
\be
ds^2_\text{near-horizon} = - \frac{r^2}{L_\mathrm{A}^2} \, dt^2 + \frac{L_\mathrm{A}^2}{r^2} \, dr^2 + L_\mathrm{S}^2 \, ds^2_{S^2} \;,
\ee
where $L_\mathrm{A}$ and $L_\mathrm{S}$ are the radii of AdS$_2$ and $S^2$, respectively.
Electric and magnetic charges are defined as appropriate integrals over $S^2$ in the near-horizon region, respectively:
\be
\label{charge_def}
q_\Lambda = \frac{g}{4\pi} \int_{S^2} 16\pi G_\text{N}^{(4)} \, \frac{\delta S_\text{4d}}{\delta F^\Lambda} \;,\qquad\qquad p^\Lambda = \frac{g}{4\pi} \int_{S^2} F^\Lambda \;.
\ee
Here $G_\text{N}^{(4)}$ is the 4d Newton constant, related to the 5d one by
\be
\frac{4\pi}{G_\text{N}^{(5)} g} = \frac1{G_\text{N}^{(4)}} \;,
\ee
while $S_\text{4d}$ is the 4d supergravity action.
The 4d black holes we are interested in have both electric and magnetic charges. The magnetic charge $p^0=1$ is equal to the first Chern class of the Hopf fibration. On the other hand, we fix the redundancy (\ref{ansatz redundancy}) by setting the remaining magnetic charges to zero. In Appendix~\ref{app: charges} we compute the relation of the 5d charges $Q_I$ and angular momentum $J$ measured at infinity, with the 4d charges measured at the horizon. We should be careful that only massless vector fields are associated to conserved charges. We indicate as $\bB\ud{I}{J}$ the matrix of linear redefinitions such that $\bB\ud{I}{J} A_\mu^J$ are the 5d mass eigenstates in the AdS$_5$ vacuum, and we take the index $\fT$ to run only over the massless vectors $\bB\ud{\fT}{J} A_\mu^J$. The corresponding conserved charges are $Q_\fT \equiv Q_J (\bB^{-1})\ud{J}{\fT}$. We find
\bea
\label{4d charges from 5d charges}
p^0 &= 1 \;,\qquad\qquad& q_0 &= 4 G_\text{N}^{(4)} g^2 J + \frac13 C_{IJK} \xi^I \xi^J \xi^K \;, \\
p^I &= 0 \;,\qquad\qquad& q_\fT &= 4G_\text{N}^{(4)} g^2 Q_\fT + \frac12 C_{\fT JK} \xi^J \xi^K \;,
\eea
where $J_1 = J_2 \equiv J$, while the ``non-conserved charges'' $q_{J \neq \fT}$ will be fixed by the equations of motion.
Notice that the conserved charges $Q_\fT$ are the same, but possibly in a different basis, as the charges $Q_a$ introduced in Sections \ref{subsec:toric} and \ref{subsec:entropyfunctional_generic}.%
\footnote{Similarly, the restriction of $C_{IJK}$ to $C_{\fT\fJ\fK}$ with curly indices is the same, but possibly in a different basis, as the 't~Hooft anomaly coefficients $C_{abc}$ previously defined.}

Using a symplectic covariant notation, electric and magnetic charges form a symplectic vector
\be
\cQ = (p^\Lambda,q_\Lambda)\;.
\ee
One also defines
\be
\vec\cP = (0, \vec P_\Lambda) \;,\qquad\qquad \vec\cQ = \langle \vec\cP , \cQ\rangle \;,
\ee
where vectors are triplets and $\langle V, W \rangle = V_\Lambda W^\Lambda - V^\Lambda W_\Lambda$ is the symplectic-invariant antisymmetric form.

To find covariantly-constant spinors, we impose the following twisting ansatz:
\be
\epsilon_i = - \vec\cQ \cdot \vec\sigma\du{i}{j} \, \Gamma^{\hat t \hat r } \epsilon_j\;,
\ee
whose square gives $\vec\cQ \cdot \vec\cQ = 1$. Here $\Gamma^{\hat t\hat r}$ is the antisymmetric product of two gamma matrices with flat indices $\hat t$ and $\hat r$. We choose a gauge in which $\cQ^1=\cQ^2=0$ and
\be
\label{first_BPS}
\cQ^3 = -1
\ee
at the horizon, as in \cite{Benini:2017oxt}.

The remaining BPS conditions are in general complicated, but they simplify at the horizon \cite{DallAgata:2010ejj, Halmagyi:2013sla, Klemm:2016wng}. First, Maxwell's equations give
\be
\label{second_BPS}
\cK^u h_{uv} \langle\cK^v,\cQ\rangle = 0\;,
\ee
where we defined
\be
\cK^u = (0,k_\Lambda^u)
\ee
because we work in a duality frame with purely electric gaugings. In fact, (\ref{second_BPS}) in this case is equivalent to
\be
\label{cond from spherical symm}
p^\Lambda k_\Lambda^u = 0
\ee
that must hold in the full solution simply because of spherical symmetry (see Appendix~\ref{app: charges}).
Second, vanishing of the hyperino variation implies
\be
\label{third_BPS}
\langle \cK^u, \cV \rangle=0 \;,
\ee
where $\cV(z,\bar z) = e^{\cK/2} (X^\Lambda, F_\Lambda)$ is the covariantly-holomorphic section defined in (\ref{cov holo sections}) and $F_\Lambda = \partial_\Lambda F(X)$.
Third, we have the attractor equations%
\footnote{There is an extra factor of $2$ in front of $L^2_\mathrm{S}$ compared to \cite{Andrianopoli:1996cm, Benini:2016rke, Benini:2017oxt} due to the different normalization of kinetic terms in the Lagrangian (\ref{4d_lagrangian}): this is noticed footnote 4 of \cite{Looyestijn:2010pb} and in footnote 10 of \cite{Hosseini:2017mds}.}
\be
\label{fourth_BPS}
\parfrac{}{z^I} \left( \frac{\cZ}{\cL} \right) = 0 \;,\qquad\qquad \frac{\cZ}{\cL} = 2i g^2 L_\mathrm{S}^2 \;,
\ee
where the derivatives are with respect to the physical scalars $z^I$ and we defined
\be
\cZ = \langle\cQ, \cV\rangle \;,\qquad\qquad \cL = \langle \cP^3, \cV \rangle \;.
\ee
The equation on the right in (\ref{fourth_BPS}) determines $L_\mathrm{S}$, and thus the horizon area.

\subsection{Example: the conifold}
\label{sec: SUGRA example conifold}

We apply the general strategy to the case of the conifold. We start with the 5d $\cN=2$ gauged supergravity with $n_V=2$ vector multiplets and $n_H=2$ hypermultiplets constructed in Section~7 of \cite{Cassani:2010na} (called the ``second model'' in that paper), obtained from a consistent reduction of 10d type IIB supergravity on $T^{1,1}$ that preserves the $SU(2)^2 \times U(1)$ isometry. In Appendix~\ref{subapp:Cassani_matching} we have recast its action as in the general formalism, and in Appendix~\ref{subapp:conifold_reduction} we have reduced it down to 4d $\cN=2$ gauged supergravity. We are now ready to look for BPS near-horizon black hole solutions.

 Using (\ref{gaugings}) and  (\ref{moment_maps}), the conditions \eqref{first_BPS} and \eqref{cond from spherical symm} take the form:
\be
\label{solution hyper BPS conditions}
\left\{ \begin{aligned}
P^3_0 &= -1\\
k_0^u &= 0
\end{aligned} \right.
\qquad\Rightarrow\qquad\qquad
b_{1,2}^\Omega = c_{1,2}^\Omega=0 \;,\qquad \xi^1 = - \xi^2 = - \frac{1}{3} \;,
\ee
where $b_{1,2}^\Omega, c_{1,2}^\Omega, a, \phi, C_0, u$ are the scalar fields in hypermultiplets. In fact, since \eqref{cond from spherical symm} must hold in the whole solution, so (\ref{solution hyper BPS conditions}) does. Using the form (\ref{moment_maps}) of the moment maps, this is consistent with $\cQ^1=\cQ^2=0$. The hyperino condition \eqref{third_BPS} then gives
\be
\label{hyperino_condition}
X^1 + X^2 = 0
\ee
at the horizon, where $X^\Lambda$ are the holomorphic sections.
The fields $C_0$ and $\phi$ are not fixed by the equations of motion. However, together they form the axiodilaton of type IIB supergravity and are thus fixed by the boundary conditions that set them in terms of the complexified gauge coupling of the boundary theory.
As apparent from the expression of $k_2^u$ in (\ref{gaugings}), $a$ is a St\"uckelberg field that breaks an Abelian gauge symmetry and is eaten up as the corresponding gauge field becomes massive via Higgs mechanism.

The remaining BPS conditions are the attractor equations (\ref{fourth_BPS}). Given $C_{IJK}$ in (\ref{Cijk}), the prepotential is
\be
F(X) =  \frac{\check X^1 \bigl( (\check X^2)^2- (\check X^3)^2 \bigr)}{X^0} \qquad\text{where}\qquad \check X^I = X^I + \xi^I X^0 \;.
\ee
Using special coordinates $z^I = X^I/X^0$ as well as homogeneity of the prepotential $F(X)$, one can easily show that the two equations in (\ref{fourth_BPS})  are equivalent to
\be
\label{fourth_BPS with X}
\partial_\Lambda \Bigl[ e^{-\cK/2} \Bigl( \cZ(X) - 2i g^2 L_\mathrm{S}^2 \, \cL(X) \Bigr) \Bigr] = 0 \;,
\ee
where the derivatives are with respect to \emph{independent} sections $X^\Lambda$. In these equations $L_\mathrm{S}$ should be regarded as one of the unknowns. Notice that (\ref{fourth_BPS}) or (\ref{fourth_BPS with X}) give, in general, isolated solutions in terms of $(z^I, L_\mathrm{S})$, however the sections $X^\Lambda$ are only fixed up to the ``gauge'' redundancy (related to K\"ahler transformations on $\cK\cM$) $X^\Lambda \to e^f X^\Lambda$. In order to remove the redundancy, we choose to fix $\cL(X)$ to a constant, which can elegantly be imposed by taking a derivative of the square bracket in (\ref{fourth_BPS with X}) with respect to $L_\mathrm{S}^2$ as well. More precisely, expanding $\cZ$ and $\cL$ using (\ref{moment_maps}), we consider the following set of equations:
\bea
\label{fourth_BPS gauge fixed}
\partial_\Lambda \biggl[ \frac{ X^1 \bigl( (X^2)^2 - (X^3)^2 \bigr) }{ (X^0)^2} + \wh q_\Lambda X^\Lambda - 2i g^2 L_\mathrm{S}^2 \Bigl( 3X^1 - X^0 - 2e^{-4u}(X^1 + X^2) - \alpha \Bigr) \biggr] &= 0 \\
\parfrac{}{L_\mathrm{S}^2} \biggl[ \frac{ X^1 \bigl( (X^2)^2 - (X^3)^2 \bigr) }{ (X^0)^2} + \wh q_\Lambda X^\Lambda - 2i g^2 L_\mathrm{S}^2 \Bigl( 3X^1 - X^0 - 2e^{-4u}(X^1 + X^2) - \alpha \Bigr) \biggr] &= 0
\eea
where
\be
\label{hatted 4d charges}
\wh q_I = q_I - \frac12 C_{IJK} \xi^J \xi^K \;,\qquad \wh q_0 = q_0 - \frac13 C_{IJK} \xi^I \xi^J \xi^K \;.
\ee
The first line is the same as (\ref{fourth_BPS with X}), except for the addition of the constant $\alpha$ that does not affect the equations. The second line fixes the gauge $\cL = \alpha$. Notice that (\ref{hyperino_condition}) should be imposed after solving (\ref{fourth_BPS gauge fixed}).

From the point of view of AdS/CFT, only massless vector fields correspond to symmetries of the boundary theory and only their charges are conserved and fixed by the boundary conditions. On the contrary, the ``charges'' under massive vector fields are not conserved, and their radial profile should be determined by the equations of motion. The spectrum of the 5d supergravity under consideration around its supersymmetric AdS$_5$ vacuum was computed in \cite{Cassani:2010na} and we report it in our conventions in (\ref{spectrum of vectors}). In the basis
\bea
\label{mass eigenstates}
A^R &\,\equiv\, A^1 - 2A^2 \;,\qquad& &A^3 \;,\qquad& A^W &\,\equiv\, A^1 + A^2 \;, \\
k_R &\,\equiv\, \tfrac13 (k_1 - k_2) \;,\qquad& &k_3 \;,\qquad& k_W &\,\equiv\, \tfrac13(2k_1 + k_2) \;,
\eea
it turns out that $A^R$ (corresponding to the R-symmetry) and $A^3$ are massless, while $A^W$ is massive because of Higgs mechanism eating up the St\"uckelberg field $a$. In (\ref{mass eigenstates}) we have indicated also the Killing vectors of the corresponding gauged isometries. On the black hole background the mass eigenstates may change (because the gauge kinetic functions have a non-trivial radial profile), however the fact that
\be
\label{vanishing of Killing vectors}
k_R = k_3 = 0
\ee
everywhere --- which follows from (\ref{solution hyper BPS conditions}) --- guarantees that there is no hypermultiplet source in the 5d Maxwell equations (\ref{5d EOMs}) and thus the Page charges $Q_R$ and $Q_3$ are conserved (while $Q_W$ is not).

Indeed, the variation in (\ref{fourth_BPS gauge fixed}) with respect to $X^2$ gives the complex equation
\be
\label{eqn massive vector charge}
2\, \frac{X^1 X^2}{(X^0)^2} + \wh q_2 + 4i g^2 L_\mathrm{S}^2 \, e^{-4u} = 0
\ee
that fixes $u$ and the ``non-conserved charge'' $q_2$ in terms of the sections and $L_\mathrm{S}$. We can then use the hyperino condition (\ref{hyperino_condition}) to eliminate $X^2$ as well. Notice that the second condition in (\ref{solution hyper BPS conditions}) implies that in 5d we cannot turn on a ``flat connection'' for $A^W$ along the circle.

We are left with the unknowns $X^0, X^1, X^3, L_\mathrm{S}^2$. One can check that, when (\ref{hyperino_condition}) and (\ref{eqn massive vector charge}) are in place, the remaining equations in (\ref{fourth_BPS gauge fixed}) are equivalent to the conditions of extremization of the function
\be
\cS = \beta \Biggl[ \frac{X^1 \bigl( (X^1)^2 - (X^3)^2 \bigr) }{ (X^0)^2 } + \wh q_0 X^0 + 3 \wh q_R X^1 + \wh q_3 X^3 - 2i g^2 L_\mathrm{S}^2 \bigl( 3X^1 - X^0 - \alpha \bigr) \Biggr]
\ee
with respect to the variables $X^0, X^1, X^3, L_\mathrm{S}^2$. Here $\beta$ is a constant included for later convenience, while $\wh q_R$ is the charge with respect to the massless vector $A_\mathrm{R}$:
\be
\wh q_R = \frac{\wh q_1 - \wh q_2}3 = \frac{g}{4\pi} \int_{S^2} 16\pi G_\text{N}^{(4)} \, \frac{\delta S_\text{4d}}{\delta F^R} - \frac16 \bigl( C_{1JK} - C_{2JK} \bigr) \xi^J \xi^K = 4g^2 G_\text{N}^{(4)} Q_R \;.
\ee
It is encouraging that we find an extremization problem in which only conserved charges appear. Since $\cS$ is homogeneous in $X^\Lambda$ of degree 1 except for the term involving $\alpha$, it follows that $\cS\big|_\text{crit} = 2i \alpha\beta g^2 L_\mathrm{S}^2$ at the critical point. With the choice
\be
\alpha \beta = \frac{\pi}{2i  G_\mathrm{N}^{(4)} g^2}
\ee
we obtain that $\cS\big|_\text{crit}$ is the black hole entropy:
\be
\cS\big|_\text{crit} = \frac{4\pi L_\mathrm{S}^2}{4G_\mathrm{N}^{(4)}} = S_\mathrm{BH} \;,
\ee
and therefore $\cS$ is the entropy function. Using (\ref{4d charges from 5d charges}) and (\ref{hatted 4d charges}) we can express the 4d charges $\wh q_0$, $\wh q_\fT$ computed at the horizon in terms of the 5d black hole charges $J$, $Q_\fT$ computed at infinity:
\begin{multline}
\label{entropy function conifold}
\cS = \frac1\alpha \Biggl[ \frac{\pi}{2iG_\text{N}^{(4)} g^2} \, \frac{(X^1)^3 - X^1(X^3)^2}{(X^0)^2} - 2\pi i \Bigl( JX^0 + 3 Q_R X^1 + Q_3 X^3 \Bigr) \\
- 2\pi i \Lambda \Bigl( 3X^1 - X^0 - \alpha \Bigr) \Biggr] \;,
\end{multline}
where we redefined the Lagrange multiplier $L_\text{S}^2 = 2i G_\text{N}^{(4)} \Lambda$ for convenience.

It remains to spell out the AdS/CFT dictionary between gravity and field theory charges. First, the gauge group ranks in field theory are determined by (see Appendix~\ref{subapp: charge quantization})
\be
\label{relation N conifold}
N^2 = \frac{8\pi}{27 \, G_\text{N}^{(5)} g^3} = \frac{2}{27 \, G_\text{N}^{(4)} g^2} \;.
\ee
This is in agreement with (\ref{relation a GNewton}) using $a = \frac{27}{64}N^2$ for the Klebanov-Witten theory.
Second, the angular momentum $J$ is the same in gravity and in field theory. Third, the electric charges are identified as
\be
\label{AdS/CFT dictionary of charges}
r = 2Q_R \;,\qquad\qquad Q_B = \frac43 Q_3 \;.
\ee
This is determined as follows. From (\ref{cov derivative SUSY parameter}) we infer that the gravitino components have charge $Q_R = \pm \frac12$. In the boundary field theory, the corresponding operators are of the schematic form $\Tr(F_{\mu\nu} \Gamma^\nu \lambda)$ (where $F$ is a field strength and $\lambda$ a gaugino) and have charge $r = \pm1$ under $U(1)_R$. We deduce the first relation in (\ref{AdS/CFT dictionary of charges}). Obtaining the second relation is more subtle because no supergravity field is charged under $A^3$: what is charged are massive particles obtained from D3-branes wrapped on the 3-cycle of $T^{1,1}$, corresponding to dibaryon operators $A_{1,2}^N$ or $B_{1,2}^N$ in field theory. The 5d supergravity gauge field $A^3$ comes from the reduction of the Ramond-Ramond field strength $F_5^\text{RR}$ of 10d type IIB supergravity on $T^{1,1}$. Therefore, from the 10d flux quantization condition we can deduce the 5d charge quantization condition $4Q_3/3N \in \bZ$ (see the details in Appendix~\ref{subapp: charge quantization}). In field theory the dibaryon operators have charge $Q_B = \pm N$, implying the second relation in (\ref{AdS/CFT dictionary of charges}). Alternatively, we could compare the Chern-Simons terms restricted to massless vector fields in the 5d Lagrangian with the 't~Hooft anomalies of the boundary theory. Taking into account the 't~Hooft anomalies $\Tr(r^3) = \frac32 N^2$ and $\Tr(r Q_B^2) = -2N^2$ at leading order in $N$, the restriction of the 5d Chern-Simons action in (\ref{5d_lagrangian}) to $A^W \to 0$ matches the general expression
\be
S_\text{CS} = \frac{g^3}{24\pi^2} \int \Tr (Q_a Q_b Q_c) \, F^a \wedge F^b \wedge A^c
\ee
after setting $A^R \to 2A_r$ and $A^3 \to \frac43 A_B$. These correspond to (\ref{AdS/CFT dictionary of charges}).

Rewriting the entropy function (\ref{entropy function conifold}) in terms of field theory charges, we find
\begin{multline}
\cS = \frac1\alpha \Biggl[  - \frac{27\pi i N^2}{4} \, \frac{ (X^1)^3 - X^1 (X^3)^2}{ (X^0)^2} - 2\pi i \biggl( JX^0 + \frac32 r X^1 + \frac34 Q_B X^3 \biggr) \\
- 2\pi i \Lambda \Bigl( 3X^1 - X^0 - \alpha \Bigr) \Biggr] \;.
\end{multline}
This exactly matches the entropy function (\ref{conifold entropy function special}) we found in field theory from the large $N$ limit of the superconformal index of the Klebanov-Witten theory, after the change of coordinates $X^0 \to 2\alpha\tau$, $X^1 \to 2\alpha X_R/3$, $X^3 \to 4\alpha X_B/3$.

%%%%%%%%%%%%%%%%%%%%%%%%%%%%%%%%%%%%%%%%%%%%%%%%%%
%%%%%%%%%%%%%%%%%%%%%%%%%%%%%%%%%%%%%%%%%%%%%%%%%%

\section*{Acknowledgements}
We thank Arash Arabi Ardehali, Seyed Morteza Hosseini, Paolo Milan, and Antoine Van Proeyen for useful conversations and correspondence.
F.B., S.S., and Z.Z. are partially supported by the MIUR-PRIN contract 2015 MP2CX4, and by the INFN ``Iniziativa Specifica ST\&FI''.
F.B. is also supported by the MIUR-SIR grant RBSI1471GJ, and by the ERC-COG grant NP-QFT no.~864583.
A.Z. is partially supported by the INFN, the ERC-STG grant 637844-HBQFTNCER, and the MIUR-PRIN contract 2017 CC72MK003.

%%%%%%%%%%%%%%%%%%%%%%%%%%%%%%%%%%%%%%%%%%%%%%%%%%
%%%%%%%%%%%%%%%%%%%%%%%%%%%%%%%%%%%%%%%%%%%%%%%%%%

\appendix

%%%%%%%%%%%%%%%%%%%%%%%%%%%%%%%%%%%%%%%%%%%%%%%%%%
%%%%%%%%%%%%%%%%%%%%%%%%%%%%%%%%%%%%%%%%%%%%%%%%%%

\section{Subleading effect of simplifications}
\label{app: simplifications}

\subsection{Simplifications of the building block}
\label{subapp: simplifications building block}

We want to show that the terms neglected in passing from \eqref{before simplification} to \eqref{after simplification} are subleading at large $N$. We will first analyze the effect of dropping the term $\omega (d-c)/N$ from the arguments of the gamma functions, in all those terms with $\gamma\neq \delta$. We will later estimate the contribution from the terms with $\gamma=\delta$ that were discarded from the sum.

Defining 
\be
\label{def function to bound}
f(z)=\sum_{\gamma \neq \delta}^{\wt N } \log \wt\Gamma \left(z + \omega \frac{\delta-\gamma}{\wt N} ; ab\omega, ab\omega \right) \;,
\ee
we want to show that
\be
\label{approx main result}
\left|f\bigl( z+C\omega/\wt N \bigr) - f(z)\right| \leq \cO(N\log N) \;,
\ee
where $C=(d-c)/ab$, $z= \Delta + \omega \bigl( d-c+as+br \bigr)$ and $c,d=1,\ldots,ab$, $r=0,\ldots,a-1$, $s=0,\ldots,b-1$. Without loss of generality we can assume $C>0$, because the case $C<0$ is analogous while $C=0$ is trivial. As in \cite{Benini:2018ywd}, we discard the Stokes lines $\Delta \in \bZ + \bR \,\omega$ except the point $\Delta=0$, because the limit we compute would be singular along those lines anyway. If $\Delta$ is not on a Stokes line, then the restriction of $f$ to a straight line in the complex plane passing through the points $z$ and $z+\omega$ is a $C^\infty$ complex function. In the case $\Delta = 0$, instead, we consider the restriction of $f$ to a straight closed segment from $z$ to $z + C\omega/\wt N$ and one can check that $f$ is $C^\infty$ along that segment, because for $\gamma \neq \delta$ the segment, suitably shifted, does not hit zeros nor poles of any of the gamma functions in (\ref{def function to bound}) (in both cases, $f$ is a holomorphic function in a neighbourhood of the restricted domain). A complex analogue of the Mean Value Theorem (MVT) then states that
\bea
\re \frac{f\bigl( z+C\omega/\wt N \bigr)-f(z)}{\omega} &= \frac{C}{\wt N } \, \re f' \bigl( z+\Bar{c}_1\omega/\wt N \bigr) \\
\im \frac{f\bigl(z+C\omega/\wt N \bigr)-f(z)}{\omega} &= \frac{C}{\wt N } \, \im f' \bigl(z+\Bar{c}_2\omega/\wt N \bigr)
\eea
with $\Bar{c}_1,\Bar{c}_2\in(0,C)$. Summing the absolute values, it follows the bound
\be
\label{complex_mvt}
\left|\frac{f\bigl( z+C\omega/\wt N \bigr)-f(z)}{\omega}\right| \leq 
\frac{1}{\wt N } \, \biggl( \left| f'\bigl( z+\Bar{c}_1\omega/\wt N \bigr) \right| + \left| f'\bigl( z+\Bar{c}_2\omega/\wt N \bigr) \right| \biggr)
\ee
where we used $|C|\leq 1- \frac1{ab} <1$. It is therefore sufficient to show that 
\be
\frac{1}{\wt N } \, \left| f'\bigl( z+\Bar{c}\omega/\wt N \bigr) \right| \leq \cO(N\log N)
\ee
for any $\Bar{c}\in(0,C)$. Notice that $0 < \bar c < 1- \frac1{ab}$. 

We reason as follows.  For $\Delta \not\in \bZ + \bR\,\omega$, the arguments of the elliptic gamma functions in (\ref{def function to bound}) remain at an $\wt N$-independent distance from the zeros and poles, that in our case are placed at the points
\be
u_{0,i} = (1+i) ab \,\omega\;, \hspace{1cm} u_{\infty,j} = (1-j) ab\,\omega \hspace{1cm} \text{for} \hspace{1cm} i,j\in\mathbb{Z}_{\geq 1}\;,
\ee
respectively.
The orders of the zeros and poles are $i$ and $j$ respectively. The ratio $\bigl| \wt\Gamma' / \wt\Gamma \bigr|$ is bounded on the range of possible arguments, therefore
\be
\label{case simple bound}
\frac{1}{\wt N } \, \left| f'\bigl( z+\Bar{c}\omega/\wt N \bigr) \right| \leq \wt N \max_{\rule{0pt}{.8em} t \in \left[ \rule{0pt}{.6em} -ab,\, 3ab-a-b\right]} \, \left| \frac{ \wt\Gamma' \bigl( \Delta+t\omega; ab\omega, ab\omega \bigr) }{ \wt\Gamma \bigl( \Delta+t\omega; ab\omega, ab\omega \bigr)} \right| = \cO(\wt N)\;.
\ee
The case $\Delta = 0$ is more subtle since, as $\wt N$ grows, the arguments of some of the gamma functions can get increasingly close to zeros or poles instead of staying at an $\wt N$-independent distance, and the $\wt N$-independent bound above does not apply. This happens when 
\be\label{def bar u}
z =\bar{u}_{0,i}\in \{u_{0,i},u_{0,i}\pm\omega\} \hspace{1cm} \text{or} \hspace{1cm} z =\bar{u}_{\infty,j}\in \{u_{\infty,j},u_{\infty,j}\pm\omega\}\;.
\ee
One can easily see that for $\Delta=0$, $z$ can range from $(1-ab)\omega$ to $(3ab-a-b-1)\omega$, so that the problematic points we may approach are the simple zero at $u_{0,1}$, the simple pole at $u_{\infty,1}$, and the double pole at $u_{\infty,2}$. 

We now introduce a few results for later use. For a meromorphic function $g$ whose zeros include $\{z_i\}$ of order $\{m_i\}$ and whose poles include $\{p_j\}$ of order $\{n_j\}$, one can write
\be
\label{weierstrass_factorization}
g(z) = \frac{\prod_i (z-z_i)^{m_i}}{\prod_j (z-p_j)^{n_j}} s(z),
\ee
where $s(z)$ is meromorphic with zeros and poles at the remaining zeros and poles of $g$ that were not included in $\{z_i\}$ and $\{p_j\}$. Taking the derivative of this expression and computing $g'/g$, one finds
\be
\label{cauchy_arg_lemma}
\frac{g'(z)}{g(z)}=\sum_i\frac{m_i}{z-z_i}-\sum_j\frac{n_j}{z-p_j}+h(z),
\ee
where $h(z)=s'(z)/s(z)$ is meromorphic with simple poles at the remaining zeros and poles of $g$ that were not included in $\{z_i\}$ and $\{p_j\}$. Therefore we can apply \eqref{cauchy_arg_lemma} to the meromorphic function $\wt\Gamma$ and say that
\begin{multline}
\label{apply_lemma_gamma}
\frac{1}{\wt N} \, \left| f' \bigl( z + \bar c \omega / \wt N \bigr) \right| \leq
\frac{1}{\wt N} \sum_{\gamma \neq \delta}^{\wt N} \left| \frac{\wt\Gamma' \bigl(z + u_{\gamma,\delta}^{\bar c} ; ab\omega, ab\omega \bigr) }{ \wt\Gamma \bigl( z + u_{\gamma,\delta}^{\bar c} ; ab\omega, ab\omega \bigr)} \right|\\
\leq\frac{1}{\wt N}\sum_{\gamma\neq\delta}^{\wt N}\left(\frac{1}{|z+u_{\gamma,\delta}^{\bar c}-2ab\omega|}+\frac{1}{|z+u_{\gamma,\delta}^{\bar c}|}+\frac{2}{|z+u_{\gamma,\delta}^{\bar c}+ab\omega|}\right)+(\wt N-1)K
\end{multline}
where we defined
\be
\label{u_gamma_delta}
u_{\gamma,\delta}^{\bar c} = \omega \, \frac{\delta - \gamma + \bar c}{\wt N} \;, \hspace{2.5cm}
K = \max_{\rule{0pt}{.8em} t\in[-ab,\, 3ab-a-b]} \, \left| h_{\wt \Gamma}(t\omega) \right|
\ee
and $h_{\wt \Gamma}$ is the meromorphic function associated to $\wt \Gamma$ in \eqref{cauchy_arg_lemma}. We can bound its value with an $\wt N$-independent quantity because it is holomorphic on the range of possible arguments. If $z\neq \bar u_{0,1},\bar u_{\infty,1}, \bar u_{\infty,2}$, the outlying sums in \eqref{apply_lemma_gamma} will be of order $\cO(\wt N)$ since $z+u_{\gamma,\delta}^{\bar c}$ will be at least at a distance $|\omega|$ away from the zeros and poles. To complete our proof when $z= \bar u_{0,1},\bar u_{\infty,1}, \bar u_{\infty,2}$, we now need to bound the quantities
\be
\label{R_x_def}
R_x = \frac{1}{N} \sum_{\gamma\neq\delta}^{N} \frac{1}{ \bigl| x + \frac{\delta-\gamma+\bar{c}}{N} \bigr|} \qquad \text{with}\quad x=0,\pm 1 \;,
\ee
where we wrote $N$ in place of $\wt N$ in order not to clutter the formulae. We recall that $0 < \bar c < 1-1/ab$. Considering $x=0$ first, we reparametrize the sum in terms of $\delta - \gamma$ so that, after some manipulations, it becomes
\be
R_0 = \sum_{M=1}^{N-1} \left( \frac{N-M}{M+\bar c} + \frac{N-M}{M-\bar c} \right) < 2 \sum_{M=1}^{N-1} \frac{ N-M}{M-\bar c} \;.
\ee
The summand on the right is a positive decreasing function of $M$, therefore it can be bound by its integral:
\be
R_0 < \frac{2(N-1)}{1-\bar c} + 2 \int_1^{N-1} \frac{N-x}{x-\bar c} \, dx = \cO(N\log N) \;.
\ee
To ensure convergence of sums and integrals it is crucial to recall that $1-\bar c > (ab)^{-1}$.
In a similar way, for $x=+1$ we can write
\be
R_1 = \sum_{M=1}^{N-1} \left( \frac{N-M}{N+M + \bar c} + \frac{N-M}{N-M+\bar c} \right) < \sum_{M=1}^{N-1} 2 = \cO(N) \;,
\ee
while for $x=-1$ we can write
\be
R_{-1} = \sum_{M=1}^{N-1} \left( \frac{N-M}{N-M-\bar c} + \frac{N-M}{N+M-\bar c} \right) < 2 \sum_{M=1}^{N-1} \frac{N-M}{N-M-\bar c} < \frac{2(N-1)}{1-\bar c} = \cO(N) \;.
\ee

It remains to show that the terms we discarded from (\ref{before simplification}) when substituting the condition $i\neq j$ with the condition $\gamma \neq \delta$ give a subleading contribution. These are the terms in (\ref{before simplification}) with $\gamma=\delta$, whose total contribution is
\be\label{contribution_gamma_eq_delta}
\Phi = \wt N \sum_{r=0}^{a-1} \sum_{s=0}^{b-1} \sum_{c \neq d}^{ab} \log \wt \Gamma \left( \Delta + \omega\, \frac{d-c}{N} + \omega \bigl( d - c + as + br \bigr);ab\omega,ab\omega\right) \;.
\ee
We need to show that this is subleading in the large $N$ limit. We will bound the absolute value of the summand for all possible $c\neq d$, $r$, $s$ and drop the sums since they give an overall order $\cO(1)$ factor. After choosing a branch of the logarithm, the phases of $\wt \Gamma$ can clearly only give an order $\wt N$ contribution to (the imaginary part of) $\Phi$.

For what concerns the absolute value of $\wt\Gamma$, reasoning in a very similar way to the $\gamma\neq\delta$ case discussed above, we see that if $\Delta$ is not on a Stokes line then $\bigl| \log |\wt\Gamma| \bigr|$ is bounded above by an $N$-independent quantity and thus $\Phi$ is of order $\cO(N)$. When $\Delta=0$, the argument of $\wt \Gamma$ can only approach zeros or poles if $z= \omega \bigl( d-c+as+br \bigr) \in\{u_{0,1},u_{\infty,1},u_{\infty,2}\}$. Using \eqref{weierstrass_factorization}, we can write
\be
\label{weierstrass_factorization_phi}
\log\left|\wt \Gamma \Bigl(z + \omega\, \frac{d-c}{N}; ab\omega, ab\omega \Bigr) \right| = \log \left| \frac{\left(z + \omega\, \frac{d-c}{N}-u_{0,1}\right)s_{\wt\Gamma}\Bigl(z + \omega\,\frac{d-c}{N} \Bigr)}{\left(z + \omega\, \frac{d-c}{N}-u_{\infty,1}\right)\left(z + \omega\, \frac{d-c}{N}-u_{\infty,2}\right)^2} \right|
\ee
where $s_{\wt\Gamma}$ is a function which is regular at $u_{\infty,1},u_{\infty,2}$ and non-zero at $u_{0,1}$. We can therefore bound $\bigl| \log|s_{\wt\Gamma}| \bigr|$ over its possible arguments with an $N$-independent constant, so that it contributes to $\Phi$ at order $\cO(N)$. When $z=u_{0,1},u_{\infty,1},u_{\infty,2}$, only one of the factors multiplying $s_{\wt \Gamma}$ is of order $\cO(\log N)$ while the other two do not approach zero and can be bounded by an $N$-independent constant. Explicitly,
\be
\wt N \, \Biggl| \log \biggl| \wt \Gamma \Bigl(z + \omega\, \frac{d-c}{N}; ab\omega, ab\omega \Bigr) \biggr| \Biggr| \leq 2\wt N \, \Biggl| \log \biggl| \omega\frac{d-c}{N} \biggr| \Biggr| + \cO(N)= \cO(N\log N)\;.
\ee

\subsection[\texorpdfstring{$\rSU(N)$}{SU(N)} \textit{vs}. \texorpdfstring{$\rU(N)$}{U(N)} holonomies]{\matht{\rSU(N)} \textit{vs}. \matht{\rU(N)} holonomies}
\label{subapp: m sum rule}

In what follows, as it is done in Section \ref{sec: N=4 SYM} and Section \ref{sec: toric}, in order to parametrize the $\rSU{(N)}$ holonomies $u^\rSU$ we introduce $\rU{(N)}$ holonomies $u^\rU$, constrained by
\be
\sum_{i=1}^{N} u_i^\rU = 0\;.
\ee
With the choice of bases for the Cartan subalgebras of $\rSU(N)$ and $\rU(N)$ required to write the BA operators as in \eqref{BA_op_SYM}, the relation between the two sets of holonomies when expressing a generic element of the Cartan subalgebra of $\rSU(N)$ is
\be\label{U(N)_param}
u_i^{\rU} = u_i^{\rSU}\quad\text{for}\ i\neq N\;, \hspace{2cm} u_N^\rU = -\sum_{j=1}^{N-1} u_j^{\rSU}\;.
\ee
Note that the holonomies are only defined modulo $\mathbb{Z}$.

The $\rSU(N)$ superconformal index defined by \eqref{BA formula} contains a sum over $\{m_i^\rSU\}$ that picks up (representatives of) solutions to the BAEs whose residue can contribute to the index, as explained in \cite{Benini:2018mlo} and made explicit in \eqref{Z_tot_def}. Under a shift $\{m_i^{\rSU}\}$ of the $\rSU{(N)}$ holonomies, the $\rU{(N)}$ holonomies shift by corresponding amounts given by
\be
\label{shift_identification}
m_i^\rU = m_i^\rSU\;, \hspace{2cm} m_N^\rU = - \sum_{j=1}^{N-1} m_j^\rSU\;.
\ee
Given these identifications for the holonomies and shifts, the $\rSU(N)$ quantities are always equal to the first $N-1$ $\rU(N)$ quantities, so that in the following we will drop the superscripts $\rSU$ and $\rU$, remembering that $u_{1,\ldots,N-1}$ and $m_{1,\ldots,N-1}$ are independent while $u_N$ and $m_N$ are determined by \eqref{U(N)_param} and \eqref{shift_identification}, respectively. 

One might then worry that the choice of $\{m_j\}$ given in \eqref{integersm} is not allowed, since the last integer $m_N$ there does not satisfy \eqref{shift_identification}. Specifically, let us choose 
\be
m_j \in \{1,\ldots, ab\} \qquad\text{such that}\qquad m_j = j \mod ab \;,\qquad \text{for } j= 1, \ldots, N-1 \;,
\ee
so that $m_N$ is fixed by \eqref{shift_identification} to be a negative integer of $\cO(N)$. To match with the choice in \eqref{integersm}, we want to replace this with $m_N=N \text{ mod } ab$ and in $\{1,\dots, ab\}$. We will show that this replacement does not affect the value of $\cZ$ to leading order in $N$. This will be done in two steps. We will first show that the function $\cZ$ evaluated on a configuration $\{u_1, \dots, u_N\}$ which is obtained from the basic solution by shifting one or more variables $u_i$ by multiples of $2ab\omega$ (or even of $ab\omega$, in many cases), is the same as $\cZ$ evaluated on the basic solution. Using this property, $\cZ$ is unaltered if evaluated on the following shifted value of $m_N$:
\be
\label{tilde_m_def}
\wt m_N \in \{1, \dots, 2ab\} \qquad\text{such that}\qquad \wt m_N = \biggl( -\sum_{i=1}^{N-1}m_i \biggr)\mod 2ab \;.
\ee
We will then show that the contribution to $\cZ$ of the single holonomy $u_{N}$ is subleading, provided $\wt m_N\in\{1,\dots,2ab\}$. Therefore, choosing instead $m_N = N \text{ mod } ab$ and in $\{1,\dots, ab\}$ as we did in (\ref{integersm}) does not change $\cZ$ at leading order in $N$. This completes the proof.
%\footnote{Assuming $N = ab\wt N$, one finds $\wt m_N = - ab \left( \frac{\wt N (ab+1)}2 \right) \text{ mod } ab$. If either $\wt N$ is even, or $a,b$ are both odd, then $\wt m_N$ is already equal to $N$ mod $ab$ and therefore no extra shift is necessary.}

As shown in \cite{Benini:2018mlo}, when evaluated on solutions to the BAEs, the function $\cZ$ for a general semi-simple gauge group is invariant under independent shifts of any gauge holonomy by $ab\omega$. This is proven assuming that gauge and global symmetries are non-anomalous. In our case, this result only allows us to shift the $u_i$'s while preserving the $\rSU(N)$ constraint. This property does not allow us to independently shift the last holonomy $u_N$, since it is always fixed by the $\rSU(N)$ constraint. We now show that an \emph{independent} shift of $u_N$ by a multiple of $ab\omega$
% of $u_N$
is also an invariance of $\cZ$ for $\cN=4$ $\rSU(N)$ SYM, when this function is evaluated on the basic solution. In order to prove this, one has to use the property
\begin{align}
\label{shift_Gamma_w_theta}
& \wt \Gamma \bigl( u+mab\omega,a\omega,b\omega \bigr) \\
&\quad = \; (-e^{2\pi i u})^{-\frac{ab}{2}m^2 + \frac{a+b-1}{2}m} \, (e^{2\pi i \omega})^{-\frac{ab}{6}m^3 + \frac{ab(a+b)}{4}m^2 - \frac{a^2+b^2+3ab-1}{12}m} \, \theta_0(u,\omega)^m \, \wt\Gamma(u,a\omega,b\omega) \nn
\end{align}
that was proven in \cite{Benini:2018mlo}, the fact that the $\rU(N)$ BA operators are periodic modulo $\omega$ in the $u_i$'s, and the explicit form of the basic solution \eqref{basic_solution}. Applying \eqref{shift_Gamma_w_theta}, we first have that
\begin{align}
\label{shift_Gamma_w_theta_product}
& \prod_{i\neq j}\wt \Gamma \Bigl( u_{ij} + \Delta + mab\omega(\delta_{iN}-\delta_{jN });a\omega,b\omega \Bigr) \\
&\;\; = e^{-\pi i ab m^2(1+2\Delta)+2\pi i (a+b-1) m \sum_i u_{iN }+\pi i ab(a+b)m^2 \omega} \prod_{i} \frac{\theta_0(u_{N i}+\Delta,\omega)^m}{\theta_0(u_{iN }+\Delta;\omega)^m} \prod_{i\neq j}\wt \Gamma (u_{ij} + \Delta;a\omega,b\omega)\;, \nn
\end{align}
and so from \eqref{BA_op_SYM}, \eqref{z_def} and \eqref{basic_solution} one obtains
\begin{align}
&\cZ(u_i+mab\omega\delta_{iN};a\omega,b\omega,\Delta) \nn \\
& =\prod_{i} \left( \frac{\theta_0( u_{N i}+\Delta_1,\omega) \, \theta_0( u_{N i}+\Delta_2,\omega) \, \theta_0( u_{iN },\omega) \, \theta_0( u_{iN }+\Delta_1+\Delta_2,\omega)}{\theta_0( u_{iN }+\Delta_1,\omega) \, \theta_0( u_{iN }+\Delta_2,\omega) \, \theta_0( u_{N i},\omega) \, \theta_0( u_{N i}+\Delta_1+\Delta_2,\omega)} \right)^m \!\!\! \cZ( u_i;a\omega,b\omega,\Delta) \nn \\
& =(-1)^{m(N-1)} \, e^{2\pi i m\lambda} \, Q_N ^{-m}( u_i;\omega,\Delta) \, \cZ( u_i;a\omega,b\omega,\Delta) \nn \\
& = \cZ( u_i;a\omega,b\omega,\Delta) \;.
\label{shift_z_w_theta_product}
\end{align}
In the steps above we also used the theta function reflection property
\be
\theta_0(u;\omega) = -e^{2\pi i u} \, \theta_0(-u;\omega)\;.
\ee

More generally, we can show that this shift invariance is true for quiver gauge theories, when $\cZ$ is evaluated on the basic solution and the chemical potentials $u^\alpha_N$ are shifted by a multiple of $2ab\omega$ (or even of $ab\omega$, in many cases) simultaneously for all gauge groups $\rSU(N)_\alpha$.
The steps are the same as in \eqref{shift_z_w_theta_product}.
We should notice that the expression for any particular Lagrange multiplier $\lambda_\alpha$ is more complicated than for $\cN=4$ SYM, but the sum of all Lagrange multipliers is simple:
\be
e^{2\pi i\sum_{\alpha=1}^{G}\lambda_\alpha}=(-1)^{n_\chi(N-1)}\;,
\ee
where $\alpha$ runs over the $G$ $\rSU(N)$ gauge group factors and $n_\chi$ is the number of chiral multiplets in the theory. Performing these steps one obtains
\begin{align}
& \cZ\bigl( u^\alpha_i + mab\omega\delta_{iN };a\omega,b\omega,\Delta \bigr) \\
&\qquad = \frac{e^{2\pi i m\sum_{\alpha=1}^{G}\lambda_\alpha} \, (-1)^{-mG(N-1)+mab(N-1)(n_\chi-G)}}{\bigl( \prod_\alpha Q_N^\alpha( u_i^\alpha;\omega,\Delta) \bigr)^m} \, \cZ( u^\alpha_i; a\omega,b\omega,\Delta) \nn \\[.4em]
&\qquad = (-1)^{m(G-n_\chi)(ab+1)(N-1)} \, \cZ( u^a_i; a\omega,b\omega,\Delta) \;. \nn
\end{align}
There are now different cases in which the sign in the last line disappears. First, in the case of toric quiver gauge theories one uses the relation \cite{Hanany:2005ve,Franco:2005rj}
\be
G - n_\chi + N_W = 0
\ee
between the number of gauge groups, of chiral multiplets, and of superpotential terms, as well as the fact that the number $N_W$ of superpotential terms is even, to show that the sign disappears. Second, if $N$ is odd then the sign disappears. Third, if the coprime integers $a,b$ are both odd%
\footnote{This restriction is quite uninfluential, because the set of pairs $\{\tau + \bZ,\sigma + \bZ\}$ such that $\tau/\sigma = a/b \in \bQ_{>0}$ with $a,b$ both odd is still dense in $\bH^2$.}
then the sign disappears. Fourth and most importantly, if we take $m$ even then the sign disappears.

We now proceed to show that the contribution to $\cZ$ of a single holonomy $u_i$ is subleading, provided that $m_{i<N}\in\{1,\dots,ab\}$ and $m_N\in\{1,\dots,2ab\}$. In the building block $\Psi$ defined in \eqref{building block def}, the contribution of a single holonomy $u_i$ consists of the two terms
\be
\label{Phi_pm_def}
\Phi^\pm_i \equiv \sum_{j (\neq i)}^{N} \log\wt\Gamma \left( z_\pm \pm \omega\;\frac{j-i}{N} ; ab\omega, ab\omega \right)\;,
\ee
where we have defined
\be
\label{m_k_N_def}
z_\pm \equiv \Delta \pm \omega\;\bigl( m_j - m_i\bigr) + \omega\;\bigl(as + br \bigr)\;.
\ee
In particular, for the case $i=N$ we will use the shift property just proven to substitute $m_N$ with $\wt m_N$ defined in \eqref{tilde_m_def}.

We will now show that $\Phi^\pm_i$ is subleading. In the case $i=N$ this will allow us to choose $\wt m_N$ as in \eqref{integersm}. In order to do this we want to bound the absolute value of the summand $\log\wt\Gamma$ in $\Phi^\pm_i$. What follows will be completely analogous to the argument used to show that \eqref{contribution_gamma_eq_delta} is subleading. After choosing a branch of the logarithm, the phases of $\wt\Gamma$ can only contribute at order $\cO(N)$ to $\Phi^\pm_i$. As before, we exclude Stokes lines and note that for $\Delta \neq 0$ we can bound $\bigl| \log|\wt\Gamma| \bigr|$ with an $N$-independent constant so that $|\Phi^\pm_i|=\cO(N)$. For $\Delta=0$, $z_\pm$ have the range
\be
z_{\pm} \in \{-2ab+1,\dots,4ab-a-b-1\}\omega \;,
\ee
and the argument of $\wt\Gamma$ may approach zeros or poles when $z_\pm=\bar u_{0,1}, \bar u_{0,2}, \bar u_{\infty,1}, \bar u_{\infty,2}, \bar u_{\infty,3}$, which are defined in \eqref{def bar u}. If this is the case, further inspection is required. Using again \eqref{weierstrass_factorization}, we can write
\be
\log \wt \Gamma \left(  z_{\pm}\pm\omega\frac{j-i}{N} ; ab\omega, ab\omega \right) = \log \left[\dfrac{\prod_{m=1}^2\left( z_{\pm}\pm\omega\frac{j-i}{N}-u_{0,m}\right)^m s_{\wt\Gamma}\left(  z_{\pm}\pm\omega\frac{j-i}{N}\right)}{\prod_{n=1}^3\left(z_{\pm}\pm\omega\frac{j-i}{N}-u_{\infty,n}\right)^n}\right]
\ee
where $s_{\wt\Gamma}$ is a function that is regular at $u_{\infty,1}$, $u_{\infty,2}$, $u_{\infty,3}$, and non-zero at $u_{0,1}$, $u_{0,2}$. This allows us to bound $\bigl| \log|s_{\wt\Gamma}| \bigr|$ with an $N$-independent constant, and its contribution to $\Phi^{\pm}_i$ is of order $\cO(N)$. When $z_\pm = \bar u_{0,1}$, $\bar u_{0,2}$, $\bar u_{\infty,1}$, $\bar u_{\infty,2}$, $\bar u_{\infty,3}$ the logarithms of the other factors are either bounded by an $N$-independent constant, or are of the form
\be
\sum_{j\neq i}^{N} \Biggl| \log\left|x\pm\frac{j-i}{N}\right| \Biggr|\leq (N-1)\log N \;,
\ee
where $x=0,\pm 1$. Notice that the use of the shift property previously proved plays a major role here. If we tried to apply this argument directly without first shifting $m_N$, we would have to consider an $\cO(N)$ number of poles or zeros whose order is also $\cO(N)$. This would lead to an $\cO(N^3\log N)$ bound, which does not help. What we did shows that a single $\Phi^{\pm}_i$ is of order $\cO(N\log N)$ for any choice of the corresponding $m_{i}$. In particular this allows us to choose $\wt m_{N} = N \text{ mod } ab\in\{1,\dots,ab\}$ as we do in \eqref{integersm}, without affecting the leading behavior of the building block $\Psi$.

\subsection[Generic \texorpdfstring{$N$}{N}]{Generic \matht{N}}
\label{subapp: generic N}

Here we generalize the computation done in Section~\ref{subsec: building block} and consider a generic $N$ which is not necessarily a multiple of $ab$. We will exploit many of the arguments in Section~\ref{subapp: simplifications building block}. Let $N=ab\wt N+q$, where $q\in\{0,\dots,ab-1\}$. We need to examine the leading order contribution of the building block
\be
\Psi = \sum_{r=0}^{a-1}\sum_{s=0}^{b-1} \sum_{i\neq j}^N \log \wt \Gamma \left( \Delta + \omega\, \frac{j-i}N + \omega\bigl( m_j - m_i + as + br \bigr); ab\omega, ab\omega \right)\;.
\ee
As shown in the final part of Section~\ref{subapp: m sum rule}, the contribution to the building block of a single holonomy $u_i$ is subleading. Therefore the contribution of the last $q$ holonomies $u_{ab\wt N+1},\dots, u_N$ is also subleading and can be discarded. Now, the sum over $i\neq j$ only goes up to $ab\wt N$, and we can decompose the indices as in \eqref{before simplification}. Neglecting the $\gamma=\delta$ terms using the same argument as after \eqref{contribution_gamma_eq_delta}, we get
\be
\Psi \simeq \sum_{r=0}^{a-1} \sum_{s=0}^{b-1}\sum_{\gamma\neq\delta=0}^{\wt N-1} \sum_{c,d=0}^{ab-1}\log \wt\Gamma \left( \Delta + \omega\: \frac{\delta-\gamma}{\wt N+\frac{q}{ab}} + \omega \frac{d-c}N + \omega\bigl( d - c + as + br \bigr); ab\omega, ab\omega \right)\;.
\ee
As in Section~\ref{subapp: simplifications building block}, we want to drop $\omega(d-c)/N$ in the argument of the elliptic gamma function, and we can use the same reasoning given there, with the minor change that \eqref{R_x_def} takes the form
\be
\wt R_x = \dfrac{1}{N+\frac{q}{ab}}\sum_{\gamma\neq\delta}^N\dfrac{1}{\left|x+\frac{\delta-\gamma+\bar c}{N+q/ab}\right|}\:,\qquad x=0,\pm1\;.
\ee
The same bounds as for $R_x$ can be used here, since one can show that
\be
\wt R_0 = R_0\;, \hspace{2cm} \wt R_{\pm 1} \leq R_{\pm 1} \;.
\ee
We can then use \eqref{product of Gamma's fund}, as we did in Section \ref{subsec: building block}, to change the moduli of the elliptic gamma function from $(ab\omega,ab\omega)$ to $(\omega,\omega)$:
\bea
\label{psi_from_abomega_to_omega}
\Psi &\simeq \sum_{r=0}^{a-1} \sum_{s=0}^{b-1}\sum_{\gamma\neq\delta=0}^{\wt N-1} \sum_{c,d=0}^{ab-1}\log \wt \Gamma \left( \Delta + \omega\: \frac{\delta-\gamma}{\wt N+\frac{q}{ab}} + \omega\bigl( d - c + as + br \bigr) ; ab\omega, ab\omega \right)\\
&=\sum_{r=0}^{a-1} \sum_{s=0}^{b-1}\sum_{\gamma\neq\delta=0}^{\wt N-1}\log \wt \Gamma \left( \Delta + \omega\: \frac{\delta-\gamma}{\wt N+\frac{q}{ab}} + \omega\bigl(1-ab+ as + br \bigr) ; \omega, \omega \right)\;\\
&=\frac{1}{(ab)^2}\sum_{r=0}^{a-1} \sum_{s=0}^{b-1}\sum_{\gamma\neq\delta=0}^{\wt N-1}\sum_{c,d=0}^{ab-1}\log \wt \Gamma \left( \Delta + \omega\: \frac{\delta-\gamma}{\wt N+\frac{q}{ab}} + \omega\bigl(1-ab+ as + br \bigr) ; \omega, \omega \right)\;.
\eea
In the last equality, to make future steps clearer, we added a sum over $c,d$ even though nothing depends on $c$ and $d$.
   
Now, in order to get the desired result we trace our steps backwards. First, we will reintroduce the term $\omega(d-c)/N$ into the argument of the elliptic gamma functions. Then we will add to the sum in \eqref{psi_from_abomega_to_omega} the $\gamma=\delta$ terms to form the sum over $i\neq j$ up to $ab\wt N$. Finally we will add terms containing the last $q$ holonomies $u_{ab\wt N+1},\dots, u_N$ in order to build the complete sum up to $N$. These are the exact same steps we just performed to express $\Psi$ as in \eqref{psi_from_abomega_to_omega} up to subleading terms, with the only difference being that the moduli of $\wt \Gamma$ are now $(\omega,\omega)$ rather than $(ab\omega,ab\omega)$. Therefore the same arguments can be used, with only slight modifications involving the number and order of zeros and poles, but since these are parametrized here by $r$ and $s$ that are $N$-independent, this is of no consequence. At this point, $\Psi$ at leading order is
\be
\Psi \simeq \frac{1}{(ab)^2}\sum_{r=0}^{a-1} \sum_{s=0}^{b-1}\sum_{i\neq j}^N\log \wt\Gamma \left( \Delta + \omega\, \frac{j-i}N + \omega\bigl(1-ab+ as + br \bigr) ;\omega, \omega \right)\;,
\ee
and using the result of \cite{Benini:2018ywd} (that is our equation \eqref{single_fugacity_index}) we obtain
\be
\Psi \simeq -\frac{\pi i N^2}{3(a\omega)(b\omega)}\:\frac{1}{ab}\sum_{r=0}^{a-1} \sum_{s=0}^{b-1} B_3\Big([\Delta]'_\omega+\omega\bigl( as + br -ab\bigr)\Big)\;.
\ee
Then, using the property of Bernoulli polynomials \eqref{bernoulli_property}, we finally get \eqref{Psi large N final}.

%%%%%%%%%%%%%%%%%%%%%%%%%%%%%%%%
%%%%%%%%%%%%%%%%%%%%%%%%%%%%%%%%

\section{5d \matht{\cN=2} Abelian gauged supergravity}
\label{app:5d_sugra}

We report here the general form of 5d $\cN=2$ Abelian gauged supergravity with $n_V$ vector multiplets and $n_H$ hypermultiplets \cite{Gunaydin:1983bi, Gunaydin:1984ak, Ceresole:2000jd}.%
\footnote{A more complete discussion was developed in \cite{Bergshoeff:2002qk}.}
The graviton multiplet contains a graviton, a gravitino and a vector; each vector multiplet contains a vector, a gaugino and a real scalar; each hypermultiplet contains four real scalars and a hyperino. All fermions are Dirac, but can conveniently be doubled with a symplectic Majorana condition. We follow the notation of \cite{Freedman:2012zz}. We use indices
\be
I,J,K=1, \dots, n_V+1 \;,\qquad i,j=1, \dots, n_V \;,\qquad u,v=1, \dots, 4n_H
\ee
for the gauge fields $A^I_\mu$, for the scalars $\phi^i$ in vector multiplets, and for the scalars $q^u$ in hypermultiplets, respectively. The data that define the theory are:
\begin{enumerate}
\item A very special real manifold $\cS\cM$ of real dimension $n_V$.
\item A quaternionic-K\"ahler manifold $\cQ\cM$ of real dimension $4n_H$.
\item A set of $n_V+1$ Killing vectors on $\cQ\cM$ compatible with the quaternionic-K\"ahler structure (if $n_H=0$, $n_V+1$ FI parameters not all vanishing).
\end{enumerate}
The Killing vectors could be linearly dependent or vanish.

The bosonic Lagrangian is given by
\begin{align}
\label{5d_lagrangian}
8\pi G^{(5)}_\text{N} e^{-1} \ccL_\text{5d} &= \frac{R_s}{2}-\frac{1}{2} \, \cG_{ij}(\phi) \, \partial_\mu \phi^i\partial^\mu\phi^j - \frac{1}{2} \, h_{uv}(q) \, \cD_\mu q^u \cD^\mu q^v -\frac{1}{4} \, G_{IJ}(\phi) \, F^I_{\mu\nu}F^J{}^{\mu\nu} \nn \\
&\quad + \frac{e^{-1}}{48} \, C_{IJK} \, \epsilon^{\mu\nu\rho\sigma\lambda} \, F^I_{\mu\nu}F^J_{\rho\sigma}A^K_\lambda-g^2V(\phi,q) \;.
\end{align}
Here $G^{(5)}_\text{N}$ is the 5d Newton constant, $e\, d^5x$ is the spacetime volume form, $R_s$ is the scalar curvature, $F^I_{\mu\nu}$ is the field strength of $A^I_\mu$, $g$ is a coupling constant, and $V$ is the scalar potential. Let us explain the other terms.

\paragraph{Very special geometry.} The scalars $\phi^i$ are real coordinates on the very special real manifold $\cS\cM$ \cite{deWit:1991nm}.
The latter is specified by the totally symmetric tensor $C_{IJK}$ (which, controlling also the Chern-Simons couplings, should be suitably quantized) as the submanifold
\be
\label{def very special real manifold}
\cS\cM = \Bigl\{ \cV(\Phi) \,\equiv\, \frac16 \, C_{IJK} \, \Phi^I \Phi^J \Phi^K = 1 \Bigr\} \,\subset\, \bR^{n_V+1} \;.
\ee
Here $\Phi^I$ are coordinates on $\bR^{n_V+1}$, and give rise to ``sections'' $\Phi^I(\phi^i)$ on $\cS\cM$. The metrics $G_{IJ}$ and $\cG_{ij}$ for vector fields and vector multiplet scalar fields are
\be
G_{IJ}(\phi) = - \frac12\, \parfrac{}{\Phi^I} \parfrac{}{\Phi^J} \log \cV \, \Big|_{\cV=1} \;,\qquad\qquad \cG_{ij}(\phi) = \partial_i \Phi^I \, \partial_j \Phi^J \, G_{IJ} \, \Big|_{\cV=1}
\ee
where $\partial_i \equiv \partial/\partial \phi^i$. We recognize that $\cG$ is the pull-back of $G$ from $\bR^{n_V+1}$ to $\cS\cM$. From (\ref{def very special real manifold}) it immediately follows
\be
C_{IJK} \, \Phi^I \Phi^J \partial_i \Phi^K \, \big|_{\cV=1} = 0 \;.
\ee
With a little bit of algebra one then obtains a more explicit expression for $G$:
\be
\label{alternative 5d metric G}
G_{IJ} = -\frac12 C_{IJK} \Phi^K + \frac18 C_{IKL} C_{JMN} \Phi^K \Phi^L \Phi^M \Phi^N \, \Big|_{\cV=1} \;.
\ee
It follows that the kinetic term for vector multiplet scalars can also be written as
\be
\label{alternative 5d kinetic term}
-\frac12 \, \cG_{ij} \, \partial_\mu \phi^i \partial^\mu \phi^j = \frac14 \, C_{IJK} \, \Phi^I \partial_\mu \Phi^J \partial^\mu \Phi^K \, \Big|_{\cV=1} \;.
\ee

One can define on $\cS\cM$ the sections with lower indices:
\be
\Phi_I \,\equiv\, \frac23 G_{IJ} \Phi^J \Big|_{\cV=1} = \frac16 C_{IJK} \Phi^J \Phi^K \Big|_{\cV=1} = \frac13 \parfrac{\cV}{\Phi^I} \, \Big|_{\cV=1} \;.
\ee
With simple algebra one can show the following identities:
\bea
\Phi_I \Phi^I &= 1 \;,\qquad\qquad& G_{IJ} &= \frac92 \Phi_I \Phi_J - \frac12 C_{IJK} \Phi^K\;, \\
\partial_i \Phi_I &= - \frac23 G_{IJ} \, \partial_i \Phi^J \;,\qquad\qquad & \Phi_I \, \partial_i \Phi^I &= \Phi^I \partial_i \Phi_I = 0 \;.
\eea
In particular, $\partial_i \Phi^I$ for $i=1, \ldots, n_V$ are the tangent vectors to $\cS\cM$ in $\bR^{n_V+1}$ while $\Phi_I$ is a 1-form orthogonal to $\cS\cM$. Another identity (and similar ones obtained by lowering one or both of the indices $I,J$ with the metric $G$) is
\be
\label{projection formula}
\cG^{ij} \, \partial_i \Phi^I \partial_j \Phi^J = G^{IJ} - \frac23 \Phi^I \Phi^J \;,
\ee
where $G^{IJ}$ is the inverse of $G_{IJ}$.
To prove it, one observes that the tensor on the LHS is the projector on $\cS\cM$, and then verifies that the expression on the RHS has the same property.

When the manifold $\cS\cM$ is a locally symmetric space, one can find a constant symmetric tensor $C^{IJK}$ with upper indices such that \cite{Gunaydin:1983bi}
\be
\label{inverse tensor Ctilde}
C^{IPQ} \, C_{P(JK} \, C_{LM)Q} = \frac43\, \delta^I_{(J} \, C_{KLM)} \;.
\ee
With some algebra, it follows that
\be
\Phi^I = \frac32 \, G^{IJ} \Phi_J = \frac92\, C^{IJK} \Phi_J \Phi_K \;,\qquad\qquad G^{IJ} = 2 \Phi^I \Phi^J - 6 C^{IJK} \Phi_K \;,
\ee
as well as
\be
C^{IJK} = \frac18 \, G^{IL} \, G^{JM} \, G^{KN} \, C_{LMN} \;.
\ee

\paragraph{Quaternionic-K\"ahler geometry.} The scalars $q^u$ are real coordinates on the quaternionic-K\"ahler manifold $\cQ\cM$ with metric $h_{uv}(q)$ \cite{Bagger:1983tt}. For $n_H \geq 2$,%
\footnote{The case $n_H=1$ is special because $SU(2)^2 /\bZ_2 \cong SO(4)$ and so the holonomy condition does not impose any constraint on (orientable) Riemannian manifolds. However, supersymmetry requires (\ref{Riemann qK}) which we can take as the definition of a quaternionic-K\"ahler manifold of dimension 4. A 4-dimensional space satisfying (\ref{Riemann qK}) is Einstein with self-dual Weyl curvature.
}
this is a $4n_H$-dimensional Riemannian manifold with holonomy $SU(2) \times Sp(n_H) / \bZ_2$. To express this fact, it is convenient to introduce local ``vielbeins'' $f\du{u}{iA}$ with $i=1,2$ (not to be confused with the index $i$ of very special geometry) in the fundamental of $SU(2)$ and $A = 1, \dots, 2n_H$ in the fundamental of $Sp(n_H)$, such that
\be
h_{uv} = f\du{u}{iA} f\du{v}{jB} \epsilon_{ij} \Omega_{AB} \;,
\ee
where $\epsilon_{ij}$ and $\Omega_{AB}$ are the invariant tensors of $SU(2)$ and $Sp(n_H)$, respectively. Regarding $(iA)$ as a composite index, the inverse of the matrix $f\du{u}{iA}$ is $f\du{iA}{u} = h^{uv} f\du{v}{jB} \epsilon_{ji} \Omega_{BA}$.
One can then construct a locally-defined triplet of almost complex structures
\be
\vec J\du{u}{v} \equiv (J^x)\du{u}{v}  = -i f\du{u}{iA} f\du{jA}{v} (\sigma^x)\du{i}{j}
\ee
where $x=1,2,3$ is in the adjoint of $SU(2)$ and $\vec\sigma$ are the Pauli matrices.
The derived triplet of almost symplectic forms is $\vec J_{uv} = \vec J\du{u}{t} \, h_{tv}$. They are antisymmetric, using that $\vec\sigma\du{i}{j} \epsilon_{jk}$ is symmetric.%
\footnote{Using the fact that a $2\times 2$ matrix can be expanded in the basis $\{\unit, \vec\sigma\}$, we also find
\be
2 f\du{u}{iA} f\du{jA}{v} = \delta_u^v \delta_j^i + i \, \vec J\du{u}{v} \cdot \vec\sigma\du{j}{i} \;.
\ee}
The almost complex structures automatically satisfy the quaternion relation
\be
\label{quaternion relation}
(J^x)\du{u}{s} (J^y)\du{s}{t} = - \delta^{xy} \delta^t_u + \epsilon^{xyz} (J^z)\du{u}{t} \;.
\ee
The Levi-Civita connection takes values in $\fsu(2) \times \fsp(n_H)$. Calling $\omega\du{uj}{i}$ and $\rho\du{uB}{A}$ the two projections, respectively, they are determined by the requirement that $f\du{u}{iA}$ be covariantly constant with respect to the full connection:
\be
\label{full covariant derivative}
0 = \nabla_v f\du{u}{iA} + f\du{u}{jA} \omega\du{vj}{i}  + f\du{u}{iB} \rho\du{vB}{A} \;.
\ee
We can alternate between the vector and bispinor notations of $SU(2)$ with%
\footnote{The $SU(2)$ connection satisfies $\epsilon^{jm} \omega\du{um}{n} \epsilon_{ni} = \omega\du{ui}{j}$, in particular $\omega\du{uj}{j} = 0$, and a similar condition is satisfied by $\rho$. This follows from the properties of the Pauli matrices. In going between the vector and bispinor notation one can use the identities
\be
\vec\sigma\du{n}{m} \cdot \vec\sigma\du{i}{j} = \delta^j_n \delta^m_i - \epsilon^{mj} \epsilon_{ni} \;,\qquad\qquad
\vec\sigma\du{i}{j} \times \vec\sigma\du{\ell}{m} = i \bigl( \vec\sigma\du{i}{m} \, \delta^j_\ell - \delta^m_i \, \vec\sigma\du{\ell}{j} \bigr) \;.
\ee}
\be
\vec\omega_u = - i\, \omega\du{ui}{j} \, \vec\sigma\du{j}{i} \;,\qquad\qquad \omega\du{ui}{j} = \frac i2 \, \vec\omega_u \cdot \vec\sigma\du{i}{j} \;.
\ee
The two connections are extracted from (\ref{full covariant derivative}) through: $\omega\du{ui}{j} \, \delta_A^B + \delta_i^j \, \rho\du{uA}{B} = - f\du{iA}{w} \nabla_u f\du{w}{jB}$.
From (\ref{full covariant derivative}) it immediately follows
\be
\wt\nabla_w \vec J\du{u}{v} \equiv \nabla_w \vec J\du{u}{v} + \vec\omega_w \times \vec J\du{u}{v} = 0 \;.
\ee
In other words, $\vec J$ is covariantly constant with respect to its natural $SU(2)$ connection $\vec \omega$. From the integrability condition of (\ref{full covariant derivative}) one also obtains (in bispinor and vector notation):
\be
R_{uv\;\,t}^{\;\;\;\: s} = \cR\du{uvi}{j} \,f\du{jA}{s} f\du{t}{iA} + \cR\du{uvA}{B} \, f\du{jB}{s} f\du{t}{jA} = -\frac12 \, \vec\cR_{uv} \cdot \vec J\du{t}{s} + \cR\du{uvA}{B} \, f\du{jB}{s} f\du{t}{jA} \;,
\ee
where $R_{uv\;\,t}^{\;\;\;\: s}$ is the Riemann tensor of $h_{uv}$ and we defined
\bea
\cR\du{uvi}{j} &\equiv 2\partial_{[u}^{\phantom{j}} \omega\du{v]i}{j} - 2\omega\du{[u | i}{k} \omega\du{v]k}{j} \qquad\qquad\text{or}\quad\qquad
\vec\cR_{uv} \equiv 2 \partial_{[u} \vec\omega_{v]} + \vec\omega_u \times \vec\omega_v \\
\cR\du{uvA}{B} &\equiv 2\partial_{[u}^{\phantom{B}} \rho\du{v]A}{B} - 2 \rho\du{[u|A}{C} \rho\du{v]C}{B} \;.
\eea
In particular
\be
R_{uv\;\,t}^{\;\;\;\: s} \, \vec J\du{s}{t} = 2n_H \vec \cR_{uv} \;,
\ee
\ie, the $SU(2)$ field strength $\vec\cR_{uv}$ is the $\fsu(2)$ projection of the Riemann curvature.

One can prove \cite{Berger:1955} (see also \cite{Bagger:1983tt, Salamon:1982})
that $SU(2) \times Sp(n_H)$ holonomy manifolds with $n_H \geq 2$ are automatically Einstein. In fact, they satisfy a stronger property: the Riemann curvature is the sum of the Riemann tensor of $\bH\bP^{\,n_H}$ and of a Weyl part,
\begin{multline}
\label{Riemann qK}
R_{uvst} = \frac{R}{8n_H(n_H+2)} \Bigl( h_{s[u} h_{v]t} + \vec J_{uv} \cdot \vec J_{st} - \vec J_{s[u} \cdot \vec J_{v]t} \Bigr) + {} \\
{} + \bigl( f\du{u}{iA} f\du{v}{jB} \epsilon_{ij} \bigr) \bigl( f\du{s}{kC} f\du{t}{\ell D} \epsilon_{k\ell} \bigr) \, \cW_{ABCD} \;.
\end{multline}
The tensor $\cW_{ABCD}$ is totally symmetric and controls the Weyl curvature, which is contained in $Sp(n_H)$: it gives rise to a traceless (and thus Ricci flat) contribution to the Riemann curvature. From that expression we obtain
\be
\label{Einstein cond and SU(2) curv}
R_{vt} = \frac{R}{4n_H} \, h_{vt} \;,\qquad\qquad\qquad \vec\cR_{uv} = \frac{R}{4n_H(n_H+2)} \, \vec J_{uv} \;.
\ee
The first equation shows that the manifold is Einstein. The second equation shows that the $SU(2)$ part of the curvature is completely fixed in terms of the triplet of complex structures. The tensor $\cW_{ABCD}$ expresses the freedom in the $Sp(n_H)$ part.

While quaternionic-K\"ahler manifolds can have any size, local supersymmetry requires%
\footnote{Had we chosen a canonical normalization for the action of hypermultiplet scalars, the scalar curvature would be fixed in terms of the Planck mass to $\lambda = - m_\text{Pl}^{-2}$ \cite{Bagger:1983tt}. This reproduces the fact that the manifold of hypermultiplet scalars is hyper-K\"ahler in rigid supersymmetry.}
\be
\label{def lambda}
\lambda \,\equiv\, \frac{R}{4n_H (n_H+2)} = - 1 \;,
\ee
fixing the scalar curvature \cite{Bagger:1983tt}. Hence the manifold of hypermultiplet scalars is a non-trivial quaternionic-K\"ahler manifold with negative scalar curvature.

\paragraph{Isometries and gauging.}
We consider gaugings of Abelian isometries of the quaternionic-K\"ahler manifold $\cQ\cM$ by the vectors $A_\mu^I$. The isometries are generated by (possibly vanishing or linearly dependent) Killing vectors $k^u_I(q)$ that also satisfy a quaternionic version of the triholomorphic condition:
\be
\label{triholo Killing vector conditions}
h_{w(u} \nabla_{v)} k^w_I = 0 \;,\qquad\qquad \vec J\du{u}{w} (\nabla_w k_I^v) - (\nabla_u k_I^w) \vec J\du{w}{v} = \lambda \, \vec J\du{u}{v} \times \vec P_I \;.
\ee
The second equation expresses the fact that the derivative of each Killing vector commutes with the triplet of complex structures, up to a rotation parametrized by the $SU(2)$ sections $\vec P_I$. Notice that the LHS can be written, after lowering $v$, as $2 \wt\nabla_{[u} \bigl( \vec J_{v]s} k_I^s \bigr)$, therefore in the hyper-K\"ahler case that $\lambda=0$ and the $SU(2)$ bundle is trivial, this reduces to the familiar condition that the three symplectic forms $\vec J_{uv}$ be preserved by the isometries. By taking the cross product of the second equation in (\ref{triholo Killing vector conditions}) with $\vec J\du{v}{u}$ we obtain
\be
\label{moment maps}
2n_H \lambda \, \vec P_I = \vec J\du{u}{v} \, \nabla_v k^u_I \;.
\ee
This shows that on quaternionic K\"ahler manifolds, the sections $\vec P_I$ are completely fixed in terms of the Killing vectors. With a little bit of work%
\footnote{We take the derivative $\wt\nabla$ of (\ref{moment maps}), recalling that $\vec J$ is covariantly constant. From the algebraic Bianchi identity we have $R_{uvst} \, \vec J^{us} = \frac12 R_{vt\;\,u}^{\;\;\;\: s} \, \vec J\du{s}{u} = n_H \vec\cR_{vt} = n_H \lambda \, \vec J_{vt}$. Then we use that the vectors are Killing, as well as the properties of quaternionic-K\"ahler manifolds.}
we obtain
\be
\label{def moment maps}
\wt\nabla_u \vec P_I = \vec J_{uw} \, k^w_I \;.
\ee
This shows that $\vec P_I$ are a triplet of moment maps for the action of $k_I^u$. Taking a derivative and using that $2\wt\nabla_{[u} \wt\nabla_{v]} \vec P_I = \vec\cR_{uv} \times \vec P_I$ we get back the second equation in (\ref{triholo Killing vector conditions}), showing that the correction term on the RHS is unavoidable. The divergence of (\ref{def moment maps}) gives
\be
\wt\nabla^u \wt\nabla_u \vec P_I = -2 n_H \lambda \vec P_I \;,
\ee
showing that the moment maps are eigenfunctions of the Laplacian.

Finally, let us consider for the moment the general case that the Killing vectors might form a non-Abelian group:
\be
\label{Killing vector algebra}
[k_I, k_J]^u = 2 k_{[I}^s \nabla_s k_{J]}^u = f\du{IJ}{K} k^u_K \;,
\ee
where on the LHS is the Lie bracket and $f\du{IJ}{K}$ are the structure constants. Multiplying (\ref{triholo Killing vector conditions}) by $\nabla_v k^u_J$ and using (\ref{moment maps}), and then exploiting the derivative $\nabla_w$ of (\ref{Killing vector algebra}), we obtain
\be
\label{equivariance relation}
k_I^u \, \vec J_{uv} \, k^v_J = f\du{IJ}{K} \vec P_K + \lambda \, \vec P_I \times \vec P_J \;.
\ee
This is called the equivariance relation. In the Abelian case we just set $f$ to zero. In the special case $n_H=0$ that there are no hypermultiplets, all Killing vectors vanish and the only remnant of the quaternionic-K\"ahler structure is the condition $\vec{P}_I \times \vec{P}_J = 0$. The solution, up to $SU(2)$ rotations, is $P_I^x = \delta^{x3} \zeta_I$ where $\zeta_I$ are the so-called Fayet-Iliopoulos (FI) parameters, which in this case are extra parameters one needs to specify.

We now have all the ingredients to write the covariant derivative
\be
\label{5d cov derivative hypers}
\cD_\mu q^u = \partial_\mu q^u + g\, A_\mu^I k_I^u \;,
\ee
as well as the scalar potential
\bea
\label{pot_terms}
V &= P_I^x P_J^x \left( \frac{1}{2} \cG^{ij} \partial_i \Phi^I \partial_j \Phi^J - \frac 23 \Phi^I \Phi^J \right) + \frac{1}{2} h_{uv} \, k_I^u k_J^v \, \Phi^I \Phi^J \\
&= P_I^x P_J^x \left( \frac12 G^{IJ} - \Phi^I \Phi^J \right) + \frac{1}{2} h_{uv} \, k_I^u k_J^v \, \Phi^I \Phi^J
\eea
that couples the scalars on $\cS\cM$ and $\cQ\cM$. To go to the second line we used (\ref{projection formula}).

The covariant derivative of the supersymmetry parameter $\epsilon_i^\text{SUSY}$ (subject to symplectic-Majorana condition, with $i=1,2$) is
\be
\label{cov derivative SUSY parameter}
D_\mu \epsilon_i^\text{SUSY} = \biggl( \nabla_\mu \delta_i^j - \frac i2 \, \vec\cV_\mu \cdot \vec\sigma\du{i}{j} \biggr) \epsilon_j^\text{SUSY}
\ee
with connection
\bea
\label{spacetime SU(2) connection from QM}
\vec\cV_\mu &= \cD_\mu q^u \, \vec\omega_u - g \, A^I_\mu \vec r_I \qquad\quad\text{and}\qquad\quad \vec r_I = k_I^u \, \vec\omega_u - \lambda \, \vec P_I \;, \\
&= \partial_\mu q^u \, \vec\omega_u + g\lambda \, A_\mu^I \vec P_I
\eea
where $\lambda$ is the constant (\ref{def lambda}). Under gauge transformations%
\footnote{The covariant derivative transforms as $\delta \, \cD_\mu q^u = g\, \alpha^I \cD_\mu k^u_I$.}
\be
\delta q^u = g\, \alpha^I k_I^u \;,\qquad\qquad \delta A_\mu^I = - \partial_\mu \alpha^I
\ee
with parameters $\alpha^I$, using (\ref{Einstein cond and SU(2) curv}), (\ref{def moment maps}) and (\ref{equivariance relation}) one can show that $\vec\cV_\mu$ transforms as an $SU(2)$ connection:
\be
\delta \vec\cV_\mu = \partial_\mu \vec\Lambda + \vec\cV_\mu \times \vec\Lambda \qquad\text{with}\qquad \vec\Lambda = g\, \alpha^I \vec r_I \;.
\ee
Therefore, $D_\mu \epsilon_i^\text{SUSY}$ is covariant if $\epsilon_i^\text{SUSY}$ transforms as
\be
\delta \epsilon_i^\text{SUSY} = \frac i2\, \vec\Lambda \cdot \vec\sigma\du{i}{j} \epsilon_j^\text{SUSY} \;.
\ee

\subsection{Conifold truncation in the general framework}
\label{subapp:Cassani_matching}

Here we embed the consistent truncation of type IIB supergravity on $T^{1,1}$ to a 5d $\cN=2$ gauged supergravity with a so-called ``Betti multiplet'', described in Section 7 of \cite{Cassani:2010na} (called the ``second model'' in that paper), in the general framework. The model has $n_V=2$ and $n_H=2$. We identify the fields
\be
\phi^i = \begin{pmatrix} u+v \\ w \end{pmatrix}_{\!\!\text{CF}\hspace{-1em}} \;,\quad
\Phi^I = \begin{pmatrix}
e^{-4(u+v)/3}\\
-e^{2 (u+v)/3} \cosh 2 w\\
-e^{2 (u+v)/3} \sinh 2 w
\end{pmatrix}_{\!\!\text{CF}\hspace{-1em}} \;,\quad
A^I = \begin{pmatrix} A \\ a_1^J \\ a_1^\Phi \end{pmatrix}_{\!\!\text{CF}\hspace{-1em}} \;,\quad
q^u=\begin{pmatrix} b^\Omega_1 \\ b^\Omega_2 \\ c^\Omega_1 \\ c^\Omega_2 \\ a \\ \phi \\ C_0 \\ u \end{pmatrix}_{\!\!\text{CF}\hspace{-1em}}
\ee
where ``CF'' indicates the notation of \cite{Cassani:2010na}. The scalar fields $b^\Omega, c^\Omega$ are complex  and we used $z_1 = \re(z)$, $z_2 = \im(z)$ to indicate their real and imaginary parts, while $u,v,w,a,\phi,C_0$ are real. The hypermultiplet scalars $C_0$ and $\phi$ together form the type IIB axiodilaton $C_0+ie^{-\phi}$. Then we identify the Chern-Simons couplings
\be
\label{Cijk}
C_{122} = -C_{133} = 2
\ee
and symmetric permutations thereof, while all other components vanish, and the very special geometry of $SO(1,1) \times SO(1,1)$:
\be
\label{5d metrics conifold}
\cG_{ij} = \begin{pmatrix} 4/3 & 0 \\ 0   & 4 \end{pmatrix} \;,\qquad
G_{IJ}= e^{-\frac{4}{3}(u+v)}\begin{pmatrix}
        \frac{1}{2}e^{4(u+v)} & 0                               & 0 \\
        0                               & \cosh(4w)  & -\sinh(4w) \\
        0                               & -\sinh(4w) & \cosh(4w)     
\end{pmatrix} \;.
\ee
The tensor $C^{IJK}$ has non-vanishing components $C^{122} = - C^{133} = 1/2$ and permutations.

The quaternionic-K\"ahler manifold is $\frac{SO(4,2)}{SO(4) \times SO(2)}$. Its metric is
\bea
\label{metric compact}
h_{uv} dq^u dq^v &= e^{-4u-\phi} db^\Omega d\wb{b^\Omega} + e^{-4u+\phi} \bigl( dc^\Omega - C_0 db^\Omega\bigr)\bigl( d\wb{c^\Omega} - C_0 d\wb{b^\Omega} \bigr) \\
&\quad + \frac12 e^{-8u} \Bigl( 2da + \re\bigl( b^\Omega d\wb{c^\Omega} - c^\Omega d\wb{b^\Omega} \bigr) \Bigr)^2
+ \frac12 d\phi^2 + \frac12 e^{2\phi} dC_0^2 + 8 du^2 \;.
\eea
In this normalization $R=-32$ and thus $\lambda=-1$.
The $SU(2)$ connection is
\bea
\omega^1 - i \omega^2 &= e^{-2u-\phi/2} db^\Omega + i\, e^{-2u+\phi/2} \bigl( dc^\Omega - C_0 db^\Omega \bigr) \\
\omega^3 &= \frac12 e^{-4u} \Bigl( 2da + \re\bigl( b^\Omega d\wb{c^\Omega} - c^\Omega d\wb{b^\Omega} \bigr) \Bigr) - \frac12 e^\phi dC_0 \;.
\eea
Finally, we identify the Killing vectors
\be
\label{gaugings}
k_1 = 3 \left( - b_2^\Omega \parfrac{}{b_1^\Omega} + b_1^\Omega \parfrac{}{b_2^\Omega} - c_2^\Omega \parfrac{}{c_1^\Omega} + c_1^\Omega \parfrac{}{c_2^\Omega} \right) + 2 \parfrac{}{a} \;,\qquad k_2 = 2 \parfrac{}{a} \;,\qquad k_3= 0
\ee
and the corresponding moment maps
\be
\label{moment_maps}
P_1^x = \begin{pmatrix}
3e^{\phi/2- 2u}(c_1^\Omega - C_0b_1^\Omega + e^{-\phi} b_2^\Omega) \\
3e^{\phi/2 - 2u }(C_0b_2^\Omega - c_2^\Omega + e^{-\phi}b_1^\Omega) \\
3 - e^{-4u} ( 2 + 3 b_2^\Omega c_1^\Omega - 3 b_1^\Omega c_2^\Omega)
\end{pmatrix} \;,\qquad
P_2^x = \begin{pmatrix} 0 \\ 0 \\ -2e^{-4u} \end{pmatrix} \;,\qquad
P_3^x = 0 \;.
\ee
The $SU(2)$ connection and the moment maps were given in \cite{Halmagyi:2011yd} and can be translated into the notation of \cite{Cassani:2010na} (up to a conventional minus sign in the gauge fields) using the identifications
\be
\hspace{-.8em}
\phi^i = \begin{pmatrix} -3u_3 \\ u_2 \end{pmatrix}_{\!\!\text{HLS}\hspace{-1.4em}} ,\;\;
A^I = \begin{pmatrix}
A_1 \\[6pt]
\dfrac{k_{11}-k_{12}}{2} \\[6pt]
\dfrac{k_{11}+k_{12}}{2}
\end{pmatrix}_{\!\!\text{HLS}\hspace{-1.4em}} ,\;\;
q^u = \biggl( 2\re b_0^1, 2\im b_0^1, 2\re b_0^2, 2\im b_0^2, \frac k2, \phi,  a, u_1 \biggr)^\sT_{\!\!\text{HLS}\hspace{-1.1em}}
\ee
where ``HLS'' indicates the notation of \cite{Halmagyi:2011yd}.

The theory has a supersymmetric AdS$_5$ vacuum at $u=v=w=b^\Omega = c^\Omega = 0$ and any value of $a,C_0, \phi$ (in particular, the axiodilaton can take any value). The potential is $V\big|_\text{AdS}=-6$ leading to AdS radius $\ell_5 = g^{-1}$. The spectrum therein was computed in \cite{Cassani:2010na} (see its Table~2). We are particularly interested in the spectrum of vector fields and the Killing vectors they couple to:
\bea
\label{spectrum of vectors}
A^R &\equiv A^1 - 2 A^2 \,,\;\;& & A^3 \,:\;\;\; m^2 = 0 \;,\qquad\qquad & A^W &\equiv A^1 + A^2 \,:\;\;\; m^2 = 24 g^2 \;. \\
k_R &= \tfrac13 (k_1 - k_2) \;,& & k_3 & k_W &= \tfrac13 (2k_1 + k_2)
\eea
The vector $A^W$ acquires a mass by Higgs mechanism, eating the St\"uckelberg scalar $a$. The mass eigenstates are
\be
\bB\ud{I}{J} A^J_\mu \qquad\text{where}\qquad \bB = \mat{ 1 & -2 & 0 \\ 0 & 0 & 1 \\ 1 & 1 & 0}
\ee
is the matrix that diagonalizes them (see also Appendix~\ref{app: charges}).

%%%%%%%%%%%%%%%%%%%%%%%%%%%%%%%%
%%%%%%%%%%%%%%%%%%%%%%%%%%%%%%%%

\section {4d \matht{\cN=2} Abelian gauged supergravity}
\label{app:4d_sugra}

We summarize the salient features of 4d $\cN=2$ Abelian gauged supergravity with $n_V$ vector multiplets and $n_H$ hypermultiplets, following \cite{Andrianopoli:1996cm, Craps:1997gp, Freedman:2012zz}. The graviton multiplet contains a graviton, two gravitini and a vector; each vector multiplet contains a vector, two gaugini and a complex scalar; each hypermultiplet contains four real scalars and two hyperini (all fermions can be taken Majorana). We use indices
\be
\Lambda, \Sigma = 0, \dots, n_V \;,\qquad i,j = 1, \dots, n_V \;,\qquad u,v = 1, \dots, 4n_H
\ee
for the gauge fields $A_\mu^\Lambda$, for the complex scalars $z^i$ in vector multiplets, and for the real scalars $q^u$ in hypermultiplets, respectively. The data that define the theory are:
\begin{enumerate}
\item A special K\"ahler manifold $\cK\cM$ of complex dimension $n_V$.
\item A quaternionic-K\"ahler manifold $\cQ\cM$ of real dimension $4n_H$.
\item A set of $n_V+1$ Killing vectors on $\cQ\cM$ compatible with the quaternionic-K\"ahler structure (if $n_H=0$, $n_V+1$ FI parameters not all vanishing).
\end{enumerate}
The Killing vectors could be linearly dependent or vanish.

It is always possible to find a duality frame in which all gaugings are purely electric. In such frames the bosonic Lagrangian is
\begin{multline}
\label{4d_lagrangian}
8 \pi G^{(4)}_\text{N} e^{-1} \ccL_\text{4d} = \frac{R_s}2 - g_{i\jb}(z,\bar z) \, \partial_\mu z^i \partial^\mu \bar z^{\jb} - \frac12 \, h_{uv}(q) \, \cD_\mu q^u \cD^\mu q^v \\
{} + \frac18 \, \im\cN_{\Lambda\Sigma}(z,\bar z) \, F_{\mu\nu}^{\Lambda} F^{\Sigma\mu\nu} + \frac{e^{-1}}{16} \, \re\cN_{\Lambda\Sigma}(z, \bar z) \, F_{\mu\nu}^\Lambda F_{\rho\sigma}^\Sigma \epsilon^{\mu\nu\rho\sigma} - g^2 V(z, \bar z,q) \;.
\end{multline}
The notation is mostly as in Appendix~\ref{app:5d_sugra}. Let us explain the other terms.

\paragraph{Special K\"ahler geometry.} The scalars $z^i$ are complex coordinates on the special K\"ahler manifold $\cK\cM$ \cite{Craps:1997gp}. This is a K\"ahler-Hodge manifold --- \ie, a K\"ahler manifold with K\"ahler potential $\cK(z,\bar z)$ and metric $g_{i\jb}(z,\bar z) = \partial_i \partial_{\jb} \cK$ as well as a line bundle (\ie, a holomorphic vector bundle of rank 1) $\cL$ such that its first Chern class coincides (up to a constant) with the K\"ahler class $\omega = i \partial \bar\partial \cK$ of the manifold%
\footnote{Because fermions are sections of the square root of $\cL$, the K\"ahler class of $\cK\cM$ equal to the first Chern class of $\cL$ is required to be an even integer cohomology class.}
 --- further endowed with a flat $Sp(n_V +1, \bR)$ symplectic bundle. The manifold comes equipped with a covariantly-holomorphic section of the tensor product of the symplectic bundle with the $U(1)$-bundle $\cU$ associated to $\cL$,
\be
\label{cov holo sections}
\cV = \mat{ L^\Lambda \\ M_\Lambda} \qquad\text{such that}\qquad \begin{aligned} D_i \cV &\,\equiv\, \partial_i \cV + \tfrac12 (\partial_i \cK) \cV \\
D_{\ib} \cV &\,\equiv\, \partial_{\ib} \cV - \tfrac12 (\partial_\ib \cK) \cV = 0 \;, \end{aligned}
\ee
obeying the constraints
\be
\label{1st constraint cov holo sections}
\langle \cV,\wb\cV \rangle \,\equiv\, M_\Lambda \wb L^\Lambda - L^\Lambda \wb M_\Lambda = - i
\ee
and
\be
\label{2nd constraint cov holo sections}
\langle \cV , D_i \cV \rangle = 0 \;,
\ee
where we introduced the $Sp$-invariant antisymmetric form $i \langle\:,\, \rangle$. Equivalently, there is a holomorphic section of the tensor product of the symplectic bundle with $\cL$,%
\footnote{In particular, $A = \partial \cK$ is the Chern connection on $\cL$. Moreover, $D_i \cV = e^{\cK/2} D_i v$ and $D_\ib \cV = e^{\cK/2} D_\ib v$.}
\be
\label{def holo section}
v(z) = e^{-\cK/2} \, \cV \,\equiv\, \mat{ X^\Lambda \\ F_\Lambda} \qquad\text{such that}\qquad \begin{aligned} D_i v &\,\equiv\, \partial_i v + (\partial_i \cK) \, v \\ D_\ib v &\,\equiv\, \partial_\ib v = 0 \;, \end{aligned}
\ee
in terms of which the constraint (\ref{1st constraint cov holo sections}) reads
\be
\cK = - \log \bigl( i \, \langle v, \bar v \rangle \bigr) = - \log \Bigl[ 2\im \bigl( X^\Lambda \wb F_\Lambda \bigr) \Bigr] \;,
\ee
while the constraint (\ref{2nd constraint cov holo sections}) becomes $\langle v, D_i v \rangle = \langle v, \partial_i v \rangle = 0$. From (\ref{cov holo sections})--(\ref{2nd constraint cov holo sections}) it is easy to prove the following properties (or equivalent ones written in terms of $v$):
\bea
\label{cov holo sections extra properties}
\langle D_i \cV, \wb\cV \rangle &= 0 \;,\qquad\qquad& D_\jb D_i \cV &= g_{i\jb} \, \cV \;,\qquad\qquad& \langle D_i \cV, D_\jb \wb \cV \rangle &= i\, g_{i\jb} \\
\langle D_i \cV, D_j \cV \rangle &= 0 \;, & D_{[i} D_{j]} \cV &= 0
\eea
from which the K\"ahler metric is extracted in a symplectic-invariant way.

The rescaling of $X^\Lambda, F_\Lambda$ under K\"ahler transformations suggests to use $X^\Lambda$ as homogeneous coordinates on $\cK\cM$. It is always possible to find symplectic frames%
\footnote{See \cite{Ceresole:1995jg} for examples of frames in which, instead, a prepotential does not exist.}
in which the Jacobian matrix $e\ud{\lambda}{i}(z) = \partial_i \bigl( X^\lambda/ X^0\bigr)$ (with $\lambda=1, \dots, n_V$) is invertible. Notice that
\be
\det \bigl( e\ud{\lambda}{i} \big) = (X^0)^{n_V+1} \det \bigl( X^\Lambda, \partial_i X^\Lambda \bigr) = (X^0)^{n_V+1} \det \bigl( X^\Lambda, D_i X^\Lambda \bigr)
\ee
where the two square matrices on the right have size $n_V+1$, therefore the matrix $\bigl( X^\Lambda, \partial_i X^\Lambda \bigr)$ is invertible as well. Invertibility of the Jacobian ensures that we can use $X^\Lambda$ as homogeneous coordinates, and regard $F_\Lambda(X)$ as homogeneous functions of degree 1, namely $X^\Sigma \partial_\Sigma F_\Lambda = F_\Lambda$. From (\ref{2nd constraint cov holo sections}) and (\ref{cov holo sections extra properties}), written as $\langle v, \partial_i v \rangle = \langle \partial_i v, \partial_j v \rangle = 0$, one obtains the equations
\be
\bigl( X^\Lambda, \partial_i X^\Lambda \bigr) \, \partial_{[\Lambda} F_{\Sigma]} \, \bigl( X^\Sigma, \partial_j X^\Sigma \bigr) = 0 \;.
\ee
Invertibility of the matrix implies $\partial_{[\Lambda} F_{\Sigma]}=0$. Hence, in these frames, the sections $F_\Lambda$ are the derivatives of a holomorphic homogeneous function $F(X)$ of degree 2, called the \emph{prepotential}, namely $F_\Lambda = \partial_\Lambda F$. In such frames, the K\"ahler potential and thus the geometry are completely specified by the prepotential. The coordinates $t^i \equiv X^i / X^0$ with $i = 1, \dots, n_V$ are called \emph{special coordinates}.

The couplings of vector fields to the scalars $z^i$ are determined by the $(n_V+1) \times (n_V+1)$ period matrix $\cN$, which is uniquely defined by the relations
\be
M_\Lambda = \cN_{\Lambda\Sigma} \, L^\Sigma \;,\qquad\qquad D_\ib \wb M_\Lambda = \cN_{\Lambda\Sigma} \, D_\ib \wb L^\Sigma \;.
\ee
Explicitly, one needs to invert the matrix relation $\bigl( F_\Lambda, D_\ib \wb F_\Lambda \bigr) = \cN_{\Lambda \Sigma} \, \bigl( X^\Sigma, D_\ib \wb X^\Sigma\bigr)$.
The requirement that $g_{i\jb}$ be positive definite guarantees that the rightmost matrix is invertible \cite{Craps:1997gp}. Indeed, introducing the square matrix $\cL\ud{\Lambda}{I} = \bigl( L^\Lambda, D_\ib \wb L{}^\Lambda\bigr)$ of size $n_V+1$, one can rewrite the scalar products in (\ref{1st constraint cov holo sections}), (\ref{2nd constraint cov holo sections}) and (\ref{cov holo sections extra properties}) as
\be
\cL^\sT \bigl( \cN - \cN^\sT \bigr) \cL = 0 \;,\qquad\qquad \cL^\dag \bigl( \cN - \cN^\dag \bigr) \cL = - i \, \mathrm{diag}\bigl( 1, g_{i\jb} \bigr) \;.
\ee
The first equation shows that $\cN_{\Lambda\Sigma}$ is a symmetric matrix, given the invertibility of $\cL$. The second equation then, assuming that $g_{i\jb}$ is positive definite, proves that $\cL$ is invertible and that $\im\cN_{\Lambda\Sigma}$ is negative definite. It also gives an expression for $\im\cN_{\Lambda\Sigma}$ that, after taking the inverse, reads
\be
\label{relation im N -1}
D_i L^\Lambda D_\jb \wb L^\Sigma g^{i\jb} + \wb L^\Lambda L^\Sigma = - \frac12 \Bigl( \bigl( \im\cN \bigr)^{-1} \Bigr)^{\Lambda\Sigma} \;.
\ee
This relation, or the equivalent one in terms of the holomorphic section, will be used to rewrite the scalar potential below. When a prepotential exists, $\cN$ is obtained from
\be
\label{matrix N}
\cN_{\Lambda\Sigma} = \wb F_{\Lambda\Sigma} + 2i \, \frac{ (\im F_{\Lambda\Gamma}) X^\Gamma \, (\im F_{\Sigma\Delta}) X^\Delta }{ X^\Omega (\im F_{\Omega\Psi}) X^\Psi }\;,
\ee
where $F_{\Lambda\Sigma} = \partial_\Lambda \partial_\Sigma F$. In this expression $\cN$ is manifestly symmetric.

Finally, one can define the tensor
\be
\wt C_{ijk} = \langle D_i D_j \cV, D_k \cV \rangle = \langle \cV, D_k D_i D_j \cV \rangle \;.
\ee
Using (\ref{cov holo sections})--(\ref{cov holo sections extra properties}) and the fact that the metric is K\"ahler, one easily proves that $\wt C_{ijk}$ is totally symmetric and covariantly holomorphic, $D_{\bar\ell} \, \wt C_{ijk} = 0$ where $\wt C$ has twice the charge of $\cV$. One can prove that $(\cV, D_i \cV, \wb\cV, D_\ib \wb\cV)$ pointwise form a basis for the symplectic bundle \cite{Craps:1997gp}, hence
\be
D_i D_j \cV = i\, \wt C_{ijk} g^{k\bar k} D_{\bar k} \wb\cV
\ee
follows by taking the product of the LHS with the basis. Among other things, $\wt C$ controls the curvature tensor: $R_{\ib j \bar k \ell} = g_{j\ib} g_{\ell\bar k} + g_{j\bar k} g_{\ell\ib} - \wt C_{j\ell m} \wt C_{\ib \bar k \bar n} g^{m\bar n}$. In special coordinates the tensor $\wt C$ takes the particularly simple form
\be
\wt C_{ijk} = e^{\cK} \, \partial_i \partial_j \partial_k \cF(t) \qquad\text{with}\qquad \cF(t) = (X^0)^{-2} F(X)
\ee
and $t^i = X^i/X^0$.

\paragraph{Hypermultiplets and gauging.}
The part of the action involving the hypermultiplets has the same features as in the 5d case, summarized in Appendix~\ref{app:5d_sugra}: the hypermultiplet scalars $q^u$ (with $u=1, \dots, 4n_H$) are coordinates on a quaternionic-K\"ahler manifold $\cQ\cM$ with metric $h_{uv}(q)$. As before, we consider gauging of Abelian isometries of $\cQ\cM$, generated by $n_V+1$ (possibly vanishing or linearly dependent) Killing vectors $k^u_\Lambda(q)$ that must be compatible with the quaternionic-K\"ahler structure, with associated triplets of moment maps $\vec P_\Lambda(q)$. In full generality one could consider both electric and magnetic gaugings, described by Killing vectors $k^u_\Lambda$ and $k^{u\Lambda}$, respectively, and transforming as a vector under $Sp(n_V+1,\bR)$ duality transformations. It is always possible to find a duality frame in which all gaugings are purely electric, and we will work in such a frame. Notice that there is no guarantee that in this frame a prepotential exists.

The scalar potential is
\bea
\label{4d potential}
V &= 2 \, P^x_\Lambda P^x_\Sigma  \, e^\cK \Bigl( g^{i\jb} D_i X^\Lambda D_\jb \wb X^\Sigma - 3 X^\Lambda \wb X^\Sigma \Bigr) + 4 \, e^\cK h_{uv} \, k^u_\Lambda k^v_\Sigma X^\Lambda \wb X^\Sigma \\
&= - P^x_\Lambda P^x_\Sigma \Bigl( \bigl(\im\cN\bigr)^{-1\, \Lambda\Sigma} + 8\, e^\cK X^\Lambda \wb X^\Sigma \Bigr) + 4 \, e^\cK h_{uv} \, k^u_\Lambda k^v_\Sigma X^\Lambda \wb X^\Sigma \;.
\eea
To go to the second line we used (\ref{relation im N -1}).

The covariant derivative of the supersymmetry parameter $\epsilon_i^\text{SUSY}$ (subject to symplectic-Majorana condition, with $i=1,2$) is
\be
D_\mu \epsilon_i^\text{SUSY} = \biggl( \nabla_\mu \delta_i^j - \frac i2 \, \cA_\mu \delta_i^j - \frac i2 \, \vec\cV_\mu \cdot \vec\sigma\du{i}{j} \biggr) \, \epsilon_j^\text{SUSY}
\ee
with connections
\bea
\vec\cV_\mu &= \partial_\mu q^u \, \vec\omega_u + g\lambda \, A_\mu^I \vec P_I \\
\cA_\mu &= \frac i2 \lambda \Bigl[ (\partial_\alpha\cK) \partial_\mu z^\alpha - (\partial_{\bar\alpha} \cK) \partial_\mu \bar z^{\bar\alpha} \Bigr] \;.
\eea
Here $\vec\cV_\mu$ is the $SU(2)$ connection that descends from the quaternionic-K\"ahler manifold $\cQ\cM$, as in the 5d case (\ref{spacetime SU(2) connection from QM}). Instead $\cA_\mu$ descends from the connection on the $U(1)$-bundle $\cU$ on the special K\"ahler manifold $\cK\cM$.

%%%%%%%%%%%%%%%%%%%%%%%%%%%%%%%%
%%%%%%%%%%%%%%%%%%%%%%%%%%%%%%%%

\section{Reduction with background gauge fields}
\label{app:SS_reduction}

Following \cite{Looyestijn:2010pb} we will now reduce, piece by piece, the bosonic Lagrangian \eqref{5d_lagrangian} of 5d $\cN=2$ gauged supergravity down to 4d. We start in 5d with $n_V$ vector multiplets and $n_H$ hypermultiplets. We use indices
\be
I,J = 1, \dots, n_V + 1 \;,\qquad \Lambda, \Sigma = 0, \dots, n_V+1 \;,\qquad u,v = 1, \dots, 4n_H \,.
\ee
We indicate the 5d vector fields as $\wh A^I_M$ (where $M,N=0, \dots, 4$ are spacetime indices) and parametrize the vector multiplet scalars in terms of sections $\Phi^I$ subject to the cubic constraint $\cV(\Phi) = 1$ in (\ref{def very special real manifold}). The hypermultiplet scalars are $q^u$. We employ the rather standard Kaluza-Klein reduction ansatz (\ref{ansatz}) and (\ref{modified_ansatz}):
\begin{align}
\wh g_{MN} &= \begin{pmatrix}
e^{2\wt \phi} g_{\mu\nu} +  e^{-4\wt\phi} A^0_\mu A^0_\nu & - e^{-4\wt\phi} A^0_\mu \\
- e^{-4\wt\phi} A^0_\nu &  e^{-4\wt\phi} \end{pmatrix} \;,\quad&
\wh g^{MN} &= \begin{pmatrix}
e^{-2\wt \phi} g^{\mu\nu} & e^{-2\wt\phi} A^0{}^\mu \\
e^{-2\wt\phi} A^0{}^\nu &  e^{4\wt\phi} + e^{-2\wt\phi}A^0_\rho A^0{}^\rho \end{pmatrix} \;, \nn \\[.5em]
e_{(5)} &= e^{2\wt\phi} \, e_{(4)} \;,\qquad\qquad \Phi^I = - e^{2\wt \phi} \, z_2^I \;,& \wh A^I_M &= \bigl( A^I_\mu- z_1^IA^0_\mu,\, z_1^I + \xi^I \bigr) \;.
\label{ansatz_app}
\end{align}
The last coordinate, that we call $y$ and whose range $\Delta y$ we leave generic for now, is compactified on a circle of length $e^{-2\wt\phi} \Delta y$, and no field depends on it.
We indicated as $\wh g_{MN}$ and $e_{(5)}$ the 5d metric and the square root of its determinant, and as $g_{\mu\nu}$ and $e_{(4)}$ (with $\mu,\nu=0, \dots, 3$ spacetime indices) their 4d counterparts. In 4d we end up with $n_V+1$ vector multiplets, and we indicate as $A_\mu^\Lambda$ the vector fields. The physical scalars in 4d vector multiplets are the complex fields $z^i$. With a useful abuse of notation, we utilize the very same index $I$ for 5d vector fields and 4d physical scalars, $z^I$, because in 4d we have one more vector field than in 5d. We also use the notation
\be
z_1^I \,\equiv\, \re z^I \;,\qquad\qquad z_2^I \,\equiv\, \im z^I \;.
\ee
Notice that the real scalar $\wt\phi$ can be eliminated with the 5d constraint,
\be
e^{-6\wt\phi} = - \cV(z_2) \;,
\ee
then the scalars $z^I$ can be treated as independent.
The real parameters $\xi^I$ represent background gauge fields along the circle, therefore, up to a gauge transformation, this ansatz is equivalent to a Scherk-Schwarz reduction.

The reduction of the Einstein term gives
\be
8\pi G^{(4)}_\text{N} \ccL_1 = e_{(5)} \, \frac{\wh R_s}2 = e_{(4)} \biggl[ \frac{R_s}2 - 3 \,\partial_\mu \wt\phi \, \partial^\mu \wt\phi - \frac{e^{-6\wt\phi}}8 \, F^0_{\mu\nu} F^{0\mu\nu} \biggr] + \text{total derivatives} \;.
\ee
Here $\wh R_s$ and $R_s$ are the 5d and 4d Ricci scalars, respectively. The 4d and 5d Newton constants are related by
\be
\label{relation Newton constants}
\frac1{G^{(4)}_\text{N}} = \frac{\Delta y}{G^{(5)}_\text{N}} \;.
\ee
In the following, for clarity, we will omit the factor $8\pi G^{(4)}_\text{N}$.
The reduction of the kinetic term of vector multiplet scalars gives
\be
\ccL_2 = - e_{(5)} \frac12 \, G_{IJ} \, \wh g^{MN} \partial_M \Phi^I \partial_N \Phi^J = e_{(4)} \biggl[ - \frac{e^{4\wt\phi}}2 \, G_{IJ} \partial_\mu z_2^I \partial^\mu z_2^J + 3 \, \partial_\mu \wt\phi \, \partial^\mu \wt\phi \biggr] \;.
\ee
The last term exactly cancels the second term in $\ccL_1$, therefore
\be
\ccL_1 + \ccL_2 = e_{(4)} \biggl[ \frac{R_s}2 - \frac{e^{4\wt\phi}}2 \, G_{IJ} \, \partial_\mu z_2^I \, \partial^\mu z_2^J - \frac{e^{-6\wt\phi}}8 \, F^0_{\mu\nu} F^{0\mu\nu} \biggr] \;.
\ee
The reduction of the kinetic term of hypermultiplet scalars gives
\bea
\ccL_3 &= - e_{(5)} \frac12 \, h_{uv} \, \wh g^{MN} \wh\cD_M q^u \wh\cD_N q^v \\
&= e_{(4)} \biggl[ - \frac12 \, h_{uv} \, \cD_\mu q^u \cD^\mu q^v - \frac{g^2 e^{6\wt\phi}}2 \, \bigl( k_0^u + z_1^I k_I^u \bigr) h_{uv} \bigl( k_0^v + z_1^J k_J^v \bigr) \biggr] \;.
\eea
Here $\wh\cD_M q^u = \partial_M q^u + g\, \wh A^I_M k_I^u$ is the 5d covariant derivative in (\ref{5d cov derivative hypers}), while
\be
\label{reduced cov derivative}
\cD_\mu q^u = \partial_\mu q^u + g \, A_\mu^I k_I^u + g \, A_\mu^0 \, \xi^I k_I^u = \partial_\mu q^u + g\, A_\mu^\Lambda k_\Lambda^u
\ee
is the 4d covariant derivative, and we defined the new Killing vector
\be
\label{new Killing vector}
k_0^u \,\equiv\, \xi^I k_I^u \;.
\ee
The reduction of the gauge kinetic term gives
\bea
\ccL_4 &= - e_{(5)} \frac14\, G_{IJ} \, \wh F_{MN}^I \wh F^{J MN} \\
&= e_{(4)} \biggl[ - \frac{e^{-2\wt\phi}}4 \, G_{IJ} \bigl( F_\mu^I - z_1^I F_{\mu\nu}^0 \bigr) \bigl( F^{J\mu\nu} - z_1^J F^{0\mu\nu} \bigr) - \frac{e^{4\wt\phi}}2 \, G_{IJ} \, \partial_\mu z_1^I \partial^\mu z_1^J \biggr] \;,
\eea
where $\wh F_{MN}$ and $F_{\mu\nu}$ are the 5d and 4d field strengths, respectively. We used $\wh F_{\mu 4}^I = \partial_\mu z_1^I$ and $\wh F_{\mu\nu}^I = F_{\mu\nu}^I - z_1^I F_{\mu\nu}^0 + 2 A_{[\mu}^0 \partial_{\nu]}^{\phantom{0}} z_1^I$.

In order to reduce the Chern-Simons term, we extend the geometry (\ref{ansatz}) to a 6d bulk whose boundary is the original 5d space. A convenient way to do that is to complete the circle parametrized by $y$ into a unit disk with radius $\rho \in [0,1]$. We extend the 5d connections $\wh A^I$ in (\ref{modified_ansatz}) to 6d connections $\wt A^I$ as follows:
\be
\wt A^I = A^I + \xi^I A^0 + \rho^2 (z_1^I + \xi^I) \bigl( dy - A^0 \bigr) \;.
\ee
We then write the Chern-Simons action term as
\be
\int_\text{5d} \ccL_5 = \int_\text{5d} \frac1{12} \, C_{IJK} \, \wh F^I \wedge \wh F^J \wedge \wh A^K = \int_\text{6d} \frac1{12} \, C_{IJK} \, \wt F^I \wedge \wt F^J \wedge \wt F^K \;.
\ee
Substituting $\wt F^I = d \wt A^I$ and performing the integrals over $d\rho^2 \wedge (dy-A^0)$, we extract the 4d reduced Lagrangian
\be
\ccL_5 = \frac1{16} \, C_{IJK} \epsilon^{\mu\nu\rho\sigma} \biggl[ \bigl( z_1^I + \xi^I \bigr) F_{\mu\nu}^J F_{\rho\sigma}^K - \bigl( z_1^I z_1^J - \xi^I \xi^J \bigr) F_{\mu\nu}^K F_{\rho\sigma}^0 + \frac{z_1^I z_1^J z_1^K + \xi^I \xi^J \xi^K}3 F_{\mu\nu}^0 F_{\rho\sigma}^0 \biggr] \;.
\ee
Notice that the contributions containing the $\xi^I$'s are standard theta terms.

Finally, the reduction of the scalar potential gives
\be
\ccL_6 = - e_{(5)} g^2 V = - e_{(4)} g^2 \biggl[ P_I^x P_J^x \biggl( \frac{e^{2\wt\phi}}2 \cG^{ij} \partial_i \Phi^I \partial_j \Phi^J - \frac{2e^{6\wt\phi}}3 z_2^I z_2^J \biggr) + \frac{e^{6\wt\phi}}2 h_{uv} k^u_I k^v_J z_2^I z_2^J \biggr] \,.
\ee

We proceed now with recasting the various pieces of the action in the general form (\ref{4d_lagrangian}) of 4d $\cN=2$ gauged supergravity with $n_V+1$ vector multiplets and $n_H$ hypermultiplets. The Einstein term receives its contribution from $\ccL_1$:
\be
\ccL_1' = e_{(4)} \, \frac{R_s}2 \;.
\ee
The kinetic term of vector multiplet scalars gets contributions from $\ccL_2$ and $\ccL_4$:
\be
\ccL_2' = - e_{(4)} \frac{e^{4\wt\phi}}2 \, G_{IJ} \Bigl( \partial_\mu z_2^I \partial^\mu z_2^J + \partial_\mu z_1^I \partial^\mu z_1^J \Bigr) = - e_{(4)}\, g_{I \bar J} \, \partial z^I \partial^\mu \bar z^{\bar J} \;,
\ee
where we defined the Hermitian metric
\be
\label{Hermitian metric reduction}
g_{I\bar J} = \frac{e^{4\wt\phi}}2 \, G_{I\bar J} \;.
\ee
The kinetic term of hypermultiplet scalars gets its contribution from $\ccL_3$,
\be
\ccL_3' = - e_{(4)} \, \frac12\, h_{uv} \, \cD_\mu q^u \cD^\mu q^v \;,
\ee
with the covariant derivative $\cD_\mu$ defined in (\ref{reduced cov derivative})-(\ref{new Killing vector}).
The gauge kinetic term receives contributions from $\ccL_1$ and $\ccL_4$:
\be
\ccL_4' = - e_{(4)} \frac{e^{-6\wt\phi}}8 \biggl[ F^0_{\mu\nu} F^{0\mu\nu} + 4 g_{IJ} \bigl( F^I_{\mu\nu} - z_1^I F^0_{\mu\nu} \bigr) \bigl( F^{J\mu\nu} - z_1^J F^{0\mu\nu} \bigr) \biggr] = e_{(4)} \, \frac18 \im \cN_{\Lambda\Sigma} \, F^\Lambda_{\mu\nu} F^{\Sigma\mu\nu}
\ee
where we defined the field-dependent matrix of gauge couplings
\be
\label{im N reduced}
\im\cN_{\Lambda\Sigma} = - e^{-6\wt\phi} \mat{ 1 + 4 g_{KL} z_1^K z_1^L & - 4 g_{KJ} z_1^K \\ - 4 g_{IK} z_1^K & 4g_{IJ} }
\ee
in which the indices $\Lambda,\Sigma$ run over 0 and then the values of $I,J$. On the other hand, the field-dependent theta terms are contained in $\ccL_5$:
\be
\ccL_5' = \ccL_5 = \frac1{16} \re \cN_{\Lambda\Sigma} \, \epsilon^{\mu\nu\rho\sigma} F^\Lambda_{\mu\nu} F^\Sigma_{\rho\sigma}
\ee
where
\be
\label{re N reduced}
\re\cN_{\Lambda\Sigma} = \mat{ \frac13 C_{KLM} \bigl(z_1^K z_1^L z_1^M + \xi^K \xi^L \xi^M \bigr) & -\frac12 C_{JKL} \bigl( z_1^K z_1^L - \xi^K \xi^L \bigr) \\ -\frac12 C_{IKL} \bigl( z_1^K z_1^L - \xi^K \xi^L \bigr) & C_{IJK} \bigl( z_1^K + \xi^K \bigr) } \;.
\ee

It turns out that $g_{I\bar J}$ and $\cN_{\Lambda\Sigma}$ descend from the following prepotential:
\bea
F(X) &= \frac16\, C_{IJK} \, \frac{\check X^I \check X^J \check X^K}{X^0} \hspace{10em}\text{with } \check X^I \equiv X^I + \xi^I X^0 \\[.5em]
&= \frac16 \, C_{IJK} \, \frac{ X^I X^J X^K}{X^0} + \frac12 C_{IJK} \biggl( \xi^I X^J X^K + \xi^I \xi^J X^K X^0 + \frac13 \xi^I \xi^J \xi^K (X^0)^2 \biggr)\;.
\eea
The terms in parenthesis involving the $\xi^I$'s only affect standard theta terms, which are topological and thus do not enter in the equations of motion.
Indeed, using special coordinates $z^I = X^I/X^0$ and in the K\"ahler frame $|X^0|^2=1$, one derives the K\"ahler potential%
\footnote{The completely covariant expression for the K\"ahler potential is $e^{-\cK} = 8 \,|X^0|^2 \, e^{-6\wt\phi}$.}
\be
\cK = - \log\biggl( \frac1{6i} \, C_{IJK} \, \bigl( z^I - \bar z^I \bigr) \bigl( z^J - \bar z^J \bigr) \bigl( z^K - \bar z^K \bigr) \biggr) = - \log\bigl( 8\, e^{-6\wt\phi} \bigr)
\ee
from which the K\"ahler metric (\ref{Hermitian metric reduction}) with (\ref{alternative 5d metric G}) follows. On the other hand
\be
F_{\Lambda\Sigma} = \mat{ \frac13 C_{KLM} \bigl( z^K z^L z^M + \xi^K \xi^L \xi^M \bigr) & -\frac12 C_{JKM} \bigl( z^K z^M - \xi^K \xi^L \bigr) \\ -\frac12 C_{IKM} \bigl( z^K z^M - \xi^K \xi^L \bigr) & C_{IJK} \bigl( z^K + \xi^K \bigr) }
\ee
from which the matrix $\cN$ in (\ref{im N reduced}) and (\ref{re N reduced}) follows.
It might be useful
\bea
(X^0)^{-2} X^\Lambda \bigl( \im F_{\Lambda\Sigma} \bigr) X^\Sigma &= 4 \, C_{IJK} \Bigl( \tfrac13 \im(z^I z^J z^K) - \tfrac12 \im(z^Iz^J) \re(z^K) \Bigr) \\
&= -\frac43\, C_{IJK} z_2^I z_2^J z_2^K = e^{-\cK} = 8\, e^{-6\wt\phi} \;,
\eea
as well as $\bigl( \im F_{I\Sigma} \bigr) X^\Sigma/X^0 = i \, C_{IKM} z_2^K z_2^M$.

Finally, the scalar potential gets contributions from $\ccL_3$ and $\ccL_6$:
\bea
\ccL_6' &= - e_{(4)} g^2 \Biggl[ P_I^x P_J^x \biggl( \frac{e^{2\wt\phi}}2 \cG^{ij} \partial_i \Phi^I \partial_j \Phi^J - \frac{2 e^{6\wt\phi}}3 z_2^I z_2^J \biggr) + {} \\
&\qquad\qquad\qquad {} + \frac{e^{6\wt\phi}}2 h_{uv} \Bigl( k_I^u k_J^v z_2^I z_2^J + (k_0^u + z_1^I k_I^u)(k_0^v + z_1^J k_J^v) \Bigr) \Biggr] \\
&= - e_{(4)} g^2 \biggl[ - P_\Lambda^x P_\Sigma^x \, \Bigl( \bigl(\im\cN)^{-1\, \Lambda\Sigma} + 8 \,e^\cK X^\Lambda \wb X^\Sigma \Bigr) + 4 \, e^\cK h_{uv} k^u_\Lambda k^v_\Sigma X^\Lambda \wb X^\Sigma \biggr] \;.
\eea
To manipulate the first line we used (\ref{projection formula}) as well as
\be
\Bigl( \bigl( \im \cN \bigr)^{-1} \Bigr)^{\Lambda\Sigma} + 8 \, e^\cK X^{(\Lambda} \wb X{}^{\Sigma)} = - e^{6\wt\phi} \mat{ 0 & 0 \\ 0 & \frac14 g^{IJ} - z_2^I z_2^J } \;,
\ee
which immediately follows from (\ref{im N reduced}). Notice in particular that $\vec P_0$ drops out of the potential and cannot be extracted from it, but it is still determined as $\vec P_0 = \xi^I \vec P_I$ from (\ref{new Killing vector}). The action $\ccL_6'$ exactly reproduces the potential in (\ref{4d potential}).

Summarizing, the compactification gives the following map from 5d to 4d data:
\be
\begin{tabular}{c} 5d \\[.4em] $n_V$ vector multiplets \\[.6em] $\cS\cM$ with $C_{IJK}$ \\[.7em] $\cQ\cM$ with $h_{uv}(q)$ \\[.4em] gauging of $k^u_I$ \end{tabular}
\qquad\xrightarrow[\text{background fields}]{\text{reduction with } \xi^I}\qquad
\begin{tabular}{c} 4d \\[.4em] $n_V+1$ vector multiplets \\[.3em] $\cK\cM$ with $\ds F = \frac16 C_{IJK} \frac{\check X^I \check X^J \check X^K}{X^0}$ \\[.7em] $\cQ\cM$ with $h_{uv}(q)$ \\[.4em] gauging of $k^u_\Lambda = \bigl( \xi^J k_J^u, k_I^u\bigr)$ \end{tabular}
\ee
where $\check X^I = X^I + \xi^I X^0$.

\subsection{Reduction of the conifold truncation}
\label{subapp:conifold_reduction}

The reduction of the 5d conifold truncation described in Appendix~\ref{subapp:Cassani_matching} gives a 4d supergravity with the following data. The prepotential is
\be
F = \frac{\check X^1 \bigl( (\check X^2)^2 - (\check X^3)^2 \bigr)}{X^0} \;.
\ee
It induces the vector multiplet scalar metric
\be
g_{I\bar J} = \frac12 \mat{ \dfrac1{2(z_2^1)^2} & 0 & 0 \\ & \dfrac{ (z_2^2)^2 + (z_2^3)^2 }{ \bigl( (z_2^2)^2 - (z_2^3)^2 \bigr)^2} & - \dfrac{ 2\, z_2^2 \, z_2^3 }{  \bigl( (z_2^2)^2 - (z_2^3)^2 \bigr)^2 } \\[1.4em] \multicolumn{2}{c}{\text{Symmetrized}} & \dfrac{ (z_2^2)^2 + (z_2^3)^2 }{ \bigl( (z_2^2)^2 - (z_2^3)^2 \bigr)^2} }
\ee
that depends on $z_2^I$, the theta terms (\ref{re N reduced}) that depend on $z_1^I$ and $\xi^I$, while the gauge coupling function $\im\cN_{\Lambda\Sigma}$ takes a lengthier expression that depends on $z_1^I$ and $z_2^I$ and can be easily derived from (\ref{im N reduced}). Since in 5d $k_3=0$, the 4d extra Killing vector is $k_0 = \xi^1 k_1 + \xi^2 k_2$.

\section{Black hole charges and their reduction}
\label{app: charges}

The electric black hole charges computed in \cite{Kunduri:2006ek} in our notation read
\be
\label{5d charge definition}
Q_\fT = - \frac1{8\pi G_\text{N}^{(5)}g} \int_{S^3_\infty} G_{\fT J} \star_5 \wh F^J\;,
\ee
where the integral is taken on the three-sphere at infinity, and are dimensionless.
We recall that only a subspace $\wh A_\mu^\fT$ of the vector fields are massless on the AdS$_5$ vacuum, and the index $\fT$ runs over them. The massless vectors are such that the hypermultiplet scalars sit at a fixed point of the gauged isometries, and are thus identified by the conditions
\be
\label{condition conserved charges}
k^u_\fT(q)=0 \;.
\ee
Indeed, let $\bB\ud{I}{J}$ be a matrix of linear redefinitions such that $\bB\ud{I}{J} \wh A^J_\mu$ are mass eigenstates. Such a matrix is characterized by $\bB\ud{I}{J} G^{JN} k^u_N h_{uv} k^v_L = \lambda^I_N \bB\ud{N}{L}$ where $\lambda$ is the diagonal matrix of squared masses (in units of $g^2$). The corresponding linear transformation of charges is $Q_I \to Q_J (\bB^{-1})\ud{J}{I}$, while the Killing vectors corresponding to the mass eigenstates are $k_J^u (\bB^{-1})\ud{J}{I}$. Now consider a massless vector and let the index $\fT$ be such that $\lambda^\fT_\fT = 0$ (not summed over $\fT$). Using non-degeneracy of the metrics $G_{IJ}$ and $h_{uv}$, one easily proves that $k^u_J (\bB^{-1})\ud{J}{\fT} = 0$, which is (\ref{condition conserved charges}).

Now, the equations of motion for the bosonic fields of 5d gauged supergravity that follow from (\ref{5d_lagrangian}) are
\bea
\label{5d EOMs}
d\Bigl( G_{IJ} \star_5 \wh F^J \Bigr) &= \frac14 C_{IJK} \wh F^J \wedge \wh F^K - g\, h_{uv} \, k^u_I \star_5 \wh \cD q^v \\
\wh R_{MN} &= G_{IJ} \biggl( \wh F^I_{MP} \wh F^{JP}_N - \frac16 \, \wh g_{MN} \wh F^I_{PQ} \wh F^{JPQ} \biggr) \\
&\quad + \cG_{ij} \, \partial_M \phi^i \partial_N \phi^j + h_{uv} \, \wh\cD_M q^u \wh\cD_N q^v + \frac23 \, \wh g_{MN} \, g^2 V \;.
\eea
Notice that (\ref{condition conserved charges}) is just the condition not to have a source in the $\fT$-th component of Maxwell's equation from the hypermultiplets.
We can express the charges $\cQ_\fT$ in terms of integrals at the horizon \cite{Hanaki:2007mb} using the EOMs (\ref{5d EOMs}):
\be
Q_\fT = - \frac1{8\pi G_\text{N}^{(5)}g} \Biggl[ \int_{S^3_r} G_{\fT J} \star_5 \wh F^J + \int_{S^3_r \times I[r,\infty]} \biggl( \frac14 \, C_{\fT JK} \, \wh F^J \wedge \wh F^K - g \, h_{uv} \, k^u_\fT \star_5 \wh\cD q^v \biggr) \Biggr] \;.
\ee
The first term is an integral evaluated at radius $r$, that we will take to be the horizon location. The second term is a correction, integrated on a cylinder $S^3 \times I$ where $I$ is the interval from $r$ to $\infty$, that leads to a Page charge. Assuming that the condition $k^u_\fT(q)=0$ remains true also on the black hole background,%
\footnote{In the case of the conifold compactification discussed in Section~\ref{sec: SUGRA example conifold}, this assumption is true, see (\ref{vanishing of Killing vectors}). We expect the assumption to be true in all cases.}
the third term vanishes.

We can apply a similar manipulation to the angular momenta $J_{a=1,2}$. Given the spacetime Killing vectors $K_a \equiv K_a^M \partial_M$, the angular momenta are defined in \cite{Kunduri:2006ek} as
\be
J_a = \frac1{16\pi G^{(5)}_\text{N}} \int_{S^3_\infty} {} \star_5 dK_a
\ee
where we have indicated with the same symbol $K_a \equiv K_{aM} dx^M$ the 1-forms dual to the Killing vectors, and the integral is evaluated once again at infinity. One can show that the Killing equation implies
\be
d\star_5 d K = 2 \wh R_{MN} K^M \star_5 dx^N \;.
\ee
We can then use the EOMs (\ref{5d EOMs}) to replace the Ricci scalar $\wh R_{MN}$. We assume that $S^3$ is invariant under the isometries generated by $K_a$, therefore, indicating as $\mathrm{i}_K$ the interior product, the integral of $\wh g_{MN} K^M \star_5 dx^N = \mathrm{i}_K (\star_5 1)$ vanishes. We also assume that $\mathrm{i}_K d\phi^i = 0$. We obtain
\be
\label{angular momentum 5d}
J_a = \frac1{16\pi G_\text{N}^{(5)}} \Biggl[ \int_{S^3_r} {} \star_5 dK_a + 2 \int_{S^3 \times I} \biggl( G_{IJ} \bigl( \mathrm{i}_{K_a} \wh F^I \bigr) \wedge \star_5 \wh F^J + h_{uv} \bigl( \mathrm{i}_{K_a} \wh\cD q^u \bigr) \star_5 \wh\cD q^v \Biggr] \;.
\ee

Now let us proceed and reduce the charges to 4d imposing the ansatz (\ref{ansatz_app}), in particular
\bea
\wh A^I &= A^I + \xi^I A^0 + (z_1^I + \xi^I)(dy - A^0) \\
\wh F^I &= F^I - z_1^I F^0 + dz_1^I \wedge (dy - A^0) \;,
\eea
and performing the integrals along the circle. Notice that because of (\ref{relation Newton constants}) and since the horizon areas in 5d and 4d are related by $\text{Area}_{(5)} = \Delta y \, \text{Area}_{(4)}$, the black hole entropy is the same in 5d and 4d. We find
\bea
\int_{S^3} G_{IJ} \star_5 \wh F^J &= \Delta y \int_{S^2} e^{-2\wt\phi} \, G_{IJ} \star_4 \bigl( F^J - z_1^J F^0 \bigr) \\
C_{IJK} \int_{S^3\times I} \wh F^J \wedge \wh F^K &= - \Delta y\, C_{IJK} \int_{S^2_r} \Bigl( 2z_1^J F^K - z_1^J z_1^K F^0 \Bigr) \;.
\eea
In the second equality we used that $z_1^I \to 0$ at infinity. The electric charges are thus
\be
\cQ_\fT = \frac1g \int_{S^2_r} \frac{\delta S_\text{4d}}{\delta F^\fT} - \frac1{8\pi G_\text{N}^{(4)}g} \, C_{\fT JK} \int_{S^2_r} \biggl( \frac12\, \xi^J F^K + \frac14\, \xi^J \xi^K F^0 \biggr) \;,
\ee
where
\be
\frac{\delta S_\text{4d}}{\delta F^\Lambda} = \frac1{16\pi G_\text{N}^{(4)}} \Bigl( \im\cN_{\Lambda\Sigma} \star_4 F^\Sigma + \re\cN_{\Lambda\Sigma} F^\Sigma \Bigr)
\ee
are the derivatives of the action obtained from (\ref{4d_lagrangian}) with (\ref{im N reduced}) and (\ref{re N reduced}).

We define the 4d dimensionless magnetic charges as
\be
\label{def magnetic charges}
p^\Lambda = \frac{g}{4\pi} \int_{S^2} F^\Lambda \;,
\ee
where the integral can be done at any radius because of the Bianchi identities. On the other hand, the first Chern class of the circle fibration --- that we take to be the Hopf fibration of $S^3$ --- is $\frac1{\Delta y} \int dA^0 = 1$. Thus, we obtain a properly quantized $p^0 = 1$ if we set
\be
\Delta y = \frac{4\pi}g \;.
\ee
We will use this normalization from now on.

Let us now reduce the angular momentum. We consider the case $J_1 = J_2$, with $J_{1,2}$ normalized such that they generate orbits of length $2\pi$, and define $J = (J_1 + J_2)/2$. The corresponding Killing vector and dual 1-form are
\be
K^M \partial_M = \frac{\Delta y}{4\pi} \, \frac{\partial}{\partial y} = \frac1g \, \frac{\partial}{\partial y} \;,\qquad\qquad K_M dx^M = \frac1g \, e^{-4\wt\phi} (dy-A^0) \;.
\ee
The first term in (\ref{angular momentum 5d}) gives
\be
\int_{S^3} {} \star_5 dK = - \frac{\Delta y}g \int_{S^2} e^{-6\wt\phi} \star_4 F^0 \;.
\ee
To reduce the second term we use $\mathrm{i}_K \wh F^I = - \frac1g dz_1^I$, integrate by parts, and use the EOMs (\ref{5d EOMs}). To reduce the third term we use $\mathrm{i}_K \wh\cD q^u = \bigl( z_1^I + \xi^I \bigr) k_I^u$ and $\mathrm{i}_\omega (\star\,1) = \star\,\omega$ for a 1-form $\omega$. Eventually
\begin{align}
\label{J reduced long expression}
J &= \frac1{8\pi G_\text{N}^{(4)}g} \Biggl\{ \int_{S^2_r} \Biggl[ -\frac12\, e^{-6\wt\phi} \star_4 F^0 + e^{-2\wt\phi} \, G_{IJ} \, z_1^I \star_4 \bigl( F^J - z_1^J F^0 \bigr) \\
&\hspace{8em} - C_{IJK} \biggl( \frac14\, z_1^I z_1^J F^K - \frac16\, z_1^I z_1^J z_1^K F^0 \biggr) \Biggr] + \int_{S^2 \times I} \!\!\! {} \star_4 g \, k^u_0 \, h_{uv}\, \cD q^v \Biggr\} \nn\\
&= \frac1g \int_{S^2_r} \frac{\delta S_\text{4d}}{\delta F^0} - \frac1{8\pi G_\text{N}^{(4)}g} \Biggl[ C_{IJK} \xi^I \xi^J \int_{S^2_r} \biggl( \frac14\, F^K + \frac16\, \xi^K F^0 \biggr) + \int_{S^2 \times I} \!\!\! {} \star_4 g \, k^u_0 \, h_{uv}\, \cD q^v \Biggr] \;. \nn
\end{align}
The four-dimensional angular momentum of the black hole solution is proportional to $J_1 - J_2$, which vanishes in the case under consideration. This implies that we can impose spherical symmetry on $S^2$. The section $\cD q^v$ is charged under the Abelian vector fields $A_\mu^\Lambda$, therefore the magnetic fluxes $p^\Lambda$ give rise to an effective spin $s$ on $S^2$. However, the spin spherical harmonics \cite{Newman:1966ub, Wu:1976ge} have total angular momentum $j\geq |s|$, which should vanish in order to have a spherically-symmetric configuration. Since the Abelian symmetries are realized non-linearly on $\cD q^v$ as soon as $k^u_\Lambda \neq 0$, we obtain the condition
\be
p^\Lambda k^u_\Lambda(q) = 0
\ee
for spherically-symmetric black hole solutions. Without loss of generality, in Section~\ref{sec: conifold} we have set $p^I = 0$ which implies $k^u_0 = 0$. We then see that the last term in (\ref{J reduced long expression}) vanishes.

The magnetic charges that appear in the attractor equations of \cite{Klemm:2016wng}, in our conventions, are (\ref{def magnetic charges}) while the electric  charges are
\be
\label{def electric charges}
q_\Lambda = \frac g{4\pi} \int_{S^2_r} G_\Lambda \qquad\text{with}\qquad G_\Lambda = 16\pi G_\text{N}^{(4)} \, \frac{\delta S_\text{4d}}{\delta F^\Lambda} \;.
\ee
Setting $p^I = 0$, we obtain the following dictionary between 5d and 4d charges:
\bea
\label{map of charges}
q_0 &= 4G_\text{N}^{(4)}g^2 \, J + \frac13\, C_{IJK} \xi^I \xi^J \xi^K p^0 \\
q_\fT &= 4G_\text{N}^{(4)}g^2 \, Q_\fT + \frac12\, C_{\fT JK} \frac12 \xi^J \xi^K p^0 \;.
\eea

\subsection{Baryonic charge quantization in the conifold theory}
\label{subapp: charge quantization}

In order to fix the exact relation between the supergravity charge $Q_3$ and the field theory baryonic charge $Q_B$, we deduce the Dirac quantization condition satisfied by $A^3_\mu$ from the consistent reduction of \cite{Cassani:2010na}.

The metric of $T^{1,1}$ is
\be
ds^2 = \frac16 \sum_{i=1,2} \Bigl( d\theta_i^2 + \sin^2\theta_i \, d\varphi_i^2 \Bigr) + \eta^2 \qquad\text{with}\qquad
\eta = - \frac13 \biggl( d\psi + \sum_{i=1,2} \cos\theta_i\, d\varphi_i \biggr) \;.
\ee
We define the 2-forms%
\footnote{The 2-form $J$ should not be confused with the angular momentum of the black hole.}
\bea
J &= \frac16 \Bigl( \sin\theta_1\, d\theta_1 \wedge d\varphi_1 +  \sin\theta_2\, d\theta_2 \wedge d\varphi_2 \Bigr) = \frac12 d\eta \\
\Phi &= \frac16 \Bigl( \sin\theta_1\, d\theta_1 \wedge d\varphi_1 -  \sin\theta_2\, d\theta_2 \wedge d\varphi_2 \Bigr) \;.
\eea
The expansion of the 10d RR field strength $F_5^\text{RR}$ in \cite{Cassani:2010na} around the AdS$_5 \times T^{1,1}$ vacuum (where $u = v = w = b^\Omega = c^\Omega = 0$), keeping only the dependence on the gauge fields and the St\"uckelberg scalar $a$, in our conventions reads
\bea
F_5^\text{RR} &= 4 g \,\star_5 1 - 2g^{-1}\, (\star_5 \, Da) \wedge (\eta - g\wh A^1) - g^{-2} \, (\star_5\, d\wh A^2) \wedge J + g^{-2} \, (\star_5\,  d \wh A^3) \wedge \Phi \\
&\quad - g^{-3} \, d\wh A^2 \wedge J \wedge (\eta - g\wh A^1) - g^{-3} \, d\wh A^3 \wedge \Phi \wedge (\eta - g\wh A^1) \\
&\quad + g^{-4} \, J \wedge J \wedge \bigl( Da + 2( \eta - g\wh A^1)\bigr)  \;,
\eea
where $\star_5$ is the Poincar\'e dual in AdS$_5$ while $Da = da + 2g(\wh A^1 + \wh A^2)$. Dirac's quantization condition reads
\be
\label{F5 quantization condition}
\frac{1}{2\sqrt\pi\, \kappa_{10}} \int_{\cC_5} F_5^\text{RR} \in \bZ
\ee
for any closed 5-cycle $\cC_5$. Here $\kappa_{10}$ is the 10d gravitational coupling, related to the 5d Newton constant by
\be
\frac{\Vol(T^{1,1})}{g^5 \kappa_{10}^2} = \frac1{8\pi G_\text{N}^{(5)}}
\ee
where $\Vol(T^{1,1}) = 16\pi^3/27$. Applying (\ref{F5 quantization condition}) to $\cC_5 = T^{1,1}$ and imposing that there are $N$ units of 5-form flux, we recover (\ref{relation N conifold}). On the other hand, let us apply (\ref{F5 quantization condition}) to the 5-cycle $X_2 \times S^3$, where $X_2$ is the non-trivial 2-cycle of $T^{1,1}$ while $S^3$ is a spatial 3-sphere in AdS$_5$. Using $\int_{X_2} J = 0$ and $\int_{X_2}\Phi = 4\pi/3$ as well as (\ref{relation N conifold}), we obtain
\be
\frac1{2\sqrt\pi\, \kappa_{10}} \int_{X_2 \times S^3} F_5^\text{RR} = \frac1{6\pi G_\text{N}^{(5)} g N} \int_{S^3} \Bigl( \star_5\, \wh F^3 + \wh F^3 \wedge \wh A^1 \Bigr) = - \frac{4}{3N} \, Q_3 \in \bZ \;,
\ee
where $\wh F^3 = d\wh A^3$. According to (\ref{5d EOMs}) and using (\ref{Cijk}) and (\ref{5d metrics conifold}), the combination in parenthesis gives the Page charge $Q_3$, which is conserved and quantized. Taking the 3-sphere to spatial infinity, it coincides with the charge defined in (\ref{5d charge definition}).

%%%%%%%%%%%%%%%%%%%%%%%%%%%%%%%%%%%%%%%%%%%%%%%%%%
%%%%%%%%%%%%%%%%%%%%%%%%%%%%%%%%%%%%%%%%%%%%%%%%%%

\bibliographystyle{ytphys}
\bibliography{BHEntropy}
\end{document}